\documentclass[12pt]{article}
\usepackage{amsfonts,amsmath}

\makeatletter \@addtoreset{equation}{section} \makeatother

\newcommand{\be}{\begin{equation}}
\newcommand{\ee}{\end{equation}}
\newcommand{\bee}{\begin{eqnarray}}
\newcommand{\beee}{\begin{array}}
\newcommand{\eee}{\end{eqnarray}}
\newcommand{\eeee}{\end{array}}

\newcommand{\un}{{{n}}}
\newcommand{\um}{{{m}}}

\newcommand{\ga}{\alpha}
\newcommand{\pa}{{\ga^\prime}}
\newcommand{\pb}{{\gb^\prime}}
\newcommand{\pga}{{\gamma^\prime}}
\newcommand{\gb}{\beta}
\newcommand{\gga}{\gamma}
\newcommand{\gla}{\lambda}

\newcommand{\M}{{\cal M}}

\newcommand{\Hh}{{\cal H}}

\newcommand{\rhs}{{\it r.h.s.} }
\newcommand{\rhss}{{\it r.h.s.} }
\newcommand{\lhs}{{\it l.h.s.} }
\newcommand{\lhss}{{\it l.h.s.} }
\newcommand{\ie}{{\it i.e.,} }
\newcommand{\ls}{\!\!\!\!\!\!}

\newcommand{\gl}{\lambda}

\newcommand{\gvep}{\varepsilon}

\newcommand{\gs}{\sigma}

\newcommand{\go}{\omega}

\newcommand{\by}{{\bar{y}}}

\newcommand{\q}{\,,\qquad}

\newcommand{\mR}{{\mathbb R}}
\newcommand{\mC}{{{\mathbb C}}}

\newcommand{\bb}{{\overline{b}}}
\newcommand{\ba}{{\overline{a}}}
\newcommand{\bnu}{{\overline{\nu}}}
\newcommand{\nn}{\nonumber}

\newcommand{\half}{\frac{1}{2}}

\newcommand{\p}{\partial}
\newcommand{\D}{{\cal D}}
\newcommand{\f}{\frac}

\begin{document}

\vspace{1.7 cm}

\begin{center}
{\large\bf On Conformal, $SL(4,{{\mathbb R}})$ and $Sp(8,{\mathbb R})$ Symmetries
of\\ $4d$
\vspace{0.4 cm}
 Massless Fields}

\vspace{1 cm}

{\bf  M.A.~Vasiliev}\\
\vspace{0.5 cm}
{\it
 I.E. Tamm Department of Theoretical Physics, Lebedev Physical Institute,\\
Leninsky prospect 53, 119991, Moscow, Russia}

\end{center}

\vspace{0.4 cm}

\begin{abstract}
\noindent
The  $sp(8,{\mathbb R})$ invariant formulation of free
field equations of massless fields of all spins in $AdS_4$
available previously in terms of gauge invariant
field strengths is extended to gauge potentials. As a by-product,
free field equations for a massless gauge field are shown to
possess both $su(2,2)\sim o(4,2)$ and $sl(4,\mR)\sim o(3,3)$
symmetry. The proposed formulation is well-defined in the $AdS_4$
background but experiences certain degeneracy in the flat limit
that does not allow conformal invariant field equations for spin
$s>1$ gauge fields in Minkowski space. The basis model
involves the doubled set of fields of all spins. It is manifestly invariant
under $U(1)$ electric-magnetic duality extended to higher spins.
Reduction to a single massless field contains the equations
that relate its electric and magnetic potentials which are
mixed by  the conformal transformations for $s>1$.
We use the unfolded formulation approach recalled
in the paper with some emphasis on the role of
Chevalley-Eilenberg cohomology of a Lie algebra $g$ in
$g$-invariant field equations. This method
makes it easy to guess a form of the $4d$
 $sp(8,{\mathbb R})$ invariant massless field equations and
 then to extend them to the ten dimensional $sp(8,{\mathbb R})$
 invariant space-time. Dynamical content of the field
 equations is analyzed in terms of $\gs_-$ cohomology.

\end{abstract}

\newpage
\tableofcontents

\section{Introduction}
\label{conserv}

It was argued by Fronsdal  in the  pioneering
work \cite{F} that the tower of free higher-spin (HS)
 fields of all spins in four dimensions admits $sp(8,{\mathbb R})$
symmetry. The conformal
symmetry $su(2,2)$, that acts individually on
a massless field of a fixed spin, extends to $sp(8,{\mathbb R})$ that mixes
 states of different spins.
 Based on this observation Fronsdal addressed two
fundamental  questions:
what is a minimal space where $sp(8,{\mathbb R})$ is geometrically realized
and what are manifestly $sp(8,{\mathbb R})$ invariant field
equations that describe $4d$ massless fields of all spins?

He answered  the first question by showing that the
relevant space is ten dimensional with real
symmetric matrices $X^{AB}=X^{BA}$ as local coordinates \cite{F}
($A,B\ldots  =1,\ldots 4$ are $4d$ Majorana spinor indices).
This is Lagrangian Grassmannian
 $\M_4\sim Sp(8,{\mathbb R})/P$
where  $P$ is the  parabolic subgroup of $Sp(8,{\mathbb R})$
that results from crossing out the right node of
the Dynkin diagram of $sp(8,{\mathbb R})$
$
\begin{picture}(80,25)(0,3)
{\linethickness{.25mm}
\put(0,10){\line(1,0){60}}%
\put(60,9){\line(1,0){30}}%
\put(60,11){\line(1,0){30}}%
\put(83,6){{ \large $\mathbf \times$}}
\put(70,6){{ \large $\mathbf <$}}
 \put(-5,6.3)  {{  \bf $\bullet$}}
\put(25,6.3) {{ \bf $\bullet$}}
\put(55,6.3) {{ 
\bf $\bullet$}} }
\end{picture}
\qquad$.
Using two-component spinor notation, $A = (\ga,\pa )$,
$\ga,\gb\ldots =1,2$, $\pa,\pb\ldots =1,2$, the ten dimensional
matrix space extends Minkowski coordinates
 $x^{\alpha\alpha^\prime}$ { to}
$X^{AB}=(X^{\alpha\alpha^\prime},X^{\alpha\beta},
\overline{X}^{\alpha^\prime\beta^\prime})$
where $X^{\alpha\beta}$ and $
\overline{X}^{\alpha^\prime\beta^\prime}$ are additional six coordinates
that form an antisymmetric Lorentz tensor.
Note that the relevance of this space to the description
of massless fields was rediscovered by
Bandos, Lukierski and  Sorokin in \cite{BLS}.
More generally $\M_M$ denotes the Lagrangian Grassmannian
with local coordinates $X^{AB}=X^{BA}$, $A,B=1,\ldots M$ (in
this paper we do not distinguish between the parabolic space
and  its big cell $R^{\f{M(M+1)}{2}}$).

The form of the dynamical variables and field equations
in $\M_4$  was obtained later in \cite{BHS}
where it was shown that  the tower of all $4d$
massless integer and half-integer spins can be described,
respectively,  by a single scalar
$C(X)$ and spinor $C_A(X)$ in $\M_4$ that satisfy the field equations
\be
\label{kg}
\left (\f{\p^2}{\p X^{AB}\p X^{CD}}
-\f{\p^2}{\p X^{CB}\p X^{AD}}\right )C(X)=0\,,
\ee
\be
\label{dir}
\left (\f{\p}{\p X^{AB}}C_C(X)
-\f{\p}{\p X^{CB}}C_A(X)\right )=0\,.
\ee
These equations possess no gauge symmetry because the fields
$C(X)$ and $C_A(X)$ describe gauge invariant objects like
scalar (spin 0), Maxwell field strength (spin 1), Weyl tensor
(spin 2) and their HS generalizations.

The infinite towers of fields that appear in the $Sp(8,{\mathbb R})$
invariant consideration are precisely the HS multiplets of
the $4d$ nonlinear HS gauge theory \cite{con,more}. This is not  accidental because
the original argument of Fronsdal in favor of $Sp(8,{\mathbb R})$ was based on
the prominent Flato-Fronsdal theorem \cite{ FF} stating that
the  tensor product of two singletons, where $Sp(8,{\mathbb R})$ acts
in a natural
way, is equivalent to the set of
massless fields of all spins  as a $sp(4,{\mathbb R})\sim o(3,2)$-module. On the other hand, the states of the HS
gauge theories  can also be understood as resulting from
tensoring singletons \cite{KV}. (Recall that singletons are
 conformal scalar and  spinor fields in three dimensions.)

The fields $C(X)$ and  $C_A(X)$ can be interpreted as
``hyperfields'' in the ``hyperspace" $\M_4$ that
allow to describe all $4d$ massless fields at once.
$\M_4$
plays for  a HS multiplet a role analogous to that of superspace
for supersymmetric theories. The concise form of the equations (\ref{kg})
and (\ref{dir}) makes it tempting to look for
a formulation of the full nonlinear theory in this formalism.
Note that, as mentioned in \cite{BHS}, the $4d$ nonlinear HS
models of \cite{con,more} can indeed be interpreted as possessing
a spontaneously broken $Sp(8,{\mathbb R})$ symmetry.

Theories in $\M_M$
 have been studied in a number of papers from different
perspectives
\cite{Mar,IB,DV,PST,s3,GV,ZU,BPST,BBAST,EL,FE,FEL,west,EI}. In particular, an
attempt to formulate a nonlinear theory in this framework
 was   undertaken in \cite{BPST}. One reason why
it was hard to go beyond free theory is that $C(X)$ and $C_A(X)$ describe
gauge invariant curvature tensors. It was not clear so far how to
describe gauge field potentials like spin two metric tensor in the
$sp(8,{\mathbb R})$ covariant way and in $\M_4$. As  gauge
potentials play the key role in any nonlinear field theory including
Yang-Mills theory,
  Einstein gravity, supergravity
and nonlinear HS gauge theories \cite{BBB,BBD,BB,FV1,con,more,non,BBC}
(see also \cite{Gol,solv,SSS} for reviews of nonlinear HS
theories and more references), to proceed towards a nonlinear
HS theory in $\M_4$ one has to introduce the gauge
potentials in the $Sp(8,{\mathbb R})$ invariant framework.
This is the primary aim of this paper.

For the first sight, apart from being  interesting, the problem
may look unsolvable. Indeed, at any rate the project is to find an
$sp(8,{\mathbb R})$ invariant formulation of free massless fields described
in terms of gauge potentials. Since $sp(8,{\mathbb R})$ contains
the conformal symmetry
$su(2,2)$, this  should result in a conformal invariant
formulation of free massless field equations in terms of potentials.
There is a lot of studies of conformal field equations
in the literature starting from the seminal work of Dirac
\cite{dcon} (see, e.g., \cite{Marn,Siegel,Mets,conf1,FerFr,AM,STV} and references therein).
It is known however that the free field equations in terms of potentials
are not conformal for spins $s>1$ in flat space.
Actually, the complete list of free conformal invariant equations
in flat space available in \cite{STV} does not contain
the $4d $ massless equations in terms of gauge potentials except for
the $4d$ Maxwell equations, {\it i.e.,} spin one.

On the other hand this looks unnatural because
the equations in terms of gauge invariant field strengths are
conformal invariant and the space of states of
$4d$ massless equations admits the action of the conformal symmetry.
Aiming at preserving the $sp(8,{\mathbb R})$
symmetry, we have to find out what goes wrong with
conformal symmetry in terms of potentials.

The results of this paper  show that the $sp(8,{\mathbb R})$ covariant
formulation is consistent in the $AdS_4$ background but
experiences certain degeneracy in the flat limit to Minkowski
space, in which either the form of the equations breaks down
leading to (anti)selfdual equations (insisting on the symmetry) or
the symmetry transformation law on the fields in Minkowski space
blows up (insisting on the full massless equations).
The same happens with the conformal symmetry
that can be defined in $AdS_4$ but not in Minkowski space.
Note that this is not the
first time when the $AdS$ curvature  resolves a no-go
statement. Analogous phenomenon occurs for HS interactions
in HS gauge theories \cite{FV1}.

Our model exhibits manifest electric-magnetic (EM)
duality symmetry extended to all HS fields as a $u(1)$
subalgebra of $sp(8,{\mathbb R})$. A closely related
property is that it contains two sets of fields of all spins related
by the EM duality symmetry. This doubling
also plays a role in the conformal transformations that
mix the two species of gauge fields. The reduction to
the undoubled set of fields in which every massless
field of spin $s>0$ appears in one copy is also possible.
In this case, the field equations relate electric and
magnetic potentials of spins $s\geq 1$. An interesting feature
of this dynamical system is that the conformal transformations
mix electric and magnetic potentials for $s>1$.

Apart from the conformal embedding of the $AdS_4$ symmetry
$sp(4,{\mathbb R})\sim o(3,2)\subset su(2,2)\sim o(4,2)\subset sp(8)$ a
different embedding $sp(4,{\mathbb R})\sim o(3,2)\subset sl(4,{\mathbb R})
\sim o(3,3)\subset sp(8,{\mathbb R})$ exists. This simple observation
has a surprising output that the gauge theories in $AdS_4$
exhibit $sl(4,{\mathbb R})\sim o(3,3)$ symmetry at the free field
level even for a single massless field of a fixed spin.
This raises an intriguing question whether the $sl(4,{\mathbb R})$
extends to nonlinear $4d$ models, including both  HS theories
 and lower spin (super)gravity-like theories.

The organization of the rest of the paper is as follows.
The Sections \ref{mfe}-\ref{star}
remind the reader some known facts about HS field
equations and unfolded dynamics approach extensively used
throughout this
paper. Namely, in Section \ref{mfe}
we recall the $Sp(8,{\mathbb R})$ invariant formulation of the
dynamics of massless fields in terms of gauge invariant
field strengths. In Section \ref{hsads} the
unfolded formulation of the field equations of
HS gauge fields in $AdS_4$ is  summarized. The  flat limit that
reproduces the standard on-shell formulation of HS dynamics
in Minkowski space is discussed in Section \ref{hsmin}. In Section
\ref{Unfolded dynamics} we summarize relevant elements of the
unfolded dynamics approach with some emphasize on the role of
Chevalley-Eilenberg cohomology. General strategy of searching
unfolded formulation of a $g$--symmetric
field-theoretical model is outlined in Section \ref{GS}.
In Section \ref{Conformal geometries} we interpret Minkowski
and $AdS_d$ geometries in terms of flat connections of $o(d-1,2)$.
In Section \ref{Generalized conformal geometries} we
extend this analysis to $Sp(8,{\mathbb R})$,
focusing main attention on the group manifold $Sp(4,{\mathbb R})$ as a
ten dimensional generalization of $AdS_4$. In Section \ref{star}
we introduce the star-product formalism underlying the
unfolded formulation of the HS dynamics and recall the
pure gauge representation of the flat $Sp(2M)$ connection
found in \cite{DV}.

The original part of the paper
starts in Section \ref{sp8mod} where the Fock modules
appropriate for the $Sp(8,{\mathbb R})$ invariant description of HS
gauge fields are introduced. Conformal invariant
unfolded field equations for massless fields of all spins
are analyzed in Section \ref{conf}, where we prove the formal consistency
of the conformal field equations, analyze their dynamical content,
global symmetries and specificities of the flat limit.
In particular we show in this Section that the proposed
field equations are invariant under the EM duality
transformation and that the conformal symmetry cannot be
preserved in Minkowski space because the special conformal
part of the field transformation
blows up in the flat limit. In Section \ref{sp8} these
results are extended to $sp(8,{\mathbb R})$ while the
$gl(4,{\mathbb R})$ symmetry of the equations is considered in Section \ref{sl4}.
The detailed study of the dynamical content of the proposed equations
within the $\gs_-$ cohomology approach is done first
for the $4d$ case in Section \ref{ceq} and then for the case of
matrix space in Section \ref{s-}. Sections \ref{ceq} and \ref{s-}
can be skipped by the reader not interested in
details of the formalism. Conclusions and perspectives are
discussed in Section \ref{con}.

\section{$4d$ massless fields and $Sp(8,{\mathbb R})$ symmetry}
\label{mfe}

The key observation is \cite{4dun,Ann}
(see also \cite{Eis}) that the generating function
$$C(b|x)=
\sum_{k=0}^\infty \f{1}{k!} C_{A_1\ldots A_{k}}(x)
b^{A_1}\ldots b^{A_k}=
\sum_{n,m=0}^\infty \f{1}{n!m!}
C_{\ga_1\ldots \ga_{n}\,, \pa_1\ldots \pa_{m}}(x)
b^{\ga_1}\ldots b^{\ga_n}\overline{b}^{\pa_1}\ldots \overline{b}^{\pa_m}
$$
can be used to describe all $4d$ massless fields by virtue of the equation
\be \label{4deq} \left ( \f{\p}{\partial x^{\ga\pa}}  +
\frac{\partial^2}{\partial b^\ga{\partial \overline{b}^\pa}}\right
)C(b|x)=0 \,, \ee where $x^{\ga\pa}$ are Hermitian coordinates of
Minkowski space-time ($x^{\ga\pa} = \sigma^{\ga\pa}_n x^n$,
$n=0,1,2,3$, $\sigma^{\ga\pa}_n$ are four $2\times 2$ Hermitian
matrices) and $b^\ga, \bar{b}^\pa $ are auxiliary commuting spinor
variables. The interpretation of  the components $ C_{\ga_1\ldots
\ga_{n}\,, \pa_1\ldots \pa_{m}}(x)$ is as follows. Those  that carry
both primed and unprimed indices  ({\it i.e.,} $mn\neq 0$) are
expressed by  (\ref{4deq}) via space-time derivatives of the
holomorphic and antiholomorphic components that carry only primed or
only unprimed indices, respectively, \ie \be \label{aux}
 C_{\ga_1\ldots \ga_{m+k}\,, \pa_1\ldots \pa_{m}}(x) = (-1)^{m}
\p_{\ga_1\pa_1}\ldots \p_{\ga_m\pa_m}
C_{\ga_{m+1}\ldots \ga_{m+k}}(x)\q
\p_{\ga\pa} = \f{\p}{\p x^{\ga\pa}}\,,
\ee
where the indices $\ga_k$ and (independently) $\pa_k$
are symmetrized. The formula in the conjugated
antiholomorphic sector is analogous.

The holomorphic and antiholomorphic fields
describe a scalar field $C(x)$ for $s=0$, selfdual and
anti-selfdual components of the spin one Maxwell field strength,
$C_{\ga\gb}(x)$ and $C_{\pa,\pb}(x)$, selfdual and
anti-selfdual components of the Weyl tensor for spin two,
$C_{\ga_1\ldots \ga_4}(x)$ and $C_{\pa_1 \ldots \pa_4}(x)$,
and so on.
The system (\ref{4deq}) decomposes into an infinite set of
subsystems for the
fields of definite helicities according to
different eigenvalues of the helicity operator
$
\Hh = \half \left (b^\ga \f{\p}{\p b^\ga} - \bar{b}^\pa\f{\p}{\p \bar{b}^\pa}
\right ).
$

Apart from expressing auxiliary fields
in terms of space-time derivatives of the (anti)holomorphic fields
via (\ref{aux}), the equation (\ref{4deq}) imposes the massless
field equations on the latter
\be
\label{hol}
\f{\p}{\p b^{[\alpha}}\f{\p}{\p x^{\beta]\alpha^\prime}} C(b,0|x)=0
\,,\qquad \f{\p}{\p \bar{b}^{[\alpha^\prime}}\f{\p}{\p
x^{\alpha\beta^{\prime ]}}} C(0,\bar{b}|x)=0
\ee
{ and}
\be
\label{kg4}
\Box C(0,0|x)=0\,.
\ee
It is easy to check that (\ref{aux}),(\ref{hol}) and (\ref{kg4})
exhaust the content of (\ref{4deq}).

The equation (\ref{4deq}) can be interpreted as the covariant
constancy condition
\be
\label{cc}
D |C(b|x)\rangle =0
\ee
for the field
\be
\label{fc}
|C(b|x)\rangle= C(b|x)|0\rangle \,,\qquad
\ee
that takes values in the Fock module generated from the vacuum
state
\be
\label{fv}
 a_A |0\rangle =0\,
\ee
of the algebra of oscillators
\be
\label{osc}
 [a_A\,, b^B ]= \delta_A^B\q [a_A\,, a_B ]=0\q[b^A\,, b^B ]=0\,.
\ee
The generators of $sp(8,{\mathbb R})$ are realized in this module as
\be
\label{osp8osc}
P_{AB} = \half a_A a_B\,,\qquad
L_A{}^B = \half (a_A b^B +b^B a_A )\,,\qquad
K^{AB} = \half b^A b^B\,.
\ee
Note that $L_A{}^B$ form $gl(4|\mathbb{R})$.

The covariant derivative
\be
\label{dcart}
D=d + dx^{\ga\pa}P_{\ga\pa}
\ee
is a particular flat $sp(8,{\mathbb R})$ connection, i.e.
\be
\label{sp8c}
D=d+w\,,\qquad w= h^{AB} P_{AB} + \omega_B{}^A L_A{}^B
+ f_{AB} K^{AB}\q
D^2 =0\,.
\ee
The connection (\ref{dcart}) is flat
because the generators $P_{\ga\pa}$ commute to themselves.
This choice of the flat connection
corresponds to Cartesian coordinates in Minkowski space.

As explained in Subsection \ref{unf},
that the $4d$ massless equations (\ref{4deq})
have the form of
a covariant constancy condition with the covariant derivative
 in a $sp(8,{\mathbb R})$-module $V$ (here $V$ is the space of functions
of $b$) implies their invariance under the global
$sp(8,{\mathbb R})$ symmetry. The generators of
$sp(8,{\mathbb R})$ act on the dynamical
fields as differential operators with coefficients polynomial in
$x$ (see \cite{BHS} and Section \ref{star}
for explicit field transformations).

The conformal symmetry $su(2,2)$ extended to $u(2,2)$ by the
helicity generator
is the subalgebra of $sp(8,{\mathbb R})$ spanned by the generators
\be
\label{cgs}
P_{\ga\pb} = a_\ga \bar{a}_{\pb}\,,\qquad
 L_\ga{}^\gb = \half\{ a_\ga\,,b^\gb\}\,,\qquad
\bar{L}_\pa{}^\pb = \half\{ \bar{a}_\pa\,,\bar{b}^\pb\}\,,\qquad
K^{\ga\pb} = b^\ga \bar{b}^{\pb}\,
\ee
with the respective  gauge fields  $h^{\ga\pb}$,
$\omega_\gb{}^\ga{}$, $\overline{\omega}_\pb{}^\pa$, $f_{\ga\pb}$.
The dilatation and helicity generators $\D$ and $\Hh$ are
\be
\label{dil}
\D = \half\left ( L_\ga{}^\ga + L_\pa{}^\pa\right )\,,
\ee
\be
\label{hel}
\Hh =\half \left ( L_\ga{}^\ga - {L}_\pa{}^\pa\right )\,.
\ee

The helicity operator $\Hh$ is  central  in $u(2,2)$.
This is expected because it takes a fixed value on any
  $su(2,2)$-module with definite  helicity. $\Hh$ is the generator of
  EM duality transformations which is the manifest
  symmetry of this formulation \cite{BHS}.

The extension of the dynamical equations to $\M_4$
is achieved by replacing the unfolded equation (\ref{4deq})
with
\be
\label{10eq}
\left (\f{\p}{\p X^{AB}} + \f{\p^2}{\p b^A \p b^B}\right ) C(b|X) =0\,,
\ee
where $X^{AB}=X^{BA}$
are symmetric matrix coordinates associated with
the generalized momentum $P_{AB}$. This is the
equivalent extension of the unfolded $4d$ massless equations
to $\M_4$. Indeed, the part of the equations (\ref{10eq})
with $A=\ga$, $B=\pb$ and $X^{\ga\pb} =x^{\ga\pb}$ is just the equation (\ref{4deq}) while
 the equations that contain extra six coordinates
$X^{\ga\gb}$ and ${X}^{\pa\pb}$ reconstruct the dependence on
these coordinates in terms of the generating function of
$4d$ massless fields $ C(b|x)$. (See also Subsection \ref{prop}.)

On the other hand, one can interpret the equation (\ref{10eq})
differently by observing that they  express all components of
$C(b|X)$ that contain two or more oscillators $b^A$ via
derivatives over the hyperspace coordinates $X^{AB}$
 of the two dynamical fields which are polynomials
of zeroth and first degree of $b^A$. Thus $C(X) +C_A (X) b^A$
are dynamical fields in $\M_4$ while
all other components are auxiliary fields expressed by (\ref{10eq})
via $X$-derivatives of the dynamical fields.
Namely, for
\be\nn
C(b|X)=\sum_{n=0}^\infty \f{1}{n!} C_{A_1 \ldots A_n}(X) b^{A_1} \ldots
b^{A_n}
\ee
\be
\label{10expx}
C_{A_1 \ldots A_n}(X) =(-1)^{[\f{n}{2}]}\p_{A_1 A_2} \ldots
\p_{A_{2[\f{n}{2}]-1}A_{2[\f{n}{2}]}}
C_{A_{n-2[\f{n}{2}]}}(X)\q \p_{AB} = \f{\p}{\p X^{AB}}\,,
\ee
where $[\f{n}{2}]$ is the integer part of $\f{n}{2}$
and
$C_{A_{n-2[\f{n}{2}]}}(X)$ is either $C(X)$ or $C_A(X)$.

The equations (\ref{kg}) and (\ref{dir}) are consequences of
(\ref{10eq}). Eqs. (\ref{kg}), (\ref{dir})
and (\ref{10expx}) exhaust all restrictions on
$C(b|X)$ imposed by (\ref{10eq}).
This proves that the equations
(\ref{kg}) and (\ref{dir}) in $\M_4$ are equivalent to the
$4d$ massless field equations for all spins.
That all $4d$ massless fields are described
by only two hyperfields is because spin is carried by the spinning
coordinates $X^{\ga\gb}$ and ${X}^{\pa\pb}$
in the hyperspace $\M_4$
(see also \cite{BBAST}.)

\section{Higher spin gauge fields in $AdS_4$}
\label{hsads}
In this Section we recall the unfolded form of $4d$ free
 HS  field equations proposed in
\cite{4dun,Ann}. It is based on the frame-like approach to HS gauge
fields \cite{frame,Fort1} where a spin $s$ HS gauge field is
describ€ed by the set of 1-forms \be\nn \omega_{\ga_1\ldots
\ga_k,\pa_1\ldots\pa_l}= dx^n \omega_{n\,\ga_1\ldots
\ga_k,\pa_1\ldots\pa_l}\q k+l=2(s-1)\,. \ee
 The HS gauge fields are self-conjugated
$\overline{\omega_{\ga_1\ldots \ga_k\,,\pb_1\ldots\pb_l}}=
\omega_{\gb_1\ldots\gb_l\,,\pa_1\ldots \pa_k}.
$ This set is equivalent to the real
1-form $\omega_{A_1\ldots A_{2(s-1)}}$
symmetric in the  Majorana indices $A$, that
carries an irreducible module of the  $AdS_4$
symmetry algebra $sp(4,{\mathbb R})\sim o(3,2)$.

 The $AdS_4$ space is
described by the Lorentz connection $\omega^{\ga\gb}$,
$\overline{\omega}^{\pa\pb}$ and vierbein $e^{\ga\pa}$.
Altogether they form the $sp(4,{\mathbb R})$ connection $w^{AB}=w^{BA}$
that satisfies the $sp(4,{\mathbb R})$ zero curvature conditions
\be
\label{ads}
R^{AB}=0 \,,\qquad R^{AB} = dw^{AB} +w^{AC}\wedge w_C{}^B\,,
\ee
where indices are raised and lowered by a $sp(4,\mR)$
invariant form $C_{AB}=-C_{BA}$
\be
\label{Cind}
A_B =A^A C_{AB}\q A^A = C^{AB} A_B\q C_{AC}C^{BC} = \delta_A^B\,.
\ee
In terms of Lorentz components
$
w^{AB} = (\go^{\ga\gb}, \overline{\go}^{\pa\pb}, \gla e^{\ga\pb},
\gla e^{\gb\pa} )
$
where $\lambda^{-1}$ is the  $AdS_4$ radius,
the  $AdS_4$ equations (\ref{ads}) read as
\be
\label{adsfl}
R_{\ga\gb}=0\,,\quad \overline{R}_{\pa\pb}=0\,,
\quad R_{\ga\pa}=0\,,
\ee
where
\be
\label{nR}
R_{\alpha \gb}=d\omega_{\alpha \gb} +\omega_{\alpha}{}^\gamma
\wedge \omega_{\gb \gamma} +\lambda^2\, e_{\alpha}{}^{{\delta}^\prime}
\wedge
e_{\gb {\delta}^\prime}\,,
\ee
\be
\nn
\overline{R}_{{\pa} {\pb}}
=d\overline{\omega}_{{\pa}
{\pb}} +\overline{\omega}_{{\pa}}{}^{{\gamma}^\prime}
\wedge \overline{\omega}_{{\pb} {\gga}^\prime} +\lambda^2\,
e^\gamma{}_{{\pa}} \wedge e_{\gamma {\pb}}\,,
\ee
\begin{equation}
\label{nr}
R_{\alpha {\pb}} =de_{\alpha{\pb}} +\omega_\alpha{}^\gamma \wedge
e_{\gamma{\pb}} +\overline{\omega}_{{\pb}}{}^{{\delta}^\prime}
\wedge e_{\alpha{\delta}^\prime}\,.
\end{equation}
(Two-component indices are raised and lowered as in (\ref{Cind})
with $C_{AB}$ replaced by the two-component symplectic
forms $\epsilon_{\ga\gb}$ or $\epsilon_{\pa\pb}$.)

The unfolded equations of motion
of a spin-$s$ massless field read as
\be
\label{CON1}
R_{\ga_1\ldots \ga_n,\pa_1\ldots\pa_m} =
\delta^0_n \overline{H}^{\pa_{2s-1}\pa_{2s}}
\overline{C}_{\pa_1\ldots\pa_{2s}}
+\delta_m^0 {H}^{\ga_{2s-1}\ga_{2s}}{C}_{\ga_1\ldots\ga_{2s}}\,,\qquad
n+m=2(s-1)
\ee
and
\be
\label{CON2}
D^{tw}C_{\ga_1\ldots \ga_n,\pa_1\ldots\pa_m}=0\,,\qquad n-m = 2s\,,\qquad
D^{tw}
\overline{C}_{\ga_1\ldots \ga_n,\pa_1\ldots\pa_m}=0\,,\qquad m-n = 2s\,.
\ee
Here the HS field strength and twisted adjoint covariant
derivative have the form
\be\nn
R_{\ga_1\ldots \ga_n,\pa_1\ldots\pa_m}=
D^L\omega_{\ga_1\ldots \ga_n,\pa_1\ldots\pa_m}
+n\lambda e_{\ga_1}{}^{\pa_{m+1}}\wedge
\omega_{\ga_2\ldots \ga_n,\pa_1\ldots\pa_{m+1}}
+m \lambda e^{\ga_n+1}{}_{\pa_1}\wedge
\omega_{\ga_1\ldots \ga_{n+1},\pa_2\ldots\pa_{m}}\,,
\ee
\be
\label{bc2}
D^{tw}C_{\ga_1\ldots\ga_n\,,\pa_1\ldots \pa_m} =
 D^L{C}_{\ga_1\ldots\ga_n\,,\pa_1\ldots \pa_m}
+\lambda
(e^{\gamma{\delta}^\prime}{C}_{\ga_1\ldots\ga_n\gamma\,,
\pa_1\ldots \pa_m\delta^\prime}
+n m  e_{\alpha_1{\alpha}^\prime_1} {C}_{\alpha_2
\ldots \ga_n,\,{\alpha}^\prime_2 \ldots \pa_m}),
\ee
\be
\label{bc2b}
D^{tw}\overline{C}_{\ga_1\ldots\ga_n\,,\pa_1\ldots \pa_m} =
 D^L\overline{C}_{\ga_1\ldots\ga_n\,,\pa_1\ldots \pa_m}
+\lambda
(e^{\gamma{\delta}^\prime}\overline{C}_{\ga_1\ldots\ga_n\gamma\,,
\pa_1\ldots \pa_m\delta^\prime}
+n m  e_{\alpha_1{\alpha}^\prime_1} \overline{C}_{\alpha_2
\ldots \ga_n,\,{\alpha}^\prime_2 \ldots \pa_m})\,,
\ee
where the indices $\ga$ and $\pa$ are (separately) symmetrized and
$D^L$ is the Lorentz covariant derivative
\be
\label{lc}
D^L \psi_\ga = d\psi_\ga + \go_\ga{}^\gb \psi_\gb\q
D^L \overline{\psi}_\pa = d\overline{\psi}_\pa + \overline{\go}_\pa{}^\pb
\overline{\psi}_\pb\,.
\ee
$H^{\ga\gb}=H^{\gb\ga}$ and $\overline{H}^{\pa\pb} =
\overline{H}^{\pb\pa}$ are the basis 2-forms
built of the vierbein 1-form $e^{\ga\pa}$
\be
\label{H}
H^{\ga\gb} = e^{\ga}{}_\pa \wedge e^\gb{}^\pa\,,\qquad
\overline {H}^{\pa\pb} = e_{\ga}{}^\pa\wedge e^{\ga\pb}\,.
\ee

Formulae simplify  in terms of the generating
functions
\bee
\label{gg}
A (y,\bar{y}\mid x)
=\sum_{n,m=0}^{\infty}
\frac{1}{n!m!}
{y}_{\alpha_1}\ldots {y}_{\alpha_n}{\bar{y}}_{{\pb}_1}\ldots
{\bar{y}}_{{\pb}_m
} A{}^{\alpha_1\ldots\alpha_n}{}_,{}^{{\pb}_1
\ldots{\pb}_m}(x)\,
\eee
with $A=\omega, C, \overline{C}, R$ etc. In particular, we have
\be
\label{RRR}
R (y,\bar{y}| x) =D^{ad}\omega(y,{\bar{y}}|x) =
D^L \omega (y,\bar{y}| x) -
\lambda e^{\ga\pb}\Big (y_\ga \frac{\partial}{\partial \bar{y}^\pb}
+ \frac{\partial}{\partial {y}^\ga}\bar{y}_\pb\Big )
\omega (y,\bar{y} | x) \,,
\ee
\be
\label{tw}
D^{tw} C(y,{\bar{y}}|x) =
D^L C (y,{\bar{y}}|x) +\lambda e^{\ga\pb}
\Big (y_\ga \bar{y}_\pb +\frac{\partial^2}{\partial y^\ga
\partial \bar{y}^\pb}\Big ) C (y,{\bar{y}}|x)\,,
\ee
where  the Lorentz covariant derivative $D^L$ takes the form
\be
\label{dlor}
D^L A (y,{\bar{y}}|x) =
d A (y,{\bar{y}}|x) -
\Big (\go^{\ga\gb}y_\ga \frac{\partial}{\partial {y}^\gb} +
\overline{\go}^{\pa\pb}\bar{y}_\pa \frac{\partial}{\partial \bar{y}^\pb} \Big )
A (y,{\bar{y}}|x)\,.
\ee

As a consequence of the zero curvature equation (\ref{ads})
which is true for $AdS_4$ geometry,
the covariant derivatives $D^{ad}$ and $D^{tw}$ are flat, \ie
\be\nn
(D^{ad})^2=(D^{tw})^2 =0\,.
\ee
These conditions are necessary for the consistency of
the equations (\ref{CON1}) and (\ref{CON2})
(i.e., the compatibility with $d^2=0$) and guarantee the
gauge invariance of the field strength (\ref{RRR})
and, therefore, the free HS field equations (\ref{CON1})
under Abelian HS gauge transformations
\be
\label{hsgau}
\delta \omega(y,\bar{y}|x) = D^{ad} \epsilon(y,\bar{y}|x)\,.
\ee
It is important that the consistency of the equations is not
spoiled by the $C$-dependent
terms in (\ref{CON1}). As explained in more detail in Subsection
\ref{s-g}, this means that these terms correspond to a
Chevalley-Eilenberg cohomology of $sp(4,{\mathbb R})$.

In the equations (\ref{CON1}) and (\ref{CON2}), a spin $s$ field
is described by the set of gauge 1-forms
$\omega{}^{\alpha_1\ldots\alpha_n}{}_,{}^{{\pb}_1
\ldots{\pb}_m}(x)$ with $n+m=2(s-1)$ (for $s\geq 1$)
and  0-forms $C{}^{\alpha_1\ldots\alpha_n}{}_,{}^{{\pb}_1
\ldots{\pb}_m}(x)$ with $n-m =2s$ along with their conjugates
$\overline{C}{}^{\alpha_1\ldots\alpha_n}{}_,{}^{{\pb}_1
\ldots{\pb}_m}(x)$ with $m-n =2s$. Indeed it is easy to see
that the field equations (\ref{CON1}) and (\ref{CON2})
for such a set of fields with some $s$ form an independent
subsystem.

The dynamical massless fields are
\begin{itemize}
\item
$C(x)$ and $\overline{C}(x)$ for two spin zero fields,
\item
$C_\ga(x)$ and $\overline{C}_\pa (x)$ for a massless spin 1/2 field,
\item
$\omega_{\ga_1\ldots\ga_{s-1},\pa_1\ldots\pa_{s-1}}(x)$
for an integer spin $s\geq 1 $ massless field,
\item
 $\omega_{\ga_1\ldots\ga_{s-3/2},\pa_1\ldots\pa_{s-1/2}}(x)$
and its complex conjugate
 $\omega_{\ga_1\ldots\ga_{s-1/2},\pa_1\ldots\pa_{s-3/2}}(x)$
for a half-integer spin $s\geq 3/2$ massless field.
\end{itemize}
All other fields  are auxiliary, being
expressed via derivatives of the dynamical massless fields
by the equations  (\ref{CON1}) and (\ref{CON2}).

The key fact of the unfolded form of free massless field equations
is the so called  Central On-Shell Theorem \cite{Ann} that states
that the content of the equations (\ref{CON1}) and (\ref{CON2})
is just  that they express all auxiliary fields in terms
of derivatives of the dynamical fields and impose the massless
field equations on the latter in the standard form
of Fronsdal \cite{Frhs} and Fang and Fronsdal \cite{Frfhs}.
To make the paper as self-contained as possible we sketch the proof of
Central On-Shell Theorem in Section \ref{ceq} using the
$\sigma_- -$cohomology technics.

The meaning of the equations (\ref{CON1}) and (\ref{CON2})
is as follows. The equations (\ref{CON2}) provide the $AdS_4$
deformation of (\ref{4deq}). They remain
independent for spins $s=0$ and $s=\half$ and partially independent
for spin one but become consequences
of (\ref{CON1}) for  $s>1$.
The equations  (\ref{CON1}) express the holomorphic
and antiholomorphic components of spin $s\geq 1$ 0-forms
$C(y,\bar{y}|x)$ via derivatives of the massless field
gauge 1-forms described by $\go(y,\bar{y}|x)$. This identifies
the spin $s\geq 1$  holomorphic
and antiholomorphic components of the 0-forms
$C(y,\bar{y}|x)$ with the Maxwell tensor,
on-shell Rarita-Schwinger curvature, Weyl tensor and their HS
generalizations. In addition, the equations (\ref{CON1}) impose
the standard field equations on the spin $s>1$ massless gauge fields
so that the field equations (\ref{hol}) become their
consequences by virtue of Bianchi identities. The dynamical equations
for spins $s\leq 1$ are still contained in the equations (\ref{CON2}).
(For more detail
 see e.g. \cite{Gol} and also Section \ref{ceq}).

Although the system (\ref{CON1}) and (\ref{CON2})
is consistent at the free field level, to extend it to the
nonlinear case  one has to double the set of
HS fields \cite{KV,Ann,con,more}
\footnote{Note that, as discussed in \cite{KV,Ann}
(see also \cite{Gol}), the full
 $N=2$ supersymmetric nonlinear HS system  can be truncated to
subsystems with  reduced sets of fields. In particular,
truncating out fermions, it is possible to consider a system
with bosonic fields of all spins in which
 every integer spin appears once and
further to the minimal  system considered
in some detail in \cite{SSmin}, in which every even spin
appears just once.}.
This can be achieved by introducing the fields
\be\nn
\omega^{ii} (y,\bar{y}| x)\,,\qquad
C^{i 1-i} (y,\bar{y}| x)\,,\qquad i=0,1
\ee
such that $\omega^{ii} (y,\bar{y}| x)$ are selfconjugated,
while $C^{01} (y,\bar{y}| x)$ and
$C^{10} (y,\bar{y}\mid x)$ are conjugated to one another,
\be
\nn
\overline{\omega^{ii} (y,\bar{y}| x)}=\omega^{ii} (\bar{y},{y}| x)\q
\overline{C^{i\,1-i} (y,\bar{y}| x)} = C^{1-i\,i} (\bar{y},{y}| x)\,.
\ee
The unfolded system for the doubled set of fields is
 \be
\label{CON12}
R^{ii}(y,\overline{y}\mid x) =  \overline{H}^{\pa\pb}
\f{\p^2}{\p \overline{y}^{\pa} \p \overline{y}^{\pb}}
{C^{1-i\,i}}(0,\overline{y}\mid x) +  H^{\ga\gb}
\f{\p^2}{\p {y}^{\ga} \p {y}^{\gb}}
{C^{i\,1-i}}(y,0\mid x)\,,
\ee
\be
\label{CON22}
D^{tw}C^{i \,1-i}(y,\overline{y}\mid x) =0\,.
\ee
Note that now all components of the expansions of
$C^{i\,1-i} (y,\bar{y}\mid x)$ contribute to the
equations (\ref{CON12}) and (\ref{CON22}),
 while in (\ref{CON1}) and (\ref{CON2})
with the single HS 1-form $\go(y,\by)$
only parts of the components of $C(y,\by)$ and
$\overline{C}(y,\by)$ contributed.

In the standard formulation of the $4d$
nonlinear HS gauge theory \cite{more} (see \cite{Gol} for
a review)
the doubling of the fields is due to the dependence on the
Klein operators $k$ and $\bar{k}$ that flip chirality \cite{FVau}
\be\nn
k^2=1\,,\qquad  k y_\ga= - y_\ga k \,,\qquad
k \overline{y}_\pa =
 \overline{y}_\pa k\,,
\ee
\be\nn
\overline{k}^2=1\,,\qquad  \overline{k} y_\ga= y_\ga
\overline{k} \,,\qquad
\overline{k} \overline{y}_\pa =
 -\overline{y}_\pa \overline{k}\q [k\,,\overline{k}]=0  \,.
\ee
The fields are 1-forms
\be\nn
\omega(k,\overline{k};y,\overline{y}\mid x)=\sum_{ij = 0,1}
(k)^i (\overline{k})^j \omega^{ij}(y,\overline{y}\mid x)\,
\ee
and 0-forms
\be\nn
C(k,\overline{k};y,\overline{y}\mid x)=\sum_{ij = 0,1}
(k)^i (\overline{k})^j C^{ij}(y,\overline{y}\mid x)\,.
\ee
Now both the adjoint and twisted adjoint covariant derivative
result from different sectors of the adjoint
covariant derivative in the Weyl algebra extended by the
Klein operators \cite{FVau}.

Massless fields are those with
\be\nn
\omega(-k,-\overline{k};y,\overline{y}\mid x)=
\omega(k,\overline{k};y,\overline{y}\mid x)\,,\qquad
C(-k,-\overline{k};y,\overline{y}\mid x)=-
C(k,\overline{k};y,\overline{y}\mid x)\,.
\ee
The fields with the opposite oddness in the Klein operators
are topological,
carrying at most a finite number of
degrees of freedom per an irreducible  subsystem
\cite{aux}.
We will see in Section \ref{conf} how this pattern of HS fields emerges in
the $sp(8,{\mathbb R})$ invariant formulation.
In particular the topological
field sector also plays a role in the
model we focus on in this paper.

\section{Flat limit}
\label{hsmin}

To take the flat limit it is necessary to perform certain rescalings.
To this end let us introduce notations
$A_\pm$ and $A_0$ so that the spectrum of the operator
$
 \left (y^\ga\f{\p}{\p y^\ga} - \overline{y}^\pa
\f{\p}{\p \overline{y}^\pa}\right )
$
is positive on  $A_+(y,\overline{y}\mid x) $,
negative on  $A_-(y,\overline{y}\mid x) $ and zero on
$A_0(y,\overline{y}\mid x)$.
Having the decomposition
\be
\label{dec}
A(y,\overline{y}\mid x) =
A_+(y,\overline{y}\mid x) + A_-(y,\overline{y}\mid x)
+A_0(y,\overline{y}\mid x)\,,
\ee
we introduce a new field
\be
\label{resc}
\tilde{A}(y,\overline{y}\mid x)=
A_+(\lambda y,\overline{y}\mid x) + A_-(y,\lambda \overline{y}\mid x)
+A_0(\lambda y,\overline{y}\mid x)\,.
 \ee
(Note that $A_0(\lambda y,\overline{y}\mid x)=
A_0(y,\lambda \overline{y}\mid x)$).
For the rescaled variables, the flat limit $\lambda \to 0$ of
the adjoint and twisted adjoint covariant derivatives (\ref{RRR})
and (\ref{tw}) gives
\be
\label{adfl}
D_{fl}^{ad}\tilde{A}(y,\bar{y} \mid x)
= D^L \tilde{A} (y,\bar{y} \mid x) -
 e^{\ga\pb}\Big (y_\ga \frac{\partial}{\partial \bar{y}^\pb}
\tilde{A}_-(y,\bar{y} \mid x)
+ \frac{\partial}{\partial {y}^\ga}\bar{y}_\pb \tilde{A}_+(y,\bar{y} \mid x)
\Big ) \,,
\ee
\be
D_{fl}^{tw} \tilde{A}(y,\bar{y} \mid x) =
D^L \tilde{A}(y,\bar{y} \mid x) + e^{\ga\pb}
\frac{\partial^2}{\partial y^\ga\partial \bar{y}^\pb}\tilde{A}(y,\bar{y} \mid x)\,.
\ee

The flat limit of the unfolded massless equations results from
(\ref{CON12}) and (\ref{CON22}) via the substitution of
$D^L$ and $e$ of Minkowski space and the
replacement of $D^{ad}$ and $D^{tw}$
by $D^{ad}_{fl}$ and $D_{fl}^{tw}$, respectively (recall that
$R^{ii}= D^{ad} \go^{ii}$). The resulting field equations
describe free HS fields in Minkowski space. Let us stress that
the flat limit prescription (\ref{resc}),
that may look somewhat unnatural
in the two-component spinor notation,
is designed just
to give rise to the theory of
Fronsdal \cite{Frhs} and Fang and Fronsdal \cite{Frfhs}
(for more detail see Section \ref{ceq}).

Note that, although  the contraction $\lambda\to 0$
with the rescaling (\ref{resc}) is consistent with the free
HS field equations, it turns out to be inconsistent
in the nonlinear HS theory because negative powers
of $\lambda$ survive in the full nonlinear equations
upon the rescaling (\ref{resc}), not allowing the
flat limit in the nonlinear theory. This is why
the Minkowski background is unreachable in the
non-linear HS gauge theories of \cite{FV1,con,non}.
The reason why conformal symmetry
blows up in the flat limit is that the translation generators
in the sector of HS gauge fields
$\go$, that respect the described limiting procedure,
disagree with their standard identification
in the conformal algebra (for more detail see Subsection \ref{gsfl}).

The flat limit of the equation $D^{tw}C =0$
just reproduces the equation (\ref{4deq}) which underlies the
original extension of the HS dynamics from four to ten dimensions.
The main question addressed in this paper
is how to include the HS gauge 1-forms into the manifestly
$sp(8,{\mathbb R})$ invariant formalism and then into the
ten dimensional formulation. This is most naturally
achieved within the unfolded formulation.
In fact, the 4d equations (\ref{CON12}) and (\ref{CON22})
do have the unfolded form. To proceed,
we now summarize  relevant properties of the unfolded
formulation  using the $4d$ HS system as the basis example.

\section{Unfolded dynamics}
\label{Unfolded dynamics}
\subsection{Unfolded equations}
\label{Unfolded equationss}

Let $M^d$ be a $d$-dimensional manifold
with coordinates $x^\un$ ($\un = 0,1,\ldots d-1$).
(For $d=4$ we use the Hermitian coordinates
$x^{\ga\pa}$.)
By unfolded formulation of a linear or nonlinear
system of differential equations and/or constraints
\footnote{\label{foot}By
{\it constraints} we mean equations like $dA=B$ that  express
{\it auxiliary fields} like $B$ in terms of derivatives of other fields
like $A$ imposing no differential equations on the latter.}
in $M^d$ we mean its
equivalent reformulation in the first-order form\footnote{The idea
of this approach was suggested and applied to the analysis
of interacting HS gauge theory in \cite{4dun,Ann} while the name
{\it unfolded formulation} was given somewhat later in \cite{un}.}
\be
\label{unf} dW^\Phi (x)= G^\Phi (W(x))\,,
\ee
where $ d=dx^\un  \frac{\p}{\p x^\un}\, $
is the exterior differential on $M^d$, $W^\Phi(x)$
is a set of degree $p_\Phi$-differential forms
and $G^\Phi (W)$ is some degree $p_\Phi +1$
function of the differential forms $W^\Phi$
\be\nn
G^\Phi (W) =
\sum_{n=1}^\infty f^\Phi{}_{\Omega_1\ldots \Omega_n} W^{\Omega_1}\wedge \ldots
\wedge W^{\Omega_n}\,,
\ee
where the coefficients $f^\Phi{}_{\Omega_1\ldots
\Omega_n}$ satisfy the (anti)symmetry condition
$
f^\Phi{}_{\Omega_1\ldots\Omega_k\Omega_{k+1} \ldots \Omega_n} =
(-1)^{p_{\Omega_{k+1}}p_{\Omega_k}}
f^\Phi{}_{\Omega_1\ldots\Omega_{k+1}\Omega_k \ldots  \Omega_n} \,
$
(extension to the supersymmetric case
with an additional boson-fermion grading is straightforward)
and $G^\Phi$ satisfies the  condition
\be \label{BI} G^\Omega (W)\wedge \f{\p
G^\Phi (W)} {\p W^\Omega}  =0\,
\ee
(the derivative $\frac{\p}{\p W^\Omega}$ is left) equivalent to the
 generalized Jacobi identity on the structure coefficients
\be \label{jid}
\sum_{n=0}^{m} (n+1) f^\Lambda{}_{[\Phi_1 \ldots
\Phi_{m-n}}  f^\Phi{}_{\Lambda\Phi_{m-n+1} \ldots \Phi_m\}} =0\,,
\ee
where
the brackets $[\,\}$ denote appropriate (anti)symmetrization of
indices $\Phi_i$. Strictly speaking, formal consistency
demands (\ref{jid}) to be satisfied only at $p_{\Phi} < d$ for
a $d$-dimensional manifold ${\cal M}^d$ where any
$d+1$-form is zero. Given solution of (\ref{jid}) it defines a
 free differential algebra (FDA) \cite{FDA1,FDA2,FDA3,FDA4}.
We  call a free differential algebra {\it
universal} if the generalized Jacobi identity is true
independently of $d$. The HS FDAs that appear
in  HS gauge theories and, in particular, those
discussed in this paper belong to the universal class.
Unfolded formulation is a covariant multidimensional extension
of the first-order reformulation of ordinary differential
equations.

Universal FDAs have the distinguished property
that the operator $\frac{\p }{\p W^\Omega}$
is well-defined irrespectively of on what it acts,
i.e. $\frac{\p F (W) }{\p W^\Omega}$ is defined for any
$F(W)$ built of wedge products of differential forms.
For non-universal FDAs this is not true. Actually,
the condition
\be
\label{Wc}
W_1 \wedge \ldots\wedge W_k =0\q p_1+\ldots p_k >d
\ee
may lead to a contradiction upon formal differentiation.
For instance, differentiating (\ref{Wc}) over all $W_i$ involved
leads to the contradiction $1=0$. In other words, for
nonuniversal FDAs the space of $W^\Omega$ is constrained
by the relations (\ref{Wc}). Correspondingly, only vector fields
tangent to the constraint surface are allowed.

For universal FDAs,
the equation (\ref{unf}) is invariant under the gauge transformation
\be \label{delw} \delta W^\Phi (x)= d \varepsilon^\Phi (x) +
\varepsilon^\Omega (x)
\frac{\p G^\Phi (W(x)) }{\p W^\Omega (x)}\,,
\ee
where the gauge parameter
$\varepsilon^\Phi (x) $ is an arbitrary  $(p_\Phi -1)$-form.  ($0$-forms
$W^\Phi (x)$ do not give rise to gauge symmetries.)
This property of universal FDAs makes the unfolded formulation
an efficient tool for the study of gauge invariant dynamical systems.
Since unfolded equations are formulated in terms of the exterior
algebra, this approach respects diffeomorphisms thus providing a
natural framework for the study of models that contain gravity.

\subsection{Vacuum}
\label{v}

An important class of universal FDAs is in the one-to-one correspondence with
  Lie algebras. Indeed, let
 $w^\alpha$  be a set of  1-forms.
If no other forms are involved (e.g., all of them are consistently
set equal to zero in a larger system) the most general expression for
$  G^\alpha (\,w)$, that has to be a 2-form,
is
$
G^\alpha (\,w)=-\half f^\alpha_{\beta\gamma}w^\beta\wedge w^\gga\,.
$
The consistency condition  (\ref{jid}) then becomes the
 Jacobi identity for the structure coefficients
$f^\alpha_{\beta\gamma}$ of a  Lie algebra $g$.
The unfolded equations (\ref{unf}) impose the
flatness condition on the connection $w^\ga$
\be
\label{MC}
 d w^\alpha +\half
f^\alpha_{\beta\gamma}w^\beta\wedge w^\gamma=0\,.
\ee

The transformation law  (\ref{delw}) gives
the usual gauge transformation of the connection $w$
\be
\label{gw}
\delta w^\ga (x) = D\gvep^\ga (x)=
 d\gvep^\alpha (x)+f^\alpha_{\beta\gamma}w^\beta(x) \gvep^\gamma (x)\,.
\ee

A flat connection $w(x)$  is invariant under
the global transformations with the covariantly constant parameters
\be
\label{glpar}
D\gvep^\alpha (x)=0\,.
\ee
This equation is consistent by
 (\ref{MC}). Therefore,  locally, it reconstructs
$\gvep^\alpha (x)$ in terms of its values
$\gvep^\alpha (x_0)$ at any given point $x_0$.
$\gvep^\alpha (x_0)$ are the moduli of the global symmetry
$g$ that is now recognized as the stability
algebra of a given flat connection $w(x)$.

This example is of key importance because this is how
$g$-invariant vacuum fields appear in the unfolded formulation.
In particular, the equation (\ref{ads})
of $AdS_4$ space-time is of this type. The same happens for the
general case. Typically, an unfolded system that contains
1-forms $w^\ga$ associated with some Lie algebra $g$ admits
a flat connection $w^\ga$ as its natural
$g$-symmetric vacuum solution.
In the perturbative analysis, $w^\ga$ is assumed to be of
the zeroth order because it contains the background metric that
is usually non-degenerate, being of order zero. The flat connection
$w^\ga$ is then referred to as {\it vacuum connection}.

Let us stress that this way of description of the
background geometry is coordinate independent.
A particular form of $w^\ga(x)$ is not needed
for the analysis unless one is
interested in explicit solutions in a specific
coordinate system. The only important condition is that
a flat vacuum connection
has to be chosen so that its part associated
with the generators of translations in $g$
(\ie vielbein sometimes
also called  soldering form, which is the 1-form $e^{\ga\pa}$
in the discussion of Section \ref{hsads}) is nondegenerate.
The ambiguity of the choice of a particular vacuum  form $w^\ga$
is up to the gauge transformations (\ref{gw}). In particular,
the coordinate choice ambiguity, which of course preserves
the flatness property, is reproduced by  the gauge transformations
(\ref{gw}). In fact, this is the general property of the unfolded
dynamics where a diffeomorphism generated by an infinitesimal vector
field $\xi^\un (x)$ can be realized as the field-dependent gauge
transformation (\ref{delw}) with the gauge parameter of the form
\be
\varepsilon^\Omega(x) = \xi^\un (x) \f{\p}{\p dx^\un} W^\Omega (dx,x)\,.
\ee

\subsection{Free fields and Chevalley-Eilenberg cohomology }
\label{ff}

Let us now linearize the unfolded equation (\ref{unf})
around some vacuum flat connection $w$ of a Lie algebra
$g$, that solves (\ref{unf}). To this end we set
\be
\label{lin1}
W^\Omega=w^\Omega+\omega^\Omega\,,
\ee
where $\go^\Omega$ are differential forms  of various
degrees that are treated as small perturbations and enter the
equations linearly.
Consider first the sector of forms $\go^i(x)$ of a given degree
$p$ ({\it e.g.}, $0$-forms)
within the set  $\go^\Omega(x)$. Then  $G^i$ is bilinear in $w$
and $\go$, \ie $
 G^i =- w^\ga(T_\ga)^i {}_j \wedge \go^j.
$
 In this case the condition
(\ref{BI}) implies that the matrices $(T_\ga)^i {}_j$ form a
representation $T$ of $g$ in a vector
space $V$ where $\go^i(x)$ takes its values (index $i$).
The corresponding equation (\ref{unf}) is the
covariant constancy condition
\be \label{covc} D_w \go^i=0
\ee
 with $D_w\equiv d+w$ being the covariant derivative
in the $g$-module $V$.

The equations (\ref{MC}) and (\ref{covc}) are invariant under the
gauge transformations (\ref{delw}) with
\be
\label{delc}
\delta \go^i (x)= -\gvep^\ga(x)(T_\ga)^i {}_j  \go^j(x)\,.
\ee
Once the vacuum connection is fixed, the system (\ref{covc})
is invariant under the global symmetry $g$ with the parameters
satisfying (\ref{glpar}).
This simple analysis allows useful applications.

Firstly, one observes that,  by unfolding, any $g$-invariant
linear dynamical system turns out to be reformulated in terms of
$g$-modules. This allows for full classification
and explicit derivation of
$g$-invariant equations by studying  various $g$-modules.
In particular, the full list of conformal invariant
equations in flat space-time of any dimension has been
obtained this way in \cite{STV}.
In this paper we will derive the manifestly $sp(8,{\mathbb R})$
covariant form of the massless field equations in terms of gauge
potentials just by guessing appropriate $sp(8,{\mathbb R})$-modules.

Secondly, it follows that if $\tilde{g}$  is a larger Lie
algebra that acts in $V$, $g\subset \tilde{g}\subset End\,\,V$,
 it is also a symmetry of (\ref{covc}) simply because any
flat $g-$connection is the same time a flat $\tilde{g}-$connection.
 As a result,
$\tilde{g}=End\,\, V$  is the maximal symmetry of (\ref{covc})
(of course, modulo possible subtleties in the  infinite dimensional case).

Thirdly, this gives an efficient tool for the
derivation of the explicit form of the symmetry transformation laws
via (\ref{glpar}) and (\ref{delc}).

{}For example, once the equation (\ref{4deq}) is
reunderstood  in the form (\ref{cc}), (\ref{dcart}), (\ref{sp8c}),
from the general analysis it follows that it is $sp(8,{\mathbb R})$
invariant. To derive the explicit form of the
global $sp(8,\mR)$ transformation we solve the covariant constancy
condition for the global symmetry parameter $\epsilon(x)$
\be
\ls
\epsilon_{gl}(a,b|x) =
\exp- [x^{\ga\pa} a_\ga a_\pa ] \epsilon_0 (a,b)
 \exp [x^{\ga\pa} a_\ga a_\pa ]\,,\quad \epsilon_0 (a,b) =
\half \epsilon_{AB} b^A b^B + \epsilon_{A}{}^B  b^A a_B
+\half \epsilon^{AB} a_A a_B\,,\nn
\ee
where $b^A=(b^\ga, \bar{b}^\pa )$ and  $a_A=(a_\ga, \bar{a}_\pa )$.
This gives
$$
\epsilon_{gl}(a,b|x) =\half \epsilon_{AB} b^A b^B +
\epsilon_{A}{}^B  b^A a_B+ \half \epsilon^{AB} a_A a_B
- x^{\ga\pa}\epsilon_{0\,\pa}{}^B a_\ga a_B
- x^{\ga\pa}\epsilon_{0\,\ga}{}^B a_\pa a_B
$$
$$
+x^{\ga\pa}\epsilon_{\ga\pa}
- b^D a_\ga \epsilon_{\pa D}  x^{\ga\pa}
- b^D a_\pa \epsilon_{\ga D}  x^{\ga\pa}\,.
$$
The desired transformation law is then obtained by
restricting the variation
\be
\label{trc}
\delta |C(b|x)\rangle =  - \epsilon_{gl}(a,b|x) |C(b|x)\rangle
\ee
to the dynamical holomorphic fields $C(b,0|x)$ or
$C(0,\bar{b}|x)$ and using (\ref{aux}) for the auxiliary fields
that appear on the \rhs of (\ref{trc}) to derive
terms with $x$--derivatives of the dynamical fields in the
transformation law. Note that this way it is possible to derive
explicit form of the transformation law of the full HS
algebra in a far more complicated
$AdS$ case (for more detail see \cite{BHS,DV} and
Section \ref{star}).

Suppose now that $\go^a(x)$ and $\go^i(x)$  are forms  of
different degrees, say, $p_a -p_i=k>0 $.
 Then, in the linearized approximation, one can consider
  functions $G^\Phi$  polylinear in the vacuum field
  $w^\ga$ but still linear in the dynamical fields $\go$
\be
\label{co}
 G^a (w, \go) =- f^a_{\ga_1\ldots \ga_{k+1}\,i}
 w^{\ga_1}\wedge\ldots \wedge w^{\ga_{k+1}}\go^i\,.
\ee
(Note that the case of $k=-1$ with
$G^a(w,\go)$ (\ref{co}) independent of $w^\ga$
is also possible. It
corresponds to the so-called contractible FDAs \cite{FDA1} and
is dynamically empty because the corresponding
unfolded equation just expresses a $p_i$-form $\go^i$ via
the lower degree forms and their derivatives.
As such it  is  not  considered
in this paper.)

Let $\go^i$ be a 0-form. The equation for
$\go^i$ is then always a covariant constancy condition (\ref{covc}).
The consistency condition (\ref{BI}) applied to (\ref{co})
then literally implies that
$ f^a_{\ga_1\ldots \ga_{k+1}\,i}w^{\ga_1}\wedge \ldots
\wedge w^{\ga_{k+1}} $ is a Chevalley-Eilenberg
cocycle of $g$ with coefficients in $V_l \otimes V^*_r$
where $V_l$ is the module where $G^a $ takes values
while $V^*_r$ is the module conjugated to that of $\go^i$.
Coboundaries are dynamically empty because, as is easy to
 see, they can be removed by a field redefinition.
Thus, in the unfolded formulation,
the Chevalley-Eilenberg cohomology
classifies possible nontrivial mixings of higher form fields
with 0-forms. This type of mixing is of most importance in
the context of known HS theories.

More generally, one can imagine the equations with the terms
of the type of (\ref{co}) that involve $p>0$ forms on the
\rhs The consistency
condition then is
that the cohomology algebra with respect to the natural product
of the elements $f(V_1, V_2^* )$ and  $\tilde{f}(\tilde{V}_1,
\tilde{V}_2^* )$ with $V_2 = \tilde{V}_1$ is nilpotent
in the  sense  that (\ref{BI}) is true.

The free HS system (\ref{ads}), (\ref{CON1}) and (\ref{CON2})
has unfolded form. It is consistent in the sense of (\ref{BI}).
 The terms on the \rhs
 of the equations  (\ref{CON1}) describe
the Chevalley-Eilenberg cohomology of $sp(4,{\mathbb R})$
with coefficients in the corresponding infinite dimensional
modules. Let us note that without these cohomological terms,
\ie relaxing the \rhs of the equation (\ref{CON1}), the sector
of 1-forms would become dynamically trivial (any solution of the
zero curvature equation is pure gauge in the topologically trivial
situation). The gluing with 0-forms via (\ref{CON1}) and (\ref{CON2})
makes the 1-form gauge potential dynamically
nontrivial and, the same time, expresses the 0-forms in terms of
derivatives of the gauge potentials (except for spin zero and spin
1/2 fields that have no associated gauge potentials
because of the second derivatives over $y$ and $\bar{y}$
on the \rhs of (\ref{CON12})).

Note that the statement that $g$ extends to a
larger symmetry algebra $\tilde{g}$ that acts in the $g$-modules $\go^i$,
which is true in the 0-form sector, may not be true in
the other sectors in presence of the cohomological
terms because it is not {\it a priory} guaranteed that a $g$-cohomology
extends to a $\tilde{g}$-cohomology. To large extent our analysis
of the $sp(8,{\mathbb R})$ symmetry in Section \ref{sp8} amounts to the
analysis  of the extension of the $sp(4,{\mathbb R})$ HS cohomology
to $sp(8,{\mathbb R})$.

It is important that in presence of the
cohomological terms (\ref{co}) the system remains
invariant under the global symmetry $g$. Indeed,
the system (\ref{MC}) along with
\be
\label{lin}
d\go^\Omega = G^\Omega{}_\Phi (w) \go^\Phi\,
\ee
is formally consistent and therefore gauge invariant.
The global symmetry $g$ is still the part of the gauge
symmetry that leaves invariant the vacuum fields $w^\ga$.
Its action on $\go^\Phi$ is deformed however by the cohomological terms
(\ref{co}) according to (\ref{delw}). In addition the system
is invariant under the Abelian gauge transformations (\ref{delw})
associated with $p>0$ form fields among $\go^\Phi$.

For example, the system (\ref{ads}), (\ref{CON1}) and (\ref{CON2})
is invariant under the
HS gauge transformations (\ref{hsgau})
associated with the 1-form connections
$\omega(y,\bar{y}|x)$, which are Abelian gauge transformations
for free massless fields of spins $s\geq 1$ whose form depends
on the vacuum $AdS$ fields $w^{AB}= (\omega^{\ga\gb},
\overline{\omega}^{\pa\pb}, e^{\ga\pb})$ that enter via
the flat covariant derivative $D^{ad}$.
Also the system (\ref{ads}), (\ref{CON1}) and (\ref{CON2}) is
invariant under the global $sp(4,{\mathbb R})$ symmetry
that leaves invariant the $AdS_4$ vacuum fields $w^{AB}$.
The formula (\ref{delw}) applied to the equation (\ref{CON1})
gives the $sp(4,{\mathbb R})$ transformation law
\bee\label{dego}\nn
\ls\delta^{gl}\omega(y,\bar{y}|x)&{}& \ls=
\Big (\epsilon^{\ga\gb}(x)y_\ga \frac{\partial}{\partial {y}^\gb} +
\overline{\epsilon}^{\pa\pb}(x)
\bar{y}_\pa \frac{\partial}{\partial \bar{y}^\pb} +
\lambda \epsilon^{\ga\pb}(x)\Big (y_\ga \frac{\partial}{\partial \bar{y}^\pb}
+ \frac{\partial}{\partial {y}^\ga}\bar{y}_\pb\Big )\Big )
\omega (y,\bar{y} | x)\\
&{}& \ls+2 \epsilon_{\ga}{}^\pa (x) e^{\ga\pb}
\f{\p^2}{\p \overline{y}^{\pa} \p \overline{y}^{\pb}}
{C^{1-i\,i}}(0,\overline{y}|x) +  2 \epsilon^{\ga}{}_\pa (x)
 e^\gb{}^\pa \f{\p^2}{\p {y}^{\ga} \p {y}^{\gb}}
{C^{i\,1-i}}(y,0|x)\,,
\eee
\be
\label{de0}
\!\!\!\delta^{gl} C(y,{\bar{y}}|x) =
\Big (\epsilon^{\ga\gb}(x)y_\ga \frac{\partial}{\partial {y}^\gb} +
\overline{\epsilon}^{\pa\pb}(x)
\bar{y}_\pa \frac{\partial}{\partial \bar{y}^\pb}
+\lambda \epsilon^{\ga\pb}(x)
\Big (y_\ga \bar{y}_\pb +\frac{\partial^2}{\partial y^\ga
\partial \bar{y}^\pb}\Big )\Big ) C (y,{\bar{y}}|x)\,,
\ee
where the parameters of global
$sp(4,\mR)$ transformations  $\epsilon^{\ga\gb} (x)$,
$\epsilon^{\pa\pb} (x)$ and $\epsilon^{\ga\pa} (x)$
 satisfy the $sp(4,\mR)$
covariant constancy conditions (\ref{glpar}).
The $C$-dependent terms in (\ref{dego})
generalize to HS fields  the
well-known formula that represents diffeomorphisms in gravity
as the deformation of the $o(d-1,2)$ (or Poincare`)
gauge transformations by the Riemann tensor dependent term
(see e.g. \cite{PVN}).
Thus, the unfolded form of field equations
reproduces necessary deformation terms automatically via
the Chevalley-Eilenberg cohomology deformation.

Finally, let us note that  the
condition that the mixture of differential forms via
Chevalley-Eilenberg cohomology is perturbatively nontrivial,
\ie that the terms (\ref{co}) do not accidentally vanish
because of unlucky choice of the vacuum connection $w$,
may impose an additional restriction on the latter. We shall see
below that it is this condition that gives preference
to $AdS_4$ geometry in the models of interest because
the relevant terms (\ref{co}) turn out to be nondegenerate
in the $AdS_4$ case but trivialize in the Minkowski case.

\subsection{Dynamical content via $\sigma_-$ cohomology}
\label{s-g}

In the unfolded dynamics approach, {\it dynamical
fields} ({\it i.e.,} those that are neither pure gauge nor
{auxiliary} (see footnote \ref{foot})), their {\it differential gauge symmetries}
({\it i.e.,} those that are not Stueckelberg (=shift) symmetries)
and {\it differential field equations} ({\it i.e.,} those that are
not constraints), are characterized by the so-called $\sigma_-$
cohomology \cite{sigma} (see also \cite{BHS,solv}).
The aim of this Subsection is to recall briefly
the main idea of this method to make it possible
to explain in Section \ref{GS} the general strategy of the
search and investigation of the $sp(8,\mR)$ invariant equations.
To keep it short, the consideration of this Subsection  is
general and formal. It will be applied to the
analysis of the unfolded HS field equations first
in Section \ref{ceq}, explaining in some detail
the dynamical content of the $4d$ unfolded HS equations and,
in particular, the Central On-Shell Theorem, and then
in Section \ref{s-}  to the case of $\M_4$.
Sections \ref{ceq} and \ref{s-} will provide  examples
clarifying the formal scheme sketched in this Subsection.

$\sigma_-$-cohomology is a perturbative
concept that emerges in the linearized analysis.
The equation (\ref{unf}) linearized by (\ref{lin1}) gives
\be
\label{d0}
{\cal D} \go^\Omega (x)=0\,,
\ee
where $\D$ is some differential built of the order zero flat vacuum
connection $w^\ga$. To fulfill  the consistency condition (\ref{BI}),
$\D$ should square to zero
\be
\label{d02}
\D^2 =0\,.
\ee
Generally, $\D$ is the generalized covariant
derivative that, in addition to usual connection-like terms
linear in the vacuum connection $w^\ga$,
contains all Chevalley-Eilenberg-type terms polylinear in the
 $w^\ga$. For a $p_\Omega$-form $\go^\Omega$
with $p_\Omega \geq 1$, the linearized gauge transformation
is
\be
\label{gtr11}
\delta \go^\Omega = \D \epsilon^\Omega\,,
\ee
where $\epsilon^\Omega (x)$ is a $(p_\Omega -1)$-form
gauge parameter.

For a meaningful dynamical interpretation of the equation (\ref{d0}),
a space $V$, where
fields $\go^\Omega $ take their values, should be endowed with
some grading $G$ such that its spectrum is bounded from below.
Typically $G$ counts a rank of a tensor (equivalently, a power of an
appropriate generating polynomial) and eventually is associated with the
order of space-time derivatives of dynamical fields.
Suppose that
\be \label{d0s} \D = \D_0
+\sigma_- +\sigma_+\,, \ee
where
$
[G\,,\sigma_- ] =
- \sigma_-\,,
$
$
[G\,,\D_0 ]= 0
$
and $\sigma_+$ is a sum
of some operators of positive grade. From (\ref{d02}) it follows that
$
\sigma_-^2 = 0\,.
$
Provided that $\sigma_-$ acts vertically ({\it i.e.,} does not
differentiate $x^n$), cohomology of $\sigma_-$ determines
the dynamical content of the dynamical
system at hand.  Namely, as shown in \cite{sigma}, for a
$p$-form $\go^\Omega$ that takes values in a vector space $V$,
$H^{p+1} (\sigma_- ,V)$, $H^{p} (\sigma_- ,V)$ and $H^{p-1}
(\sigma_- ,V)$ characterize, respectively, differential equations,
dynamical fields and differential gauge symmetries encoded by the
equation (\ref{d0})\footnote{ Let us note that higher
$H^{p} (\sigma_- ,V)$ describe so called syzygies of the field
equations.}.

The meaning of this statement is quite simple.
From the level-by-level analysis of the equations (\ref{d0}) and
(\ref{d0s})
it follows that all fields that do not belong to $Ker\,\gs_-$
are auxiliary, being expressed by (\ref{d0})
via derivatives of the lower grade fields.
Those that are $\gs_-$ exact can be gauged away by the Stueckelberg
part of the gauge transformation (\ref{gtr11}), associated with the
$\gs_-$ part of $\D$ in (\ref{d0s}). The fields that remain
 belong to the cohomology of $\gs_-$. These are {\it dynamical
fields}.
Analogously one analyzes the dynamical content of the gauge
transformations and field equations.
(For more detail  see e.g. \cite{solv}.)

 Usually $\gs_-$ originates from
the part of the covariant derivative of  a
space-time symmetry algebra, that contains vielbein
required to be nondegenerate thus providing a frame of 1-forms
at any point $x_0$ of space-time. The nondegeneracy of the
vielbein allows to express as many as possible
auxiliary fields  via space-time derivatives of the dynamical fields.

The role of the Chevalley-Eilenberg cohomology terms in the
unfolded equations is that  they fill in unwanted $\sigma_-$
cohomologies with auxiliary Weyl tensor-like variables
to avoid too strong differential consequences of the
unfolded system. For example, relaxing the $C$-dependent terms on
the \rhs of (\ref{CON12}) would imply that all HS gauge 1-forms
are pure gauge, which condition is too strong to describe
nontrivial dynamics.

In principle, the leftover $\gs_-$ cohomology responsible for the
HS field equations can also be glued with the additional
0-form fields. This corresponds to the situation
with $H^{p+1} (\sigma_- ,V)=0$ where
 no differential equations  on the dynamical
variables are imposed at all. In this case,
 the unfolded equation (\ref{unf}) just expresses the Bianchi identities
for some constraints on auxiliary fields. Unfolded systems of this type
are referred to as {\it off-shell}.
As discussed in \cite{act}, they are useful for the
Lagrangian formulation of a dynamical system.

Let  us stress that the $\gs_-$ cohomology analysis
applies both to
linear and to  non-linear systems treated perturbatively.
In particular, nonlinear equations are
off-shell once their linearization is off-shell.

\subsection{Properties}
\label{prop}

Let us summarize briefly the main  properties of
unfolded dynamics. First of all,
the method is universal in the sense that any dynamical system can in
principle be unfolded. This statement is analogous to the text-book fact
that any system of ordinary differential equations can be
reformulated in the first-order form.

Indeed, let $w=e_0^a\,P_a+\frac12 \go_0^{ab}M_{ab}$ be a vacuum
gravitational gauge field taking values in some space-time
symmetry algebra $g$. Let $C^{(0)}(x)$ be a given space-time field
satisfying some dynamical equations to be unfolded. Consider for
simplicity the case where $C^{(0)}(x)$ is a 0-form. The general
procedure of unfolding  goes schematically as
follows. For a start, one writes the equation
$D_0^L C^{(0)} \,=\,
e_0^a\,\,\, C_a^{(1)},
$
 where $D_0^L$ is the
covariant Lorentz derivative and the field $C_a^{(1)}$ is auxiliary.
Next, one checks whether the original field equations for $C^{(0)}$
impose any restrictions on the first derivatives of $C^{(0)}$. More
precisely, some part of $D^L_{0\um} C^{(0)}$ might vanish
on-mass-shell ({\it e.g.} for Dirac spinors). These restrictions in
turn impose some restrictions on the auxiliary fields $C_a^{(1)}$.
If these constraints are satisfied by $C_a^{(1)}$, then these fields
parameterize all on-mass-shell nontrivial components of first
derivatives.
One continues by writing analogous equation
for the first-level auxiliary  fields
$ D_0^L C^{(1)}_a =  e_0^b\,\, C^{(2)}_{a,b},
$
where the new fields $C^{(2)}_{a,b}$ parameterize the second
derivatives of $C^{(0)}$. Once again one checks (taking into
account the Bianchi identities) which components of  the second
level fields $C^{(2)}_{a,b}$ are nonvanishing provided that the
original equations of motion are satisfied.
This process continues indefinitely, leading to a chain of
equations having the form of some covariant constancy condition
for the chain of fields $C^{(m)}_{a_1,a_2,\ldots,a_m}$
($m\in\mathbb N$) parameterizing all  on-mass-shell nontrivial
derivatives of the original dynamical field. By construction, this
leads to a particular unfolded equation.  The set
of fields $C^{(m)}_{a_1,a_2,\ldots,a_m}$ realizes some $g$--module
$V$. The full infinite chain of equations becomes a
single covariant constancy condition $D_0C=0$, where $D_0$ is the
$g$-covariant derivative in $V$.

If one starts with some gauge field like, for example, the
fluctuational part of the metric tensor, analogous analysis
determines a form of the Stueckelberg shift gauge
transformations that subtract Stueckelberg field components
to be introduced to describe a system in terms of differential
forms. (For instance in gravity, the local Lorentz symmetry results
this way as the Stueckelberg symmetry that removes the extra
components of the vielbein 1-form compared
to the metric tensor.) The correspondence
between $p\geq 1$ forms and gauge symmetries in the unfolded dynamics
approach uncovers the pattern of
local and global symmetries associated with a given gauge field.
In particular, the pattern of the linearized $4d$ HS algebras was
deduced this way in \cite{Fort1}. These results were then used in
\cite{FVA,Fort2,FVau} to find infinite dimensional non-Abelian HS
algebras that underly the nonlinear $4d$ HS theories and
in \cite{Ann,con,more} to construct full nonlinear HS field
equations as a nonlinear deformation of the unfolded system
(\ref{CON12}) and (\ref{CON22}).

Other important properties of the unfolded formulation include:

\begin{itemize}

\item
Manifest gauge invariance and
invariance under diffeomorphisms (\ie coordinate
independence) due to using the exterior algebra formalism
is perfectly suited for the study of gauge invariant theories
in the framework of gravity and, in particular, HS gauge theories.
\item
In the topologically trivial situation,
degrees of freedom are concentrated in 0-forms $\go^i_0(x_0)$
at any $x=x_0$. This is a consequence of the Poincare'
lemma: the unfolded equations express all exterior derivatives
in terms of the values of fields themselves modulo exact forms
that can be gauged away by the gauge transformation
(\ref{delw}). What is left is the ``constant part" of the 0-forms.

This simple observation has a consequence that, to describe a system with
an infinite number of degrees of freedom, it is necessary to work with
an infinite set of
0-forms that form an infinite dimensional module of the
space-time symmetry $g$. In fact, the module carried by 0-forms
turns out to be dual (complex equivalent)  to the space of
single-particle states in the respective QFT.

On the other hand, if  the unfolded formulation of
a system operates with a finite set of 0-forms,
the system is topological, describing at most
a finite number of degrees of freedom. In particular, the topological
dynamical systems of \cite{aux} mentioned in the end of
Section \ref{hsads} are just of this type. A typical example of
such a system is
the equation (\ref{glpar}) on the global symmetry parameters.
In fact, the covariant constancy condition on the 0-forms
in the topological sector of Section \ref{hsads} coincides with the
equation (\ref{glpar}) on the HS global symmetry parameters.

\item
Unfolded formulation admits natural
realization of higher derivative infinite symmetries
as endomorphisms of  the infinite--dimensional
modules of 0-forms.

\item
Chevalley-Eilenberg cohomology of a Lie algebra $g$ underlying
the unfolded formulation of one or another system is responsible
for nontrivial mixture of differential forms of different degrees, thus
making gauge fields associated with $(p>0)$-forms dynamically nontrivial.
Note that the corresponding cohomology turns out to be
nontrivial even for simple Lie algebras associated with the space-time
symmetries like $o(d,2)$, $sp(M,\mR)$, etc. just because
it  has coefficients in infinite--dimensional $g$-modules.

\item
Unfolded formulation unifies various dual versions of the
same system. The difference results from
the ambiguity in what is chosen  to be dynamical
or auxiliary fields, the nomenclature governed by the
choice of the grading $G$ and $\sigma_-$.
Different
gradings lead to different interpretations of the same
unfolded  system in terms of different dynamical fields
that satisfy seemingly unrelated differential equations.
The key point is that if two dynamical systems
give rise to the same unfolded
system, they are equivalent\footnote{It is worth to note that,
as shown in \cite{Mat}, differential duality relations between
dual systems in flat space may sometimes become algebraic  in
$AdS$ geometry. This phenomenon is analogous to the $AdS$ resolution of
the flat space degeneracy discussed below: some operators
to be interpreted as components of  $\sigma_-$ that degenerate in flat
background may be non-degenerate in the $AdS$ background.}.
We conjecture that all dual descriptions of a given dynamical system
$D$ are contained in its maximally extended {\it projective}
unfolded version $P(D)$. By a $P(D)$-projective unfolded system
we mean such a maximal unfolded formulation of $D$ that
({\it i}) any unfolded description of $D$ is a
subsystem of $P(D)$ and ({\it ii}) $P(D)$ does not decompose
into two independent subsystems one of which is an unfolded
formulation of $D$. Note that $P(D)$ may require
 a larger set of differential form variables.

For example, the extension of the set of Weyl
0-forms $C(y,\by|x)$ by the HS gauge 1-forms $\go(y,\by|x)$ leads
to the unfolded system (\ref{CON12}), (\ref{CON22}) that
extends the HS equations (\ref{hol}) in terms $C(y,\by|x)$ to
those in terms of HS gauge potentials.
The unfolded system (\ref{CON12}), (\ref{CON22}) is not
projective however. It can be further extended without
changing its dynamical content by replacing 1-forms $\go^{ii}$
and 0-forms $C^{i\,1-i}$ by forms of all odd and even degrees,
respectively. The resulting system is likely to be projective
for the doubled set of $4d$ massless fields of all spins.

 The concept of $P(D)$-projective unfolded system is homological
in nature. General analysis of this interesting and important
issue lies beyond the scope of this paper and will be
given elsewhere.

\item
One of the striking features of the
unfolded formulation based on universal FDAs
is that, to some extent, it makes the space-time $x$-dependence
artificial. The dynamics is entirely encoded in the form
of the function $G^i(W)$. In particular, unfolded formulation
allows one to extend space-time without changing dynamics simply by
letting the differential $d$ and differential forms $W^\Phi$
to live in a larger space
\be\nn
d=dx^n\f{\p}{\p x^n}  \rightarrow \hat{d}=dx^n\f{\p}{\p x^n}
+dx^{\hat{n}} \f{\p}{\p \hat{x}^{\hat{n}}}\q
dx^n w_n \to
dx^n w_n + d\hat{x}^{\hat{n}} \hat{w}_{\hat{n}}\,,
\ee
where $\hat{x}^{\hat{n}}$ are some additional coordinates.
For a universal unfolded system such a substitution
neither spoils the consistency nor changes the
local dynamics
still determined by the 0-forms at any point of
(any) space-time. Alternatively, one observes that
the unfolded system
in the $x$ space remains a subsystem of that in the enlarged space
while the additional equations reconstruct the dependence on
the additional coordinates in terms of solutions of the original
system (of course, this consideration is local).

This property is not only practically useful allowing to
introduce easily appropriate hyperspaces \cite{BHS,ESS}, but
is likely to have deep meaning encouraging  to reconsider a role
of such fundamental concepts as local event and metric tensor
in a fundamental theory. An illuminating manifestation
of this issue comes from the analysis of $4d$ HS physics formulated
in $\M_4$ in \cite{Mar}, where it was argued
that the concepts of local event, space-time dimension and
metric tensor have dynamical origin.
In this respect, the unfolded dynamics \cite{Ann}
has some similarity with the matrix model  approach
to string theory \cite{BFSS,IKKT,S}, having however the advantage of being
covariant. (See also \cite{HKK,saitou} where
matrix models are linked to  HS theory
in a somewhat different fashion.)

\item
The unfolded formulation approach provides an efficient tool
for the analysis of gauge invariant interactions.
The problem  reduces to the unification of the
zero-order vacuum field $w$ and first-order
dynamical gauge fields $\omega$ into a single field
$W^\Omega$ by (\ref{lin1})
and to searching for a nontrivial deformation of $G^\alpha(W)$
that respects the consistency condition (\ref{BI})
and reproduces correctly the linearized dynamics.
In fact, the results of this paper form a basis for
the future search of nonlinear $sp(8,\mR)$ invariant HS gauge theories.
\end{itemize}

Note that there is a great similarity between
the unfolded formulation approach and the prolongation technics in the analysis
of partial differential equations (see e.g.,  text book \cite{olver})
that operates with jet spaces designed to describe higher derivatives
 to make it possible to rewrite a partial differential
system in the first order form. The important novelty is due to the
extension to differential forms as dynamical variables in the unfolded
dynamics approach, that results in the properties listed above.
Note also that the unfolded formulation admits a nice
interpretation \cite{act,BG} in terms of $L_\infty$ strong
homotopy algebra \cite{Linf}.

\section{Strategy}
\label{GS}

Equipped with the unfolded dynamics approach,
the strategy of the analysis of free $g$--symmetric
dynamical equations may be as follows.
\begin{itemize}

\item
Fix a flat (vacuum) connection $w$ of $g$ that gives a nondegenerate
vielbein as the 1-form connection associated with the
generators of translations in the respective space-time
symmetry subalgebra $s\subset g$ (e.g., Poincare',
$(A)dS_d$, or conformal).

\item
Guess a set of $g$-modules where the
variables $\go^\Phi$, that are differential forms of
different degrees, take their values and find
Chevalley-Eilenberg cohomology (\ref{co}) of $g$
with the coefficients in the respective $g$-modules.
It is this step that fixes a particular dynamical system
encoded by the unfolded equations (\ref{d0}).

\item
Introduce a grading $G$ of $g$-modules where the
$\go^\Phi$ take their values, that gives rise to the decomposition
(\ref{d0s}) of the generalized covariant derivative.
Since, the ambiguity in the choice of $G$ results in
 dual descriptions of the same model, to describe
a system in terms of some preferable dynamical variables
one has to choose $G$ appropriately.
Analyze the $\gs_-$ cohomology to figure out
what are differential gauge symmetries, dynamical fields and
differential
field equations encoded by the unfolded equations (\ref{lin}).

\item
Let $M$ be some $s$-invariant manifold and
 $w$ be a vacuum connection of $s$ with a nondegenerate
vielbein. A larger symmetry $g$ of an unfolded system
at hand may admit no geometric interpretation in $M$.
To find an equivalent formulation of the same unfolded dynamics
in a larger space-time $\M$ where $g$ acts geometrically
it suffices to extend the exterior differential $d$ from
$M$ to $\M$ and to replace $w$ by a flat connection of $g$
that contains nondegenerate basis 1-forms (generalized
vielbein) in $\M$. Provided that the Chevalley-Eilenberg
cohomology terms were defined with respect to $g$, the resulting unfolded equations in $\M$ remain
consistent and describe the same dynamics. Keeping
the grading $G$ unchanged, one finds the new $\sigma_-$
operator in $\M$. The dynamical content
(\ie symmetries, field variables and field equations) of the
unfolded system in $\M$ is uncovered via the analysis of the
cohomology of the new $\gs_-$.
Although the resulting dynamical variables and
field equations in $\M$ may differ from those of the original system
in $M$ (pretty much as  superfields and superfield equations
of supersymmetric theories look differently from their
component counterparts), the resulting system in $\M$ is
guaranteed to be $g$-invariant and (locally) equivalent to
that in $M$.

\end{itemize}

In the rest of the paper we systematically implement this approach.
Although it still has some research freedom
in guessing appropriate $g$-modules,
for distinguished systems like the one explored in this paper
 this part of the project is not too hard
because the choice is usually quite limited if not obvious. The
benefit is that the $g$-symmetry is guaranteed and the
rest of the analysis is straightforward. Most notably,
the $g$-invariant differential equations are derived (rather than
guessed) via the analysis of $\gs_-$ cohomology and
the equivalence of the equations in different spaces
is automatic.


\section{Conformal geometry}
\label{Conformal geometries}

To describe a conformal system in $d$ dimensions in the unfolded
dynamics approach one should first of all fix a nondegenerate flat
connection of the conformal algebra $o(d,2)$. In terms of Lorentz
(\ie $o(d-1,1))$ irreducible components, the $o(d,2)$ has the
generators $ P_n\,, L_{nm}=-L_{mn}\,, D\,, K^n\,. $
The $o(d,2)$ connection
1-form $w$ and curvature 2-form $R$ are
\be \label{congau} w= h^n
P_n +\go^{nm} L_{nm} + f_n K^n + b D\q R= R^n P_n +R^{nm} L_{nm} +
r_n K^n + r D\,,
\ee
where
\be\nn R^n = dh^n
+\go^n{}_m\wedge h^m -b\wedge h^n\,, \ee \be\nn R^{nm} = d\go^{nm}
+\go^n{}_k\wedge \go^{km} -h^n\wedge f^m+ h^m\wedge f^n\,, \ee
\be\nn r= db +h^n \wedge f_n\,, \ee \be\nn r^n = df^n
+\go^n{}_m\wedge f^m + b\wedge f^n\,. \ee Here the 1-form
$h^n=dx^\um h_\um{}^n$ is identified with vielbein. It is required
to be nondegenerate in the sense that $det  | h_\um{}^n |\neq 0$.
$\go^{nm}$ is Lorentz connection. $f_n$ and $b$ are gauge fields for
special conformal transformations and dilatation, respectively.

The conformal gauge transformations are
\be\nn
\delta h^n = D^L\epsilon^n -\epsilon^n{}_m h^m
+\epsilon  h^n -\epsilon^n b\,,
\ee
\be\nn
\delta\go^{nm} = D^L\epsilon^{nm}
-h^n \tilde{\epsilon}^m +\epsilon^n f^m
+h^m \tilde{\epsilon}^n -\epsilon^m f^n\,,
\ee
\be\nn
\delta b = d \epsilon + h^n \wedge \tilde{\epsilon}_n -
\epsilon^n f_n\,,
\ee
\be\nn
\delta f^n = D^L \tilde{\epsilon}^n -\epsilon^n{}_m h^m
-\epsilon f^n +\tilde{\epsilon}^n b \,,
\ee
where $\epsilon^n(x)$, $\epsilon^{mn}(x)$, $\tilde{\epsilon}_n(x)$ and
$\epsilon(x)$ are gauge parameters of translations, Lorentz
transformations, special conformal transformations and dilatations,
respectively.

 The interpretation of these fields
is as follows (for more detail see, e.g., \cite{conf,conf1}).
The 1-form $b$ can be gauge fixed to zero
\be
\label{b0}
b=0
\ee
 by a special conformal gauge transformation with the parameter
 $\tilde{\epsilon}^n (x)$. (Here one uses that $h^n$ is nondegenerate.)
 The leftover gauge symmetries are local translations,
 Lorentz transformations and dilatations.

Imposing the condition that the $o(d,2)$
curvatures are all zero
\be
\label{r0}
R=0
\ee
has the following consequences.
$R^n=0$ in the gauge (\ref{b0}) is the usual zero torsion
condition that expresses the Lorentz connection $\go^{nm}$
in terms of the vielbein $h^n$. The condition $R^{nm}=0$
requires the Weyl tensor $C^{nm,kl}(h)$ to be zero
and expresses the symmetric part $f_{(nm)}$ of
$f^n= dx^m f_m{}^n$ in terms of the Ricci tensor of
$h$. The antisymmetric part of $f_\um{}^n$
is zero by virtue of $r=0$ in the gauge (\ref{b0}).
 $r_n=0$   holds  by virtue of Bianchi
identities. Thus, in terms of the vielbein, (\ref{r0})
implies that Weyl tensor is zero. All other equations contained
in (\ref{r0}) are either constraints or consequences of the other
equations.

The translation of these results into
the $\gs_-$ cohomology language is as follows. The grading $G$ is
just the conformal dimension in $o(d,2)$ induced by $D$, \ie
$h^n$ has grading $-1$, $  \go^{nm}$ and $b$  have grading zero
and $f_n$ has grading $+1$.
$\gs_-$ is the $h$--dependent part the covariant derivative
where $h^n$ is treated as the respective part of the
vacuum connection. ($h^n = dx^n$ in
 Cartesian coordinates.)
$H^0(\gs_-)$ describes linearized diffeomorphisms
(all other gauge transformatons
are Stueckelberg). $H^1(\gs_-)$ describes the
conformal class of metrics (perturbatively, second rank traceless
tensors). $H^2(\gs_-)$ describes the Weyl tensor (\ie the only nontrivial
differental equation in (\ref{r0}) is that the Weyl tensor is zero).
We leave it to
the reader to check details of this correspondence as a useful exercise.

Taking into account that, locally,
any two $o(d,2)$ flat connections are related by a $o(d,2)$ gauge
transformation and that $o(d,2)$ gauge transformations contain local
dilatations of the metric, a simple consequence of
this analysis is that the metric tensor is conformally flat
iff the  Weyl tensor is zero.

In Cartesian coordinates, the Minkowski space solution of (\ref{r0}) is
$
w=dx^n P_n.
$
Another important example of conformally flat space is provided
by the anti-de Sitter geometry.
Indeed, the $AdS_{d}$ algebra $o(d-1,2)$ can be realized as
the subalgebra of $o(d,2)$ spanned by the generators
$
{\cal P}_n = P_n +\lambda^2 K_n
$
and $L_{nm}$.
Choosing a flat connection of $o(d-1,2)$, namely $e^n$ and $\go^{nm}$,
gives us a flat connection of $o(d,2)$ with
$
h^n = \lambda e^n$, $ f_n = \lambda e_n$,
$ b=0$
and the same Lorentz connection $\go^{nm}$.
This Ansatz solves (\ref{r0})
for the conformal algebra once $e^n$ and $\go^{mn}$
solve the zero curvature equations for $o(d-1,2)$.
As a by-product, this gives a coordinate
independent proof of the fact that $AdS_d$ is conformally flat.

In the $4d$ case one can use
two-component spinor notation. In these terms,
the $su(2,2)\sim o(4,2)$ connections   are
$
h^{\ga\pa}\,, \omega_{\ga}{}^{\gb}\,,
\overline{\omega}_{\pa}{}^{\pb}\,, b$
and $f_{\ga\pa}.$
Extending $su(2,2)$ to $u(2,2)$ by adding a central helicity
generator  with the gauge connection $\tilde{b}$, the $u(2,2)$
flatness conditions read as
\be
\label{h4}
R^{\ga\pb}=
dh^{\ga\pb} -\omega_\gga{}^\ga\wedge h^{\gga\pb}-
\overline{\omega}_\pga{}^\pb\wedge h^{\ga\pga} =0\,,
\ee
\be
\label{b4}
R_{\ga\pb}=
df_{\ga\pb} +\omega_\ga{}^\gga\wedge f_{\gga\pb}+
\overline{\omega}_\pb{}^\pga\wedge f_{\ga\pga} =0\,,
\ee
\be
\label{o4}
R_\ga{}^\gb =
d\omega_\ga{}^\gb +\omega_\ga{}^\gga\wedge \omega_\gga{}^\gb -
 f_{\ga\pga}\wedge h^{\pga\gb}=0\,,
\ee
\be
\label{oc4}
\overline{R}_\pa{}^\pb =
d\overline{\omega}_\pa{}^\pb +
\overline{\omega}_\pa{}^\pga\wedge
\overline{\omega}_\pga{}^\pb - f_{\gga\pa}\wedge h^{\gga\pb}=0\,.
\ee
Here the traceless parts $\go^{L}{}_{\ga}{}^\gb$ and
$\overline{\go}^{L}{}_{ \pa}{}^\pb$
of $\go_\ga{}^\gb$ and $\overline{\go}_\pa{}^\pb$
describe the Lorentz connection while their traces contain the gauge
fields $b$ and $\tilde{b}$ according to
\be
\label{btb}
b= \half \Big (\go_\ga{}^\ga + \overline{\go}_\pa{}^\pa\Big )\q
\tilde{b}= \half \Big (\go_\ga{}^\ga -\overline{\go}_\pa{}^\pa\Big )\,.
\ee

The $AdS_4$ geometry is described by
\be
\label{adsc}
 h^{\ga\pa}=  \lambda e^{\ga\pa} \q
 f_{\ga\pa} =   \lambda  e_{\ga\pa}\q b=\tilde{b}=0\,.
\ee
The $u(2,2)$ flatness conditions
are solved by the vierbein $e^{\ga\pa}$ and Lorentz connection
$\omega^{\ga\gb}$ and $\overline{\omega}^{\pa\pb}$ that
satisfy the zero curvature conditions (\ref{adsfl}).
This Ansatz for the vacuum connection will be used later
on to describe $4d$ conformal invariant systems in the unfolded
dynamics approach.

\section{Generalized conformal geometry}
\label{Generalized conformal geometries}

The example of $o(d,2)$ conformal symmetry admits a
natural generalization to $sp(2M,{\mathbb R})$ treated as
generalized conformal symmetry with the group manifold
$Sp(M,{\mathbb R})$ as an analog of $AdS_d$.
(Note that the case of
$M=2$ reproduces the usual $3d$ case with
$AdS_3\sim Sp(2,{\mathbb R})$ and $sp(4,{\mathbb R})\sim so(3,2)$.)
This interpretation of $Sp(M,{\mathbb R})$ was discussed in
\cite{BLPS,BHS,DV,PST}.

The components of the $sp(2M,{\mathbb R})$ gauge connection
(\ref{sp8c}) are real.
Generalized ``conformally flat" background geometry
is described by a $sp(2M,{\mathbb R})$ flat connection that satisfies
\be
\label{h}
R^{AB}=
dh^{AB} -\omega_C{}^A\wedge h^{CB}-\omega_C{}^B\wedge h^{CA} =0\,,
\ee
\be
\label{b}
R_{AB}=
df_{AB} +\omega_A{}^C\wedge f_{CB}+\omega_B{}^C\wedge f_{CA} =0\,,
\ee
\be
\label{o}
R_A{}^B =
d\omega_A{}^B +\omega_A{}^C\wedge \omega_C{}^B- f_{AC}\wedge h^{CB}=0\,.
\ee

The ``Cartesian coordinates" in $\M_M$ are associated with
the particular flat connection
\be\nn
\label{cart10}
h^{AB}=dX^{AB}\,,\qquad  \omega_A{}^B=0 \,,\qquad  f_{AB}=0\,.
\ee

A different choice of the $sp(2M,{\mathbb R})$ flat connection
describes the group manifold $Sp(M,{\mathbb R})$. The group of motions
$Sp(M,{\mathbb R})\times Sp(M,{\mathbb R})$ of $Sp(M,{\mathbb R})$
results from the left and
right action of the group on itself. Let two flat $sp(M,{\mathbb R})$
connections satisfy
\be
\label{rpm}
R^\pm_{AB}=dw^\pm_{AB} \mp w^\pm_{A}{}^C \wedge w^\pm_{CB}=0\,,
\ee
where indices are raised and lowered by a $Sp(M,{\mathbb R})$
invariant symplectic form $C_{AB}=-C_{BA}$ according to
(\ref{Cind}).
Locally, they admit the pure gauge
representation
\be\nn
\label{upm}
w^\pm_{AB}(X) =\mp U^\pm_{AC} (X) d U^\pm_{B}{}^C(X)\,,
\ee
where $U^\pm_A{}^B (X)$ is an arbitrary $Sp(M,{\mathbb R})$ valued
matrix that satisfies
\be\nn
U^\pm_{A}{}^C (X) U^\pm_B{}^D (X) C_{CD} = C_{AB}\q
U^\pm_{A}{}^C (X) U^\pm_B{}^D (X) C^{AB} = C^{CD}\,.
\ee

Setting
\be\nn
\label{wpm}
\omega_{AB} = \half \left (w^-_{AB} - w^+_{AB}\right )\q
\lambda e_{AB} = \f{1}{2} \left ( w^+_{AB} + w^-_{AB}\right )\,
\ee
and using (\ref{rpm}), we observe that
\be
\label{Rab}
R_{AB}=d\omega_{AB} +\omega_{A}{}^C\wedge \omega_{CB}+\lambda^2
e_{A}{}^C\wedge e_{CB}=0\,,
\ee
\be
\label{rab}
r_{AB}= de_{AB} +\omega_{A}{}^C  \wedge e_{CB}+
\omega_{B}{}^C  \wedge e_{CA}=0\,.
\ee

To embed $sp(M,{\mathbb R})\oplus sp(M,{\mathbb R})$ into $sp(2M,{\mathbb R})$
we express the connections of $sp(2M,{\mathbb R})$ in terms of
those of $sp(M,{\mathbb R})\oplus sp(M,{\mathbb R})$ as follows
\be
\label{sp48}
h^{AB}= \lambda e^{AB}\,,\qquad f_{AB} =\lambda e_{AB}\q
\omega_A{}^B =\omega_A{}^B\,.
\ee
Then the $sp(2M,{\mathbb R})$ flatness  conditions (\ref{h})-(\ref{o})
hold as a consequence of (\ref{Rab}) and (\ref{rab}).

We will use the $sp(2M,{\mathbb R})$ flat vacuum connection (\ref{sp48})
in the analysis of the $Sp(2M,{\mathbb R})$ invariant field equations in
$Sp(M,{\mathbb R})$. The important difference between the Minkowski-like
geometries and their $AdS$-like counterparts
is that in the former case the special conformal connections
$f^n$ and $f^{AB}$ vanish while in the latter case they are
nondegenerate by virtue of (\ref{adsc}) and (\ref{sp48}). This property
will be of crucial importance for the physical interpretation of
the usual and generalized HS conformal equations because they contain
some $f$-dependent terms that should be non-degenerate for a
consistent perturbative interpretation.

\section{Star-product and vacuum symmetry}
\label{star}

Instead of working in terms of oscillators (\ref{osc}),
it is convenient to use the star-product in the algebra of
polynomials of commuting variables $a_{A}$ and $b^{A}$
\be\label{old}
(f\star g)(a,b)=\frac{1}{\pi^{2M}}\int
f(a+u,b+t)g(a+s,b+v)e^{2(s_{A}
t^{A}-u_{A}v^{A})} \,
d^{M}u\,d^{M}t\,d^{M}s\,d^{M}v\,.
\ee
The star-product defined this way, often called Moyal product,
describes the product of
symmetrized (i.e., Weyl ordered) polynomials of oscillators in terms
of their {\it symbols}. The integral is normalized so that $1$ is the
unit element of the algebra
\be\nn
\frac{1}{\pi^{2M}}\int e^{2(s_{A}
t^{A}-u_{A}v^{A})} \,
d^{M}u\,d^{M}t\,d^{M}s\,d^{M}v=1 \,.
\ee

Eq.(\ref{old})
defines the associative algebra with the defining relations
\be\nn
[a_{A},b^{B}]_{\star}= \delta_{A}{}^{B}\,,\qquad
{[}a_{A},a_{B}]_{\star}=0\,,\qquad
{[}b^{A},b^{B}]_{\star}=0
\ee
($[a,b]_{\star} = a\star b - b\star a$). The following useful formulae
hold
\be
\label{ustar}
a_A\star = a_A +\half \f{\p}{\p b^A}\q
b^A\star = b^A - \half \f{\p}{\p a_A}\,,
\ee
\be
\label{staru}
\star a_A = a_A -\half \f{\overleftarrow{\p}}{\p b^A}\q
\star b^A = b^A + \half \f{\overleftarrow{\p}}{\p a_A}\,.
\ee

The star-product realization of the generators of
$sp(2M,{\mathbb R})$ is
\be\label{8star}
L_{A}{}^{B}=a_{A}b^{B} \,,\qquad
P_{AB}=\half a_{A}a_{B} \,,\qquad
K^{AB}=\half b^{A}b^{B}\,.
\ee
$sp(2M,{\mathbb R})$ extends to the superalgebra
$osp(1|2M,{\mathbb R})$ by adding the supergenerators
\be
\label{osp}
Q_A = a_A\q S^B= b^B\,.
\ee

Using the oscillator realization of
$sp(M,{\mathbb R})\oplus sp(M,{\mathbb R}) \subset sp(2M,{\mathbb R})$
we can set
\be
\label{stacon} w(a,b|X) =\omega_B{}^{A}(X) a_{A}b^{B}
+\half e_{AB}(X) (a^{A}a^{B} +\gl^{2}b^{A}b^{B})\,,
\ee
where $\go_{AB}$ and $e_{AB}$ satisfy the flatness conditions (\ref{Rab})
and (\ref{rab}) to ensure that $w(a,b|X)$ satisfies the vacuum flatness
condition
\be
\label{0cu}
d w+w\star\wedge w=0\,.
\ee

 Let us introduce the oscillators
$
\alpha^\pm_A = a_A \pm b^C C_{CA}
$
with the commutation relations
\be\nn
[\alpha^\pm_A \,,\alpha^\pm_B]_\star = \pm 2C_{AB}\,,\qquad
[\alpha^\pm_A \,,\alpha^\mp_B]_\star =0\,.
\ee
Then
$
T^\pm_{AB} = \half \ga^\pm_A\ga^\pm_B
$
are the generators of $sp^+(M,{\mathbb R})\oplus sp^-(M,{\mathbb R})
 \subset sp(2M,{\mathbb R})$.

A useful viewpoint is that $w(a,b|X)$ takes values in the
infinite dimensional star product algebra of various polynomials of
$a_A$ and $b^B$, which is the HS symmetry algebra as a Lie superalgebra.
(In accordance with the spin-statistics relationship, the boson-fermion
$Z_2$ grading $\pi$ counts  the oddness of a number of spinor indices, \ie
$w(a,b|X)=(-1)^{\pi(w)}w(-a,-b|X)$.)
 As explained in Subsection \ref{v}, any fixed vacuum solution $w_{0}$ of
(\ref{0cu}) breaks the local HS symmetry to its global
stability subalgebra with the infinitesimal parameters
$\epsilon_{0}(a,b|X)$ that satisfy the equation
\be\label{eps}
d\epsilon_{0}+[w_{0},\epsilon_{0}]_{\star}=0.
\ee

Solving (\ref{0cu}) in the pure gauge form
\be\label{gdg}
w_{0}(a,b|X)=g^{-1}(a,b|X)\star dg(a,b|X)\,,
\ee
where $g(a,b|X)$ is some invertible element of the star-product
algebra, $g\star g^{-1}=g^{-1}\star g=1$, we solve (\ref{eps}) as
\be\label{epss}
\epsilon_{0}(a,b|X) = g^{-1}(a,b|X)\star\xi(a,b)\star g(a,b|X)\,,
\ee
where an arbitrary $X$-independent star-product element $\xi(a,b)$
 describes free parameters of the global HS symmetry.
Note that the explicit form of the HS transformations depends on a chosen
coordinate system encoded by $w_0 (a,b|X)$.

As shown in \cite{DV}, the star product realization of the pure gauge
representation (\ref{upm}) is given  by the formula
\be\nn
g(X) = g^+ (X)\star g^- (X) = g^+ (X) g^- (X)\,,
\ee
where
\be
\label{main}
g^\pm (X) = \frac{2^{\f{M}{2}}}{\sqrt{\det\|U^\pm +1\|}}\exp \left( -2
f^{AB} [U^\pm ]\alpha_A^\pm
\alpha_B^\pm \right )\,,
\ee
\be
(g^\pm )^{-1}(X) = \frac{2^{\f{M}{2}}}{\sqrt{\det\|U^\pm +1\|}}\exp \left( 2
f^{AB}[U^\pm ]\alpha_A^\pm \alpha_B^\pm \right )\,
\ee
and we have set $\lambda =1$ for simplicity.
Here $U^{AB}(X)$ is some $sp(M,{\mathbb R})$ valued function of
local coordinates and we use the notations
\be\nn
f^{AB}[U]=\left (\frac{U -1}{U +1}\right )^{AB}\q
U^{AB}[f]=\left (\frac{1+f}{1-f}\right )^{AB}\,.
\ee

Note that the gauge functions $g^\pm (X)$ (\ref{main}) are
chosen so \cite{DV} that the corresponding flat connection $w^\pm$
is bilinear in the oscillators $a_A$, $b^B$, {\it i.e.,} it
indeed takes values in $sp^\pm (M,{\mathbb R})$.
A particularly useful choice is that with $U^+ =(U^-)^{-1} = U$,
giving
\be\nn
g (X) = \frac{2^{M}}{{\det\|U +1\|}}\exp \left[ -
f^{AB}[U]
(a_A a_B+ b_A b_B)\right ]\,.
\ee
As noted in \cite{DV}, the ambiguity in  the function
$f^{AB}(X)$ parameterizes the ambiguity in the choice of local
coordinates of $Sp(M,{\mathbb R})$. (For particular
coordinate choices see \cite{DV}.)

 Once the vacuum solution $w_{0}$ is fixed in the pure gauge form
(\ref{gdg}) with some gauge function $g$, it is easy to find the
gauge parameter $\epsilon_{0}(a,b|X)$ of the leftover global
symmetry. Indeed, let the generating parameter $\xi (a,b;\mu,\eta)$
in (\ref{epss}) be of the form
$
\xi=\xi_{0}\exp(a_{A}\mu^{A}-b^{A}\eta_{A})
$
where $\xi_{0}$ is an infinitesimal constant while $\mu^{A}$ and
$\eta_{A}$ are constant parameters.
 Substitution of (\ref{main}) into (\ref{epss}) gives \cite{DV}
\be\label{gauge}
\epsilon_{0}(a,b;\mu,\eta|X)=g^{-1}\star\xi\star g=
\xi_{0}\exp(a_{A}\hat{\mu}^{A}-b^{A}\hat{\eta}_{A})\,,
\ee
where
\be\nn
\label{mu}
\hat{\mu}_{A}=\Big (\f{1+\gla^{2}f^{2}(X)}{1-\gla^{2}f^{2}(X)}\Big
)_{A}{}^{B}\mu_{B}-\Big (\f{2f(X)}{1-\gla^{2}f^{2}(X)}\Big
)_{A}{}^{B}\eta_{B}\,,
\ee
\be\nn
\hat{\eta}_{A}=\Big
(\f{1+\gla^{2}f^{2}(X)}{1-\gla^{2}f^{2}(X)}\Big
)_{A}{}^{B}\eta_{B}-\gla^{2}\Big
(\f{2f(X)}{1-\gla^{2}f^{2}(X)}\Big )_{A}{}^{B}\mu_{B}\,.
\ee
 Now, any  HS global symmetry parameter, which is
 some $x$--dependent star-product polynomial, can be obtained by
 differentiation of $\epsilon_{0}(a,b;\mu,\eta|X)$ over $\mu^{A}$ or/and $\eta_{A}$.

This simple procedure demonstrates the
efficiency of the unfolded dynamics approach.
Analogously to the flat space example (\ref{trc}),
to obtain the form
of the HS transformation of massless fields
in $Sp(M,\mR)$ one has to act by the
parameter $\epsilon_{0}(a,b;\mu,\eta|X)$ on a module where the
fields take their values.
The corresponding $sp(8,{\mathbb R})$-modules, which the same time
 form the HS symmetry-modules, are introduced in the next section.

\section{$Sp(8,{\mathbb R})$ Fock modules}
\label{sp8mod}

To extend the free $sp(8,{\mathbb R})$ invariant equation (\ref{10eq})
on HS Weyl field strengths
to the sector of gauge fields we have to identify a
$sp(8,{\mathbb R})$-module where $4d$ HS gauge fields of Section
\ref{hsads} take their values. Since the HS Weyl 0-forms were
described in Section \ref{mfe} in terms of the Fock module
(\ref{fc}), (\ref{fv}) in which the $sp(8,{\mathbb R})$ is
realized by bilinears of oscillators, a natural option is
to use the same realization of $sp(8,{\mathbb R})$, changing
however the Fock module by changing its vacuum.

Fock vacua projectors can be realized in terms of the star-product
algebra. For example, with the help of (\ref{ustar}) and
(\ref{staru}) one finds that the Fock vacuum $|0,0|$ defined by
$
a_A \star|0,0| =
 |0,0| \star b^A =0
$
is realized as the exponential
$
|0,0|=2^{M} \exp{2a_A b^A},
$
where
the normalization factor is chosen so that
$
|0,0|\star |0,0|=|0,0|.
$
Alternatively, one can consider Fock vacua projectors defined with
respect to different sets of creation and annihilation
operators. Demanding their Lorentz
invariance and definite scaling dimension with respect to
the Lorentz generators in (\ref{cgs}) and
the dilatation generator (\ref{dil}), respectively, we set
\be
\label{1|0}
|1\rangle\langle 0|=4 \exp{-2a_\ga b^\ga}\,:\qquad
b^\gb  \star|1\rangle\langle 0| =
 |1\rangle\langle 0| \star  a_\gb   = 0\,,\qquad
\ee
\be
\label{0|1}
|0\rangle\langle 1|=4 \exp{2a_\ga b^\ga}\,:\qquad\quad
a_\gb  \star|1\rangle\langle 0| =
 |1\rangle\langle 0| \star  b^\gb   = 0\,,\qquad
\ee
\be
\label{1b|0}
|\bar{1}\rangle\langle \bar{0}|=
4 \exp{-2\bar{a}_\pa \bar{b}^\pa}\,:\qquad
\bar{b}^\pb  \star|\bar{1}\rangle\langle \bar{0}| =
 |\bar{1}\rangle\langle \bar{0}| \star  \bar{a}_\pb
 = 0\,,
\ee
\be
\label{0b|1}
|\bar{0}\rangle\langle \bar{1}|=4\exp{2\bar{a}_\pa \bar{b}^\pa}\,
:\qquad
\bar{a}_\pb  \star|\bar{1}\rangle\langle \bar{0}| =
 |\bar{1}\rangle\langle \bar{0}| \star  \bar{b}^\pb   = 0\,.
\ee
We have
\be
\label{00}
|0,0|=|{0}\rangle\langle{1}|\star
|\bar{0}\rangle\langle\bar{1}|= 16\exp{2a_A b^A}\,:\qquad
a_A  \star|0,0|  =
 |0,0|\star b^B =0\,,
\ee
\be
\label{10}
\ls|1,0|=|{1}\rangle\langle{0}|\star
|\bar{0}\rangle\langle\bar{1}|=16
\exp{-2(a_\ga b^\ga- \bar{a}_\pa \bar{b}^\pa})\,:\qquad
b^\gb  \star|1,0| = \ba_\pb \star |1,0| =
 |1,0|\star a_\gb   =   |1,0|\star \bb^\pb = 0\,,\qquad
\ee
\be
\label{01}
\ls|0,1|=|{0}\rangle\langle{1}|\star
|\bar{1}\rangle\langle\bar{0}|=16
\exp{2(a_\ga b^\ga- \bar{a}_\pa \bar{b}^\pa})\,:\qquad
a_\gb \star |0,1| = \bb^\pb \star |0,1| =
|0,1|\star b^\gb =  |0,1|\star \ba_\pb = 0\,,
\ee
\be
\label{11}
|1,1|=|{1}\rangle\langle{0}|\star
|\bar{1}\rangle\langle\bar{0}|=16
\exp{-2a_A b^A}\,:\qquad
b^B \star |1,1| = |1,1| \star a_A = 0\,.
\ee

Correspondingly, we introduce two-component
oscillators $\ga_{\ga\,i}$, $\overline{\ga}_\pa{}_j$ and
$\gb^\ga_{i}$,  $\overline{\gb}^{\pa}_j$ that are,
respectively, the annihilation and creation operators
of the vacuum $|i,j|$
\be
\label{ai}
\ga_{0\,\ga} = a_\ga \,,\qquad
\ga_{1\,\ga} = b^\ga \,,\qquad
\overline{\ga}_{0\,\pa} =  \overline{a}_\pa \,,\qquad
\overline{\ga}_{ 1\,\pa} =  \overline{b}^\pa  \,,
\ee
\be
\label{bi}
\gb^\ga_{0}=b^\ga\,,\qquad \gb^\ga_{1}=a_\ga\,,\qquad
\overline{\gb}^\pa_{0}=\overline{b}^\pa\,,\qquad
\overline{\gb}^\pa_{1}=\overline{a}_\pa\,.
\ee
Note that
\be
\label{ab}
\gb^\ga_{i}= \ga_{1-i \,\ga}\,,\qquad
\overline{\gb}^\pa_{i}= \overline{\ga}_{ 1-i\,\pa}\,.
\ee

It should be noted that the Fock vacua $|i,j|$
cannot be star-multiplied with $|i^\prime,j^\prime|$
in the class of regular functions if $i \neq i^\prime$
and/or $j\neq j^\prime$.
This is not accidental. Indeed, if,
 say, $T=|0,0|\star |1,1|$ existed, it would
satisfy $a_A \star T= T\star a_A =0$. Taking into account
(\ref{ustar}) and (\ref{staru}), from here it follows
that $T=\delta(a)$. (Of course, this can be
directly derived from (\ref{old}).) However $T$ does not belong
to the star-product algebra because $T\star T=\delta(0)$.

On the other hand, different Fock spaces originating
from $|i,j|$ form well-defined modules of the star-product algebra.
This is sufficient for the analysis of the
free HS dynamics of this paper. To go beyond the free field level
 one has to handle potential
difficulties of co-existence of different sectors of the star-product
algebra associated with the different Fock modules. We hope to
come back to this interesting question elsewhere.

Now we introduce four 0-form modules
\be
\label{cij}
|C_{ij}(\gb_{i}\,,\overline{\gb}_j |X)\rangle =
C_{ij}(\gb_{i}\,,\overline{\gb}_j|X)\star |i,j|\,
\ee
and four 1-form modules
\be
|\omega_{ij}(\gb_{i}\,,\overline{\gb}_j|X) \rangle =
\omega_{ij}(\gb_{i}\,,\overline{\gb}_j|X)\star |i,j|\,,\qquad
\omega_{ij}(\gb_{i}\,,\overline{\gb}_j|X) =
dX^U \omega_{U\,ij}(\gb_{i}\,,\overline{\gb}_j|X)\,,
\ee
where the meaning of the coordinates $X^U$ will
be specified later on depending on the problem under study.
More precisely, $|C_{ij}(\gb_{i}\,,\overline{\gb}_j |X)\rangle$ and
$|\omega_{ij}(\gb_{i}\,,\overline{\gb}_j|X) \rangle$ are sections of the
Fock fiber bundles over a space-time manifold with local coordinates
$X$, that can be either a $4d$ space-time or
one of the ten dimensional space-times $\M_4$ or $Sp(4,{\mathbb R})$.

The generating fields
$|C_{ij}(\gb_{i}\,,\overline{\gb}_j |X)\rangle $ and
$|\omega_{ij}(\gb_{i}\,,\overline{\gb}_j|X)\rangle$
form $sp(8,{\mathbb R})$-modules  with the generators (\ref{8star}).
This allows us to define  the $sp(8,{\mathbb R})$ covariant
derivatives  $D_{ij}$  in the module
induced from the vacuum $|i,j|$. Note that
$|C_{ij}(\gb_{i}\,,\overline{\gb}_j |X)\rangle $ and
$|\omega_{ij}(\gb_{i}\,,\overline{\gb}_j|X)\rangle$
also form modules of $su(2,2)$ with the generators (\ref{dil}), (\ref{cgs}),
of  $osp(1|8,{\mathbb R})$ with the
supergenerators (\ref{osp})
and of the infinite dimensional HS superalgebra whose
generators are various (i.e., not only bilinear)
polynomials of  $a_A$ and $b^A$.
Let us stress that, because the vacua $|i,j|$
are Lorentz invariant and have definite scaling dimensions,
so defined $sp(8,{\mathbb R})$-modules consist of towers of Lorentz
multispinor fields with definite scaling dimensions.
(If vacua were not Lorentz invariant, the resulting Lorentz
algebra--modules could be infinite dimensional.)

In our construction we postulate that the oscillators
$a_\ga$ and $b^\ga$ are complex conjugated to $\bar{a}_\pa$ and
$\bar{b}^\pa$, respectively.
With this convention the conjugation $\sigma$, that singles
out the real form $sp(8|{\mathbb R})$ of $sp(8|{\mathbb C})$,
 acts as follows
\be\nn
\sigma(P_{AB})= P_{AB}\,,\qquad
\sigma(L_A{}^B ) = L_A{}^B\,,\qquad
\sigma(K^{AB})= K^{AB}\,.
\ee
{}From the definition of the Fock modules $|i,j|$ it follows
then that $
\overline{|i,j|} = |j,i|$,
$\overline{\ga_{i\ga}} = \overline{\ga}_{i\pa}$,
$\overline{\gb_i{}^{\ga}} = \overline{\gb}_{i}{}^{\pa}\,.
$
Correspondingly, the following reality conditions are imposed
\be\nn
\overline{\go_{ij}(\gb_{i}\,,\overline{\gb}_j|X)}=
\go_{ji}(\gb_{j}\,,\overline{\gb}_i|X)\q
\overline{C_{ij}(\gb_{i}\,,\overline{\gb}_j|X)}=
C_{ji}(\gb_{j}\,,\overline{\gb}_i|X)\,.
\ee

The original
$sp(8,{\mathbb R})$ invariant form of the HS equations (\ref{10eq}) is the
covariant constancy condition
$
D_{00}C_{00}(b|X)=0\,
$
which is the analogue of
flat space limit of the $4d$ equation
(\ref{CON22}) in Cartesian coordinates. Our aim is to
extend the equations (\ref{CON12}) and (\ref{CON22})
first to  $su(2,2)$ and then to $sp(8,{\mathbb R})$ symmetric
formulations.
 To make it possible to use
general properties of the unfolded formulation,
this will be done for generic $su(2,2)$ and
$sp(8,{\mathbb R})$ flat connections in Sections \ref{conf}
and \ref{sp8}, respectively.
As a result, the obtained systems will be proven to
have global $su(2,2)$ and  $sp(8,{\mathbb R})$
symmetries that
act both on the 1-form HS gauge fields and on the 0-form field strengths.

\section{Conformal invariant unfolded massless equations}
\label{conf}

\subsection{Consistent equations}

The $u(2,2)\subset sp(8, \mR)$ covariant derivatives in
the various Fock modules are defined by
$D^{con} |C_{ij}(\gb_{ij} |X)\rangle$
where
\be\nn
\label{mcov}
D^{con} = d + w \q w= h^{\ga\pa} a_\ga \bar{a}_\pa +\go_\gb{}^\ga
a_\ga b^\gb +\overline{\go}_\pb{}^\pa \bar{a}_\pa \bar{b}^\pb
 + f_{\ga\pb} b^\ga \bar{b}^\pb\,.
\ee
Here the traceless parts $\go^{L}{}_{\ga}{}^\gb$ and
$\overline{\go}^{L}{}_{\pa}{}^\pb$
of $\go_\ga{}^\gb$ and $\overline{\go}_\pa{}^\pb$, respectively,
describe the Lorentz connection while the traces describe the gauge
fields $b$ and $\tilde{b}$ (\ref{btb}).
For the generating functions (\ref{cij}) this defines the
 the covariant derivatives
$D_{ij}^{con}$,
\be\nn
(D_{ij}^{con}C_{ij}(\gb_{i},\overline{\gb}_j|X))\star |i,j|=
D^{con} |C_{ij}(\gb_{i},\overline{\gb}_j |X)\rangle \,,
\ee
which have the form
\be
\label{dcon10}
D^{con}_{10} = D^L - \half \omega_\ga{}^\ga (a_\gb \f{\p}{\p a_\gb}+1)+
\half \overline{\omega}_\pa{}^\pa (\bar{b}^\pb \f{\p}{\p \bar{b}^\pb}+1)
 + h^{\ga\pb}
a_\ga \f{\p}{\p \bar{b}^\pb} -f_{\ga\pb} \f{\p}{\p a_\ga} \bar{b}^\pb\,,
\ee
\be
\label{dcon01}
D^{con}_{01} = D^L +  \half\omega_\ga{}^\ga (b^\gb \f{\p}{\p b^\gb}+1)-
\half \overline{\omega}_\pa{}^\pa (\bar{a}_\pb \f{\p}{\p \bar{a}_\pb}+1)
 +  h^{\ga\pb}
\bar{a}_\pb \f{\p}{\p {b}^\ga} -f_{\ga\pb} \f{\p}{\p \bar{a}_\pb}
{b}^\ga\,,
\ee
\be
\label{dcon00}
D^{con}_{00} = D^L + \half\omega_\ga{}^\ga (b^\gb \f{\p}{\p b^\gb}+1)+
\half \overline{\omega}_\pa{}^\pa (\bar{b}^\pb \f{\p}{\p \bar{b}^\pb}+1)
 +h^{\ga\pb}
\f{\p^2}{\p b^\ga\p\bar{b}^\pb} +f_{\ga\pb}b^\ga\bar{b}^\pb\,,
\ee
\be
\label{dcon11}
D^{con}_{11} = D^L - \half\omega_\ga{}^\ga (a_\gb \f{\p}{\p a_\gb}+1)-
\half \overline{\omega}_\pa{}^\pa (\bar{a}_\pb \f{\p}{\p \bar{a}_\pb}+1)
 +h^{\ga\pb}a_\ga \bar{a}_\pb +f_{\ga\pb}
\f{\p^2}{\p a_\ga \p\bar{a}_\pb} \,,
\ee
where $D^L$ is the Lorentz covariant derivative (\ref{dlor}).

{}From the form of covariant derivatives $D^{con}_{ij}$ it follows
in particular that the operators of helicity $\Hh$  (\ref{hel})
and dilatation $\D$ (\ref{dil}) act on the respective modules
$\phi_{ij}=\go_{ij}$ or $\phi_{ij}=C_{ij}$ as follows
\be
\label{h10}
\Hh \phi_{10} (a,\bar{b}) = -\half \left ( a_\ga \f{\p}{\p a_\ga} +
\bar{b}^\pa \f{\p}{\p \bar{b}^\pa} +2\right ) \phi_{10} (a,\bar{b})\,,
\ee
\be
\label{h01}
\Hh \phi_{01} (b,\bar{a}) = \half \left ( b^\ga \f{\p}{\p b^\ga} +
\bar{a}_\pa \f{\p}{\p \bar{a}_\pa} +2\right ) \phi_{01} (b,\bar{a})\,,
\ee
\be
\label{h00}
\Hh \phi_{00} (b,\bar{b}) = \half \left ( b^\ga \f{\p}{\p b^\ga} -
\bar{b}^\pa \f{\p}{\p \bar{b}^\pa} \right ) \phi_{00} (b,\bar{b})\,,
\ee
\be
\label{h11}
\Hh \phi_{11} (a,\bar{a}) = -\half \left ( a_\ga \f{\p}{\p a_\ga} -
\bar{a}_\pa \f{\p}{\p \bar{a}_\pa} \right ) \phi_{11} (a,\bar{a})\,,
\ee
\be
\label{d10}
{\cal D} \phi_{10} (a,\bar{b}) = -\half \left ( a_\ga \f{\p}{\p a_\ga} -
\bar{b}^\pa \f{\p}{\p \bar{b}^\pa}\right)\phi_{10} (a,\bar{b})\,,
\ee
\be
\label{d01}
{\cal D} \phi_{01} (b,\bar{a}) = \half \left ( b^\ga \f{\p}{\p b^\ga} -
\bar{a}_\pa \f{\p}{\p \bar{a}_\pa} \right ) \phi_{01} (b,\bar{a})\,,
\ee
\be
\label{d00}
{\cal D} \phi_{00} (b,\bar{b}) = \half \left ( b^\ga \f{\p}{\p b^\ga} +
\bar{b}^\pa \f{\p}{\p \bar{b}^\pa} +2\right ) \phi_{00} (b,\bar{b})\,,
\ee
\be
\label{d11}
{\cal D} \phi_{11} (a,\bar{a}) = -\half \left ( a_\ga \f{\p}{\p a_\ga} +
\bar{a}_\pa \f{\p}{\p \bar{a}_\pa}+2 \right ) \phi_{11} (a,\bar{a})\,.
\ee

Now we are in a position to write the
conformal invariant unfolded system of equations
\be
\label{cunf10}
D^{con}_{10}\omega_{10} (a,\bar{b}) =
h_\ga{}^\pa\wedge h^{\ga \pb}\f{\p^2}{\p \bar{b}^\pa \p \bar{b}^\pb}
C_{00}(0,\bar{b})+ f_\ga{}_\pa \wedge f_{\gb}{}^\pa
\f{\p^2}{\p  a_\ga \p a_\gb} C_{11}(a,0)\,,
\ee
\be
\label{cunf01}
D^{con}_{01}\omega_{01} (b,\bar{a}) =
h^\ga{}_\pa\wedge h^{\gb \pa}\f{\p^2}{\p {b}^\ga \p {b}^\gb}
C_{00}({b},0)+ f_\ga{}_\pa \wedge f^{\ga}{}_\pb
\f{\p^2}{\p  \bar{a}_\pa \p \bar{a}_\pb} C_{11}(0,\bar{a})\,,
\ee
\be
\label{cunf00}
D^{con}_{00}\omega_{00} (b,\bar{b}) =
f_\ga{}_\pa\wedge f^{\ga}{}_{\pb}\bar{b}^\pa \bar{b}^\pb
C_{10}(0,\bar{b})+
 f_\ga{}_\pa \wedge f_{\gb}{}^\pa
b^\ga b^\gb C_{01}(b,0)\,,
\ee
\be
\label{cunf11}
D^{con}_{11}\omega_{11} (a,\bar{a}) =
h_\ga{}^\pa\wedge h^{\ga}{}^{\pb}\bar{a}_\pa \bar{a}_\pb
C_{10}(0,\bar{a})+
 h^\ga{}_\pa \wedge h^{\gb}{}^\pa
a_\ga a_\gb C_{01}(a,0)\,,
\ee
\be
\label{dc}
D^{con}_{ij} C_{ij} =0\,.
\ee
This system decomposes into two independent
subsystems. One contains the 1-forms $\go_{ii}$ and
0-forms $C_{i\,1-i}$ while another one contains
$\go_{i\,1-i}$ and $C_{ii}$. As will be explained in Subsection \ref{dyncon},
the subsystem  (\ref{cunf00}), (\ref{cunf11}) and
(\ref{dc}) with $i=j$ is topological while the subsystem (\ref{cunf10}),
(\ref{cunf01})  and (\ref{dc}) with $i+j=1$
describes massless fields of all spins.

The important property of the system (\ref{cunf10})-(\ref{dc})
is that it is  consistent
for any flat  $u(2,2)$ connection.
Let us for definiteness  consider the sector of
$\omega_{10}(a ,\bar{b})$, \ie the
equation (\ref{cunf10}) along with
\be
\label{cex1}
D^{con}_{00}C_{00}(b,\bar{b})=0\,,\qquad D^{con}_{11}C_{11}(a,\bar{a})=0\,.
\ee
Because vacuum connections are such that $D^{con}_{ij}$
squares to zero, the proof of consistency  is equivalent to checking
that the application of $D^{con}_{10}$ to the \rhs
of ({\ref{cunf10}) gives zero by virtue of the equations (\ref{cex1}).
The analysis of the
two terms on the \rhs of ({\ref{cunf10}) is independent
of each other. Since they are exchanged by the
Chevalley automorphism
that exchanges translations and special conformal transformations,
we only consider the $h$--dependent term in ({\ref{cunf10}), which
is not a coboundary  because the
$h$--dependent terms in $D^{con}_{10}$ (\ref{dcon10}) are proportional
 to $a_\ga$ while the $h$--dependent terms on the \rhs of
 (\ref{cunf10}) are $a$--independent.
The proof of cocyclicity is elementary and consists of the
following observations:
\begin{itemize}
\item
Since the whole setup is Lorentz covariant,
the $\go^L, \overline{\go}^L$-dependent terms in the consistency conditions
cancel out.

\item
The $f_{\ga\pb}$--dependent  terms cancel out because
the one in $D^{con}_{10}$ contains derivative over $a_\ga $
of the $a_\ga$-independent expression on the \rhs of (\ref{cunf10})
while the other one in $D^{con}_{00} C_{00}(b,\overline{b})$
disappears upon setting $b=0$.

\item
The $h^3$- terms vanish  because
\be
\label{iden}
h^{\ga\pa}\wedge h^{\gb\pb}\wedge h^{\gga{\gamma^\prime}}
\f{\p^3}{\p \bar{b}^\pa \p \bar{b}^\pb \p \bar{b}^{\gamma^\prime}} =0\,
\ee
as a result of antisymmetrization of the
three two-component indices $\ga, \gb,\gga$.

\item
The terms with $\overline{\omega}_\pa{}^\pa$ are the same in
$D^{con}_{10}$ and $D^{con}_{00}$ and cancel each other while
 the terms that result from the differentiation of
$h_\ga{}^\pa\wedge h^{\ga \pb}$ by virtue of
(\ref{h4})
compensate those that result from the commutator of the covariant derivative
with $\f{\p^2}{\p \bar{b}^\pa \p \bar{b}^\pb}$ (equivalently,
the
 $\overline{L}_\pa{}^\pa$ weight of $h_\ga{}^\pa\wedge h^{\ga \pb}$
 compensates that of
$\f{\p^2}{\p \bar{b}^\pa \p \bar{b}^\pb}$).

\item
Finally, the terms
with   ${\omega}_\ga{}^\ga$ also cancel out. Namely the differential parts trivialize
either because of  differentiating a constant ($D^{con}_{1,0}$)
or setting $b^\gb=0$ ($D^{con}_{0,0}$). The constant terms, which are different,
then exactly cancel
the result of differentiation of $h_\ga{}^\pa\wedge h^{\ga \pb}$
(Equivalently, the ${L}_\ga{}^\ga$ weight of
$h_\ga{}^\pa\wedge h^{\ga \pb}$ is compensated by the difference of
the weights of the Fock vacua $|1,0|$ and $|0,0|$).

\end{itemize}

The cocyclicity of all other terms on the \rhss of
(\ref{cunf10})-(\ref{cunf11}) is checked analogously.

The essentials of the construction include

\noindent
{\it (i)}
The matching between the number of values of two-component indices
and the form degree on the \rhss of
(\ref{cunf10})-(\ref{cunf11}).
This allows us to use  the identity  (\ref{iden}).

\noindent
{\it (ii)}
The twist to $C_{1-i\,j}$ or $C_{i\,1-j}$ of the  modules
glued to $\omega_{ij}$ via the \rhss of
(\ref{cunf10})-(\ref{cunf11}) leads
to the shifts of the vacuum helicities and conformal dimensions
(\ie the constant terms in the operators acting on $\phi_{ij}$ in
(\ref{h10})-(\ref{d11})) of the respective Fock vacua that compensate
those carried by  $a$, $b$ or $\f{\p}{\p a}$ and $\f{\p}{\p b}$ on the
\rhs of (\ref{cunf10})-(\ref{cunf11}).

\subsection{Dynamical content}
\label{dyncon}

First of all we observe that the covariant derivatives
$D^{con}_{i\,1-i}$ (\ref{dcon10}) and (\ref{dcon01})
preserve a degree of a polynomial on which it acts.
This means that the Fock modules
induced from the vacua $|i,1-i|$ decompose into
infinite sums of finite dimensional $u(2,2)$-modules
carried by homogeneous polynomials.
On the other hand, from the form of covariant derivatives
$D^{con}_{ii}$ (\ref{dcon00}) and (\ref{dcon11}) it follows
that the Fock modules induced from the vacua $|i,i|$ decompose
into infinite sums of infinite dimensional $u(2,2)$-modules.

{}As explained in Subsection \ref{prop}, local degrees
of freedom of a system are carried by 0-forms. We conclude that,
the fields
$C_{i1-i}$ and, therefore,  by virtue of (\ref{cunf00}) and
(\ref{cunf11}), $\go_{ii}$ describe an infinite set of topological
systems each carrying at most a finite number of degrees
of freedom equal to the dimension of the space of
polynomials of an appropriate degree. Note that the
sector of gauge fields
of this type was originally analyzed by far more complicated
Hamiltonian methods in \cite{aux} with the same conclusion that
these fields are of topological type. (In \cite{aux} these fields were
called auxiliary to emphasize that they do not carry
field-theoretical degrees of freedom.)

On the other hand, each irreducible subsystem in the sector of $C_{ii}$
and $\go_{i1-i}$ describes an infinite number of degrees of freedom.
These are massless fields of various spins. Since
the gauge massless fields $\omega_{10}(a_\ga ,\bar{b}^\pa)$
are complex conjugated to $\omega_{01}(b^\ga ,\bar{a}_\pa)$,
 the system of equations
(\ref{cunf10}),(\ref{cunf01}) and (\ref{dc}) at $i=j$
describes the set of massless fields in which every spin
appears twice. This pattern matches that of the $AdS_4$ HS
theories as discussed in Section \ref{mfe}, although the
mechanism of the doubling is different.

The realization of the helicity operator
$\Hh$ on different Fock modules is given
in (\ref{h10})-(\ref{h11}). Correspondingly,
the eigenvalues of the helicity operator
on the HS gauge 1-forms are
\be\nn
\Hh \go_{10} = -s \go_{10} \q
\Hh \go_{01} = s \go_{01}\,,
\ee
where spin $s$ is defined according to (\ref{CON1}), {\it i.e.,}
\be\nn
\left ( y^\ga\f{\p}{\p y^\ga} + \bar{y}^\pa \f{\p}{\p \bar{y}^\pa} \right )
\go(y,\bar{y} |x) = 2(s-1) \go(y,\bar{y} |x)\,.
\ee
(We assume that spin is non-negative while helicities $\pm s$
may have any sign.)
On the HS Weyl 0-forms, the eigenvalues of the helicity operator are
\be\nn
\Hh C_{00} = \pm s C_{00} \q
\Hh C_{11} = \mp s C_{11}\,,
\ee
where signs are determined by those
of the corresponding eigenvalues of the operators (\ref{h00}) and
(\ref{h11}) (cf. (\ref{CON1}), (\ref{CON2})).
We see that the $u(1)$ symmetry generated by
$\Hh$ rotates two species of fields of all spins
with the spin-dependent phases.

Let us expand $\go_{i\,1-i}(y,\bar{y}|x) $ into the real and imaginary
parts
\be
\label{go1}
\go_{10}(y,\bar{y}|x) = \go_{1}(y,\bar{y}|x)+i \go_{2}(y,\bar{y}|x)\q
\go_{01}(y,\bar{y}|x) = \go_{1}(y,\bar{y}|x)-i \go_{2}(y,\bar{y}|x)\,,
\ee
where $\go_{i}$ are real in the sense
$
\overline{\go_{i}(y,\bar{z}|x)} =\go_{i}(z,\bar{y}|x)\,.
$
In the $AdS_4$ case, where the $u(2,2)$ connections
are realized by those of $sp(4,\mR)\subset u(2,2)$ according
to (\ref{adsc}), both $h^{\ga\pa}\neq 0$ and $f_{\ga\pa}\neq 0$,
allowing to express the 0-forms $C_{ii}$  via the
space-time derivatives of the dynamical massless fields.
In this case, the equations (\ref{cunf10})
and (\ref{cunf01}) take the form
\bee
\label{r1}
R_{1}(y,\by|x) =
\overline{H}^{\pa \pb}\f{\p^2}{\p \bar{y}^\pa \p \bar{y}^\pb}
C_{1}(0,\bar{y}|x)
+H^{\ga\gb}
\f{\p^2}{\p {y}^\ga \p {y}^\gb}
C_{1}({y},0|x)\,,
\eee
\bee
\label{r2}
R_{2}(y,\by|x) =
\overline{H}^{\pa \pb}\f{\p^2}{\p \bar{y}^\pa \p \bar{y}^\pb}
C_{2}(0,\bar{y}|x)
-H^{\ga\gb}
\f{\p^2}{\p {y}^\ga \p {y}^\gb}
C_{2}({y},0|x)\,,
\eee
where $R_{1,2}(y,\by|x)$ have the form of the linearized HS curvatures
(\ref{RRR}) and
\be
\label{c1}
C_1(y,\bar{y}|x) =\frac{\lambda^2}{2}
(C_{00}(y,\bar{y}|x)+C_{11}(y,\bar{y}|x))\q
C_2(y,\bar{y}|x) =\frac{\lambda^2}{2i}
(C_{00}(y,\bar{y}|x)- C_{11}(y,\bar{y}|x))
\ee
satisfy  the twisted adjoint covariant constancy condition
\be
\label{tw12}
D^{tw}C_{i}(y,\bar{y}|x) =0\,
\ee
with $D^{tw}$  (\ref{tw}) and the reality conditions
$
\overline{C_1(y,\bar{z})}=C_1(z,\bar{y})\,,\quad
\overline{C_2(y,\bar{z})}=-C_2(z,\bar{y})\,.
$

The equations (\ref{r1}), (\ref{r2}) and (\ref{tw12})
are equivalent to the unfolded massless equations (\ref{CON12}) and
(\ref{CON22}) in the $AdS_4$ background with the identification
\be\nn
\go^{11}= \go_{2}\q\go^{00}= \go_{1}
\q C^{1\,0} = C_{1\,-}-C_{2\,+}+C_{1\,0}-C_{2\,0}\q
C^{0\,1} = C_{1\,+}+C_{2\,-}+C_{1\,0}+C_{2\,0}\,,
\ee
where the labels $+,-$ and 0 refer to the decomposition (\ref{dec}).
Thus, in the $AdS_4$ background,
the conformal invariant equations (\ref{cunf10}), (\ref{cunf01})
and (\ref{dc}) with $i=j$ amount to the standard unfolded
 field equations for the doubled set of free massless fields of
all spins.

\subsection{Global symmetries and EM duality}
\label{gsfl}
As a consequence of general properties of unfolded
equations, the massless
equations (\ref{cunf10}), (\ref{cunf01}) and (\ref{dc}) are
invariant under the $u(2,2)$ global symmetry that consists
of the $su(2,2)$ conformal symmetry and $u(1)$
EM duality transformation generated by the
helicity operator $\Hh$. Since $\Hh$ is  central in $u(2,2)$,
from (\ref{glpar}) it
follows that the global symmetry parameter  of EM
duality transformation remains $x$-independent.

The transformation law of the gauge 1-forms
consists of the Lie-algebraic transformation
 in the module carried by the gauge fields and
 the additional terms resulting from the Chevalley-Eilenberg
 cohomology terms via  (\ref{delw}). {F}rom the
equations (\ref{cunf10}) and (\ref{cunf01}) it follows that
the transformation law is
\bee
\label{del10}
\ls\ls\delta |\go_{10}(a,\bar{b})\rangle &=& -
\epsilon_{gl}(a,\bar{a},b,\bar{b}|x)\star
|\go_{10}(a,\bar{b})\rangle \nn\\
 &\ls\ls+&\ls\!\!2\Big (
\epsilon_\ga{}^\pa (x) h^{\ga \pb}\f{\p^2}{\p \bar{b}^\pa \p \bar{b}^\pb}
C_{00}(0,\bar{b})+ \tilde{\epsilon}_\ga{}_\pa (x)  f_{\gb}{}^\pa
\f{\p^2}{\p  a_\ga \p a_\gb} C_{11}(a,0)\Big )\star |1,0| \,,
\eee
\bee
\label{del01}
\ls\ls\delta |\omega_{01} (b,\bar{a})\rangle  &=&
-\epsilon_{gl}(a,\bar{a},b,\bar{b}|x)\star
|\go_{01}(a,\bar{b})\rangle\nn\\
&\ls\ls+&\ls\!\!2\Big (
\epsilon^\ga{}_\pa (x) h^{\gb \pa}\f{\p^2}{\p {b}^\ga \p {b}^\gb}
C_{00}({b},0)+ \tilde{\epsilon}_\ga{}_\pa (x) f^{\ga}{}_\pb
\f{\p^2}{\p  \bar{a}_\pa \p \bar{a}_\pb} C_{11}(0,\bar{a})\Big )
\star |1,0\rangle\,,
\eee
where $\epsilon^\ga{}_\pa (x)$ and $\tilde{\epsilon}_\ga{}_\pa (x)$
are  parameters of global translations and special conformal
transformations, respectively. All other symmetry parameters of the
conformal algebra enter only through the original (\ie $C$--independent)
module transformation law $\epsilon\star |\go \rangle$.
The transformation
law of the  Weyl 0-forms does not deform
\be\nn
\label{delc00}
\ls \delta |C_{00}(b,\bar{b}|x)\rangle= -
\epsilon_{gl}(a,\bar{a},b,\bar{b}|x)
\star |C_{00}(b,\bar{b}|x)\rangle\q
\delta |C_{11}(a,\bar{a}|x)\rangle= -
\epsilon_{gl}(a,\bar{a},b,\bar{b}|x)
\star |C_{11}(a,\bar{a}|x)\rangle\,.
\ee

The precise form of the transformation law depends on a
chosen vacuum connection and, in particular, on a coordinate system
parameterized by the function $U^{AB}(X)$
of Section \ref{star}. For a broad class
of vacuum connections, the global symmetry parameter
$\epsilon_{gl}(a,\bar{a},b,\bar{b}|X)$ can be
obtained from (\ref{gauge}).

It is important to note that the full conformal symmetry
does not act individually on every massless field, mixing
together the two copies of fields of equal spins.
This happens because of the gauge field sector where the
complex conjugated gauge fields $\go_{i\,1-i}$ with $i=0$,$1$
transform differently. This in particular implies that the
real fields $\go_i$ (\ref{go1}) are mixed by the
$su(2,2)$ transformations as well as by the EM
duality transformations. Correspondingly, the 0-forms
$C_i$ (\ref{c1}) are also mixed by the $u(2,2)$
transformations. This agrees with the fact
that the EM duality transformations cannot act locally on a single
gauge potential. On the other hand, that the $AdS_4$ symmetry
$sp(4,\mR)$ acts individually on the each of
the two subsets of $\go_i$ and $C_i$ with $i=1$ or 2
leads to two decoupled $sp(4,\mR)$ invariant HS systems in
(\ref{r1}), (\ref{r2}) and (\ref{tw12}).
 Somewhat surprisingly, we shall
see in Subsection \ref{sl4} that, for any spin, the $sp(4,\mR)$ symmetry
of each of these subsystems extends  to
$sl(4,{\mathbb R})\sim o(3,3) \subset sp(8,\mR)$.

Conformal invariant  truncation to the undoubled set of
massless fields can be obtained as follows. Since the
generalized Weyl tensors
$C_{00}$ and $C_{11}$ are self-conjugated, each carrying
a $u(2,2)$ (in fact, $sp(8,{\mathbb R})$) module, it is
consistent with the $u(2,2)$ symmetry
to set $C_{ii}=0$ for some $i$. Let us, for example, set
\be
\label{c0}
C_{11}=0\,.
\ee
This implies that the fields $C_i$ (\ref{c1}) are linearly
dependent, namely
$
C_2(y,\bar{y}) = -i C_1 (y,\bar{y})\,.
$

Either of the equations (\ref{r1}) or (\ref{r2})
along with the respective covariant constancy condition
(\ref{tw12}) for $C_1$ or $C_2$ describes the set of massless
fields of all spins. However, the other one then has  different
interpretation. Let, say, $\omega_{1}(y,\bar{y}|x)$ be chosen as
independent {\it electric} HS gauge connection. The equation (\ref{r1})
expresses the generalized Weyl tensors
in terms of derivatives of $\omega_{1}(y,\bar{y}|x)$.
Then the meaning of the equation (\ref{r2}) with $C_{11}=0$ is
that it defines the gauge field $\omega_2(y,\bar{y}|x)$ in terms of
$C_1(y,\bar{y}|x)$. If the form of
the \rhss of the equations (\ref{r1}) and (\ref{r2})
was the same, the corresponding equations  meant that a linear
combination of $\omega_1(y,\bar{y}|x)$ and $\omega_2(y,\bar{y}|x)$
would be pure gauge. However,
the \rhss of the equations (\ref{r1}) and (\ref{r2})
have different relative signs of the holomorphic and antiholomorphic
parts. As a result, (\ref{r2}) expresses $\omega_2(y,\bar{y}|x)$ as
the EM dual of $\omega_1(y,\bar{y}|x)$,
\ie the potential $\omega_2(y,\bar{y}|x)$ is {\it magnetic}.

In terms of the complex gauge fields $\omega_{i\,1-i}$,
the (anti)holomorphic  components ($C_{11}(0,\bar{a})$) $C_{11}(a,0)$
 describe (anti)selfdual components of the
complexified $s=1$ Maxwell field
strength ($\bar{C}_{\pa\pb}$) $C_{\ga\gb}$, $s=2$ Weyl tensor
($\bar{C}_{\pa_1\ldots \pa_4}$) $C_{\ga_1\ldots \ga_4}$
and their HS counterparts. As a result, the condition
(\ref{c0}) implies that the HS field strength is (anti)selfdual.
The (anti)selfduality condition imposed in the Minkowski signature
is consistent  because the gauge field $\go_{10}$ is complex,
having $\go_{01}$ as its complex conjugate.
This complex (anti)selfduality condition relates the field
strength of one of the two spin $s$ real fields contained in $\go_{10}$
to the dual field strength of the other one. Let us stress that this
relationship is non-local in terms of gauge potentials,
being expressed by a differential duality
equation that follows from (\ref{c0}).

For example, for the spin one case in tensor notation,
(\ref{c0})  implies the condition
\be
\label{d1}
F^i_{nm} =\pm \half \epsilon^{ij} \varepsilon_{nm}{}^{pq} F^j_{pq}\q
F^i_{nm} = \partial_n \go^i_m -\partial_m \go^i_n\,,
\ee
where $i,j=1,2$ label two real components of the
complex spin one potential part $\omega_{10}(0,0)$ of
$\omega_{10}(a,\bar{b})$ ($\epsilon^{ij} = - \epsilon^{ji}$,
$\epsilon^{12} = 1$). The relationship  (\ref{d1})
between $F^1_{nm}$ and ${}^*F^2_{nm}$ is consistent
in Minkowski case due to the factor of $\epsilon^{ij}$
which becomes the imaginary unit in complex notation.

Thus, the unfolded equations (\ref{cunf10}),
(\ref{cunf01}) and (\ref{dc}) with $C_{11}=0$ or $C_{00}=0$
describe the undoubled set of all spins in conformal and
duality invariant way. In this case, the unfolded HS equations
describe both electric and magnetic HS potentials.
The conformal transformations resulting from the
$\go$-dependent terms in (\ref{del10}) and (\ref{del01})
mix the HS electric and magnetic potentials.
In agreement with the known conformal invariance of
(distinguished) $4d$ spin one models,
the  spin one case is degenerate, allowing
conformal transformation that acts only on the electric
potential.
Technically, this happens because  the spin one connection
carries trivial $su(2,2)$--module so that the
$\go$-dependent part of the transformation
(\ref{del10}) and (\ref{del01}) is absent in this case, \ie the
conformal transformation of spin one
results only from the $C$--dependent terms in (\ref{del10})
and (\ref{del01}).

It is tempting to speculate that the restrictions $C_{11}=0$ or
$C_{00}=0$ may result from some dynamical mechanism in a full
theory with spontaneously broken HS and $sp(8,{\mathbb R})$
symmetries, that freezes degrees of freedom of the magnetic phase
at least below some mass scale, leaving us only with the half of
degrees of freedom corresponding to the electric phase. The EM duality
can then be expected to be a true local symmetry
of a full $sp(8,{\mathbb R})$ invariant nonlinear HS gauge theory.
This can even be true in the HS models with nonAbelian Yang-Mills
symmetries considered in \cite{KV,Ann} and in the sector of spin
two massless fields that can appear in the HS theories in many copies,
 carrying color indices\footnote{The no-go statements
 on the existence of
 models with several interacting spin two fields \cite{CW,BDGH}
are avoided analogously to the no-go statements
 on HS interactions \cite{DA,BHWN} due to using
the $AdS_4$ background and infinite sets of HS fields.}. It would be interesting to
 interpret
 from this perspective the gravitational duality studied e.g.
 in \cite{DT,Nieto,HT,JLR,DS,LP}.

\subsection{Flat limit}

{}From (\ref{d10})-(\ref{d11}) it follows that the full chain of
field variables in the unfolded formulation of a massless field
of definite spin contains components with the conformal
weights from $-\infty$ to $\infty$.
The dynamical HS fields are those with
minimal absolute values of conformal weights, namely
$\D=\pm 1$ for spin zero, $\D=\pm 3/2$ for spin one-half,
 $\D=0$  for spin $s\geq 1$ bosonic fields and
 $D=\pm \half$ for spin $s\geq 3/2$ fermionic gauge
 fields (for more detail see the $\gs_-$--cohomology
 analysis of Section \ref{ceq}). All other fields in the chain are
 auxiliary, being expressed via derivatives
 of the dynamical fields by the unfolded equations.
Namely, the difference of the modules of conformal dimensions of
a certain auxiliary field $A$ and related
dynamical field $D$ equals to the highest number of derivatives
in the resulting expression $A(\p^k (D))$. The property that the module of the
conformal dimension rather than the conformal dimension itself
matters, is intimately  related to the relevance of the
$AdS$ background: the mismatch of dimension is compensated by
the powers of the $AdS$ radius $\lambda^{-1}$.

Let us consider more closely
the translation and special conformal transformations
of  the gauge field $\go_{10}$. {}From (\ref{dcon10}) it follows that,
discarding the $C$--dependent terms,
\be\nn
\delta \omega_{10} (a,\bar{b}|x)=
 -2(\epsilon_{gl}^{\ga\pb} (x)
a_\ga \f{\p}{\p \bar{b}^\pb} -\tilde{\epsilon}_{gl\,\ga\pb} (x)
\f{\p}{\p a_\ga} \bar{b}^\pb ) \omega_{10} (a,\bar{b}|x)+ O(C)\,.
\ee
In terms of components (\ref{gg}) this gives
\be
\label{gltrans}
\delta \omega_{10}{}^{\ga_1\ldots\ga_n\,,\pb_1\ldots \pb_m} (x)=
 -2\left (n \epsilon_{gl}{}^{\ga_n}{}_{\pga} (x)
 \omega_{10}{}^{\ga_1\ldots\ga_{n-1}\,,\pga\pb_1\ldots \pb_{m}} (x)
 +m
\tilde{\epsilon}_{gl\,\gga}{}^{\pb_m} (x)
\omega_{10}{}^{\gga\ga_1\ldots\ga_n\,,\pb_1\ldots \pb_{m-1}} (x)\right )
 \,.
\ee
With the identification (\ref{adsc}),
the \lhs of the unfolded equation (\ref{cunf10})
has the form (\ref{RRR}). Abusing notation, the expressions
for auxiliary fields resulting from
 the unfolded system are
\be\nn
\omega_{10} (n,m) (x)=
(\lambda^{-1}D^L )^{[\frac{n-m}{2}]} \go_{10}^{dyn}(n_0 , m_0 )(x)\q
|n_0-m_0|\leq 1\q n_0+m_0 = n+m\,,
\ee
where $\go(n,m)(x)$ symbolizes
$\omega_{10}{}^{\ga_1\ldots\ga_n\,,\pb_1\ldots \pb_m} (x)$
and we only keep track of the highest derivative terms.
In particular, in the case of spin two, the vierbein $\go^{\ga\pb}$
is the dynamical field while the Lorentz connection
$\go^{\ga\gb}$, $\go^{\pa\pb}$ is auxiliary.

The transformation law (\ref{gltrans}) then
implies
\be
\label{delstr}
\delta \omega^{dyn}_{10}(x)\sim
 \lambda^{-1}  (\epsilon_{gl}(x) D^L \omega^{dyn}_{10}(x)+
\tilde{\epsilon}_{gl} (x) D^L \omega^{dyn}_{10}(x)) \,.
\ee
This transformation law  is ill-defined in the
flat limit $\lambda\to 0$.
It is possible to rescale  the generators of
the conformal algebra
$
P_{\ga\pa}\to \mu^{-1} P_{\ga\pa}
$,
$
K^{\ga\pa}\to \mu K^{\ga\pa}
$
without affecting their commutation relations. This
induces the  rescaling of  the parameters
$
\epsilon^{\ga\pa}\to \mu \epsilon^{\ga\pa}
$,
$
\tilde{\epsilon}_{\ga\pa}\to \mu^{-1} \tilde{\epsilon}_{\ga\pa}
$
as well as of the corresponding vacuum connections.
This ambiguity can be used
to compensate the factor of $\lambda^{-1}$ in front of
one of the two terms on the \rhs of (\ref{delstr}).
In particular, choosing $\mu = \lambda$ we obtain
instead of
(\ref{delstr})
\be
\label{delres}
\delta \omega^{dyn}_{10}(x)\sim
  \epsilon_{gl}(x) D^L \omega^{dyn}_{10}(x)+ \lambda^{-2}
\tilde{\epsilon}_{gl} (x) D^L \omega^{dyn}_{10}(x) \,.
\ee
In the flat limit, this transformation law is well defined in
the sector of translations but blows up in the sector of special
conformal transformations. This is why spin $s>1$
massless field equations
formulated in terms of gauge fields in Minkowski space
(in particular, linearized gravity) respect Poincare' symmetry
but are not conformal invariant.

A closely related fact is that the
rescalings procedure of Section \ref{hsmin},
that leads to the conventional flat limit description of HS fields
in terms of potentials, breaks down conformal invariance.
The algebraic reason for this is that the
Poincare' covariant derivative in (\ref{adfl}) is defined so that
Poincare' translations on $\tilde{A}_+$ and on $\tilde{A}_-$
are generated, respectively, by the translation and special conformal transformations of the
original conformal module. Clearly, this realization of
 translations is incompatible  with the standard realization
of the Poincare' algebra as a proper
subalgebra of the conformal algebra.
On the other hand, the Lorentz, duality and dilatation
transformations survive in so defined flat limit
because they act on homogeneous polynomials of the oscillators and
commute with the rescalings.

On the other hand,  the naive flat limit of the equations
(\ref{cunf10}), (\ref{cunf01}) and (\ref{dc}) with $f_{\ga\pa}= 0$
at  $\lambda \to 0$ gives rise to conformal invariant
equations
that is hard to interpret. Actually, in this system
$C_{11}$ is not any longer expressed via the
gauge fields $\go_{i\,1-i}$, thus becoming an independent field.
As a result, the system decomposes into two parts. One contains the 1-forms $\go_{i\,1-i}$ along with
the 0-forms $C_{00}$ while another one is the covariant constancy
condition (\ref{dc}) on the 0-form $C_{11}$. The latter
however does not make sense in terms of any conventional
formulation of massless fields because it does not express
higher components of the expansion of $C_{11}(a,\bar{a})$ via the
space-time derivatives of the lower ones, imposing  the
condition that the lowest component is a constant.
From perspective of $\gs_-$--cohomology analysis, this
awkward picture results from the degeneracy of the operator
$\gs_-$ which is zero in this case. As a result, the system
gets a form of an infinite set of topological-like field
equations which, in fact, is hard to interpret. This makes
the naive limit ill-defined eventhough
the subsystem that contains $\go_{i\,1-i}$ and $C_{00}$ is more
tractable if  the antiholomorphic field
$\go_{10}(0,\bar{a}|x)$ (which is the antiholomorphic part of the
Lorentz connection in the spin two case)
is chosen as dynamical.

As mentioned in Subsection \ref{gsfl},
the spin one case is degenerate because there
are no auxiliary gauge connections, \ie both terms
on the \rhs of (\ref{gltrans}) are absent.
The nontrivial transformation originates
 from the $C$-dependent terms on the \rhs of
(\ref{del10}) and (\ref{del01}) where the spin one components of the
0-forms $C_{ii}$ identify with the Maxwell tensor.
These can be rescaled independently to get rid of
the  $\lambda$--dependence, preserving
conformal invariance in the flat limit in agreement with the
fact of conformal invariance of the $4d$ spin one gauge theory
in Minkowski space.

\section{$sp(8,\mR)$ invariant massless equations}
\label{sp8}

The $u(2,2)$ invariant unfolded equations (\ref{cunf10})-(\ref{dc})
admit an extension to the $sp(8,\mR)$ invariant form. The fields are
still described in terms of the modules
$|\go_{i\,1-i}(x)\rangle$ and $|C_{i\,i}(x)\rangle$.
The vacuum covariant derivative (\ref{sp8c}) is defined
in  star-product notation by (\ref{8star}), \ie
\be
\label{spv}
D = d +W\q W=\half h^{AB} a_A a_B +\go_B{}^A a_A b^B + \half
f_{AB} b^A b^B\,.
\ee

A particular form of the $sp(8,\mR)$ covariant derivative depends
on a chosen Fock vacuum
\be\nn
D |C_{ij}(\gb_i,\overline{\gb}_j |X)\rangle =
(D_{ij} C_{ij}(\gb_i,\overline{\gb}_j|X))\star |i,j|\,.
\ee
For example,
\be
\label{d00sp}
D_{00}= d+\half h^{AB} \f{\p^2}{\p b^A \p b^B } +
\go_B{}^A b^B \f{\p}{\p b^A} +\half
\go_C{}^C   + \half f_{AB} b^A b^B\,.
\ee
Here $\go_B{}^A $ are gauge fields of $gl(4,{\mathbb R})$ that act
on homogeneous polynomials of $b^A$. The
vacuum $|0,0|$ forms a one dimensional $gl(4,{\mathbb R})$-module.
The $gl(4,{\mathbb R})\subset sp(8,{\mathbb R})$
is an extension of the conformal embedding
$gl(2,{\mathbb C})\subset u(2,2)$, where $gl(2,{\mathbb C})$
contains Lorentz transformations, dilatations and duality
transformations.

Note that, every vacuum $|i,j|$ has the $gl_{ij}(4,{\mathbb R})$
invariance
with the generators bilinear in the respective creation and
annihilation generators.
The subalgebras $gl_{ij}(4,{\mathbb R})\subset sp(8,{\mathbb R})$ are
different for different $i,j$, having $gl(2,{\mathbb C})$ as
the maximal common subalgebra. As a result, $gl(2,{\mathbb C})$
remains the maximal manifest symmetry of the construction.

The explicit form of the derivatives $D_{ij}$, that extend the conformal covariant derivatives
(\ref{dcon10})-(\ref{dcon11}) to the $sp(8,{\mathbb R})$ case,
 can be obtained from (\ref{d00sp}) by renaming  the oscillators
 according to (\ref{ai}) and (\ref{bi}).
The equation (\ref{dc}) still has the covariant constancy form
\be
\label{CON32}
D_{ij} C_{ij} =0\,.
\ee
The extension of (\ref{cunf10})-(\ref{cunf11}) is
\be
\label{CON31}
D\omega_{ij}(\beta_{i},\overline{\gb}_j) =( \overline{\Delta}_{i\, {j}}\otimes
C_{i\, 1-j}(\beta_{i},\overline{\gb}_{1-j}))
\Big |_{\overline{\beta}_{1-j}=0}
 +
(\Delta_{{i}\,j} \otimes
C_{1-i\,j}(\beta_{1-i},\overline{\gb}_j))\Big |_{\beta_{1-i} =0}\,,
\ee
where
\be
\label{de}
\Delta_{{i}\,j} =\f{1}{4}\epsilon^{\ga\gb}
\f{\p}{\p \gb_{i}^\ga}
W(0,\beta_{i},\overline{\alpha}_{j}, \overline{\beta}_{j})
\wedge \f{\p}{\p \gb_{i }^\gb } W
(0,\beta_{i},\overline{\alpha}_{j}, \overline{\beta}_{j})\,,
\ee
\be
\label{dep}
\overline{\Delta}_{{i}\,{j}} =\f{1}{4}\epsilon^{\pa\pb}
\f{\p}{\p \overline{\gb}_{ j}^\pa}
W(\alpha_{i},\beta_{i},0, \overline{\beta}_{j})
\wedge \f{\p}{\p \overline{\gb}_{ j}^\pb }
W (\ga_{i},\beta_{i},0,
\overline{\beta}_{j})\,
\ee
with $W(a_A,b^B)$ (\ref{spv}) represented as
$W(\alpha_{i},\beta_{i},\overline{\alpha}_{j},\overline{\beta}_{j})$.
The tensor product symbol  $\otimes$ in  (\ref{CON31}) means that
$ W(\ga_i,\beta_{i},\overline{\alpha}_{j}, \overline{\beta}_{j})$
is realized as a sum of operators that
act both on the $(i,j)$-module and on the $(i,1-j)$-module for
$\overline{\Delta}_{ij}$ or
both on the $(i,j)$-module and on the $(1-i,j)$-module for
${\Delta}_{ij}$. Namely, by  (\ref{ab}),
\be\nn
\overline{\beta}_{ j} \otimes = \overline{\beta}_{ j} +
\overline{\alpha}_{ 1-j}\,,
\qquad
\beta_{i} \otimes =  2\beta_{i}\,,\qquad
\alpha_{i} \otimes = 2\alpha_{i}\,
\ee
in the $\overline{\Delta}_{ij}$ term
and
\be\nn
\beta_{i} \otimes = \beta_{i} +\alpha_{1-i}\,,
\qquad
\overline{\beta}_{j} \otimes =  2\overline{\beta}_{j}\,,\qquad
\overline{\alpha}_{j} \otimes = 2\overline{\alpha}_{j}\,
\ee
in the
${\Delta_{ij}}$ term in (\ref{CON31}).
Here $\alpha_k$ and $\overline{\ga}_l$ are understood as
derivatives over $\gb_k$ and $\overline{\gb}_l$
(precise signs follow from the definitions (\ref{ai})
and (\ref{bi})) with the convention that the differentiation
is done before $\overline{\beta}_{1-j}$ or $\gb_{1-i}$
are set  to zero in (\ref{CON31}).

The proof of consistency of the  unfolded system
(\ref{CON32}) and (\ref{CON31}) is relatively simple but still
miraculous.

First of all we observe that the operator
$\Delta_{{i}\,j}$ ($\overline{\Delta}_{{i}\,{j}}$) (\ref{de})
contains connections that have positive $\gb_{i}
- \ga_{i}$ ($\overline{\gb}_{j} - \overline{\ga}_{j}$) grading.
 Let us for definiteness, consider the
case $i=1,j=0$. Using (\ref{ai}) and (\ref{bi})
we find that the connections that contribute to
 $\Delta_{{1}\,0}$ are
\be
\label{neg}
\half h^{\ga\gb}a_\ga a_\gb\q h^{\ga\pb}a_\ga \overline{a}_\pb \q
\omega_\pb{}^\ga a_\ga \overline{b}^\pb \,,
\ee
and
\bee
\label{d100}
\Delta_{{1}\,0}\otimes C_{00}(b, \bar{b})\Big |_{b=0}
 &{}&= \Big ( \half h^{\gga \gb} (a_\gb +\f{\p}{\p b^\gb})
+h^{\gga\pa} \f{\p}{\p \overline{b}^\pa } +\omega_\pb{}^\gga{} \overline{b}^\pb
\Big )\nn\\
&{}&\wedge
\Big ( \half h_{\gga}{}^\ga (a_\ga +\f{\p}{\p b^\ga})
+h_{\gga}{}^{\pb} \f{\p}{\p \overline{b}^\pb } +\omega_{\pa\gga}
\overline{b}^\pa\Big ) C_{00}(b,\overline{b}) \Big \vert_{b=0}\,.
\eee

The analysis is greatly simplified in terms of the
antiholomorphic Fock space vector
\be
\label{unfock}
|\Delta_{{1}\,0}\rangle =
\left (\Delta_{{1}\,0}\otimes C_{00}(b, \bar{b})\Big |_{b=0}\right )
 \overline{*} |\bar{0}\rangle
\langle \bar{1}|\,,
\ee
where $\overline{*}$ denotes the restriction of the star-product
to the subalgebra generated by the oscillators carrying primed indices.
It can be equivalently rewritten as
\be
\label{symco}
|\Delta_{{1}\,0}\rangle =
 ( g^\gga
\wedge
g_{\gga} ) \overline{*} C_{00}(b,\overline{b})
\overline{*}
|\bar{0}\rangle
\langle \bar{1}|\Big |_{b=0}\,,
\ee
where
\be
\label{hc}
g^\gga
= \half h^{\gga \gb} (a_\gb +\f{\p}{\p b^\gb}) +h^\gga\q
h^\gga =h^{\gga\pa} \overline{a}_\pa  +
\omega_\pb{}^\gga \overline{b}^\pb\,.
\ee
Note that, because of the wedge product, the
antisymmetrization
with respect to the indices $\gga$ implies that
$g^\gga \wedge \overline{*} g_\gga$ is symmetrized
 in the primed indices carried by $h^\gga$ (\ref{hc}).
This allowed us to replace $g^\gga \wedge \overline{*} g_\gga$
by its $\overline{*}$--Weyl symbol $ g^\gga \wedge g_{\gga}$
in (\ref{symco}).

In the sector where only the fields (\ref{neg}) are present,
the consistency requires that
\bee
\label{cond}
H&=&\left (\half h^{\ga\gb}\wedge (a_\ga a_\gb - \f{\p}{\p b^\ga}\f{\p}{\p b^\gb})
g^\gga \wedge g_{\gga}
+h^\gga a_\gga\wedge  \overline{*} ( g^\gga
\wedge
g_{\gga} )
 -( g^\gga
\wedge
g_{\gga} ) \overline{*}\wedge h^\gga \f{\p}{\p b^\gga}
\right )
\overline{*} C_{00}(b,\overline{b})
\nn\\
&+&\Big ( dh^{\gga\gb} (a_\gb + \f{\p}{\p b^\gb})
 h_{\gga}\Big )\overline{*} C_{00}(b,\overline{b})\,
\eee
should vanish (the terms resulting from $d h^\gga$ do not contribute to this
sector).  Taking into account that
$
dh^{\ga\gb} =- 2h^{(\ga}\wedge \overline{*} h^{\gb)}+\ldots
$,
where dots denote terms that contain fields of non-positive
$a - b$ grading, we find that the terms
resulting from the noncommutativity of the  star-product in the
second and third terms in the first line
of (\ref{cond}) compensate the last term.
As a result, we obtain that
\be\nn
H\sim  \left ( g^\ga
\wedge g^\gga\wedge g_\gga (a_\ga -\f{\p}{\p b^\ga})\right )
\overline{*} C_{00}(b,\overline{b})\, =0\,
\ee
by antisymmetrization of three two-component indices
of $g^\gb$ due to the wedge product.

It remains to consider the part of the consistency condition
that contains vacuum connections of non-positive
$a - b$ grading. These include
\be\nn
\omega_\gb{}^\ga{} a_\ga b^\gb\,,\quad
\half h^{\pa\pb}\overline{a}_{\pa}\overline{a}_{\pb}\,,\quad
\overline{\omega}_\pb{}^\pa{} \overline{a}_\pa \overline{b}^\pb\,,\quad
\half f_{\pa\pb}\overline{b}^{\pa}\overline{b}^{\pb}\,,\quad
f_{\ga\pb}{b}^{\ga}\overline{b}^{\pb}\,,\quad
\omega_\ga{}^\pa \overline{a}_\pa b^\ga\,,\quad
\half f_{\ga\gb}{b}^{\ga}{b}^{\gb}.
\ee
Since (\ref{d100}) is independent of these fields, they
contribute to the consistency condition
at most linearly. All such terms cancel out trivially   except
for the dilatation field contained in $\omega_\gb{}^\ga a_\ga b^\gb$
which, however, cancels out just  as in the  conformal
algebra case considered in Section \ref{conf}.
This concludes the analysis of the consistency of
the term with $\Delta_{{1},0}$.
The analysis of the terms with other $\Delta_{{i}j}$ and
 $\overline{\Delta}_{i{j}}$
is analogous modulo  renaming  the oscillators.

Since, the consistency of the unfolded equations (\ref{CON31}) and
(\ref{CON32}) has been verified for arbitrary flat $sp(8,\mR)$
 connection, from the general argument of Subsection
 \ref{ff} it follows that the  system (\ref{CON32}), (\ref{CON31})
is invariant under the global $sp(8,\mR)$ transformations.

If  vacuum connection is chosen to belong to
$su(2,2)\subset sp(8,\mR)$, the  system (\ref{CON32}), (\ref{CON31})
amounts to the conformal system (\ref{cunf10})-(\ref{dc})
of Section \ref{conf}, which is therefore also shown to be $sp(8,\mR)$
invariant. As shown in Section \ref{conf}, with this choice
of the vacuum fields it describes
the doubled set of field equations for all massless fields
described by $\go_{i\,1-i}$ and $C_{ii}$ plus an infinite set of
topological fields, each carrying a finite number of degrees of
freedom, described by $\go_{ii}$ and $C_{i\,1-i}$.
The interpretation of the  $sp(8,\mR)$
invariant equations (\ref{CON32}), (\ref{CON31})
depends, however, on the choice
and interpretation of the vacuum fields
associated with a chosen flat $sp(8,\mR)$ connection.
In Section \ref{sl4} we show that the roles
of the massless and topological fields are
exchanged if the nonzero vacuum fields are
associated with the subalgebra $sl(4,{\mathbb R})
\sim o(3,3)\subset sp(8,{\mathbb R})$.

By the general argument of Subsection \ref{prop},
to extend the obtained $sp(8,\mR)$ invariant unfolded equations
to $\M_4$ one has to replace the four dimensional exterior
differential by the ten dimensional one
\be\nn
dx^{\ga\pb}\f{\p}{\p x^{\ga\pb}} \to dX^{AB}\f{\p}{\p X^{AB}}
\ee
simultaneously extending a $4d$ flat $sp(8,\mR)$ connection 1-form
to $\M_4$. As in  the $4d$ case, the flat limit degeneracy
in the ``Cartesian coordinates" with $f_{AB}=0$ is resolved
in the $AdS$-like ten dimensional space $Sp(4,\mR)$.
The dynamical interpretation of the resulting
equations in  $Sp(4,\mR)$, \ie which
field components are dynamical, auxiliary, Stueckelberg etc,
is most conveniently elucidated by the analysis of $\sigma_-$
cohomology. This is done in Section \ref{s-}.

As discussed in the beginning of this section, the manifest
({\it i.e.,} linearly acting) symmetry of the  system (\ref{CON32}),
(\ref{CON31}) consists of the
usual Lorentz symmetry plus dilatation and duality transformations
which altogether form $gl(2,{\mathbb C})=sl(2,{\mathbb C})\oplus
{\mathbb R} \oplus u(1)$. Because it is small enough,
 Minkowski coordinates $X^{\ga\pa}$ and spinning coordinates
$X^{\ga\gb}$ and $X^{\pa\pb}$
have different appearance in the full $sp(8,\mR)$ invariant
system lifted to $\M_4$ or $Sp(4,\mR)$, {\it i.e.,} the equations
for the gauge fields break the manifest $gl(4,{\mathbb R})$
 symmetry of the equation on the generalized Weyl tensor $C_{00}$
 down to $gl(2,{\mathbb C})$. A related point
is that the analysis of the role
of different $sp(8,\mR)$ curvatures along the lines of the analysis
of conformal field strengths sketched in Section
\ref{Conformal geometries}, which answers the question which of the
component of the HS field strengths can be set to zero as constraints
and which are zero by virtue of field equations or/and Bianchi
identities, turns out to be more complicated
in $\M_4$. (This is analyzed  in Section \ref{s-} in terms of $\gs_-$
cohomology.)  Correspondingly,
an $\M_4$ (or $Sp(4,\mR)$) analog of the holonomy group
is not expected to be larger than $GL(2,{\mathbb C})$.

\section{$gl(4,{\mathbb R})$ invariant massless equations}
\label{sl4}

The generators $L_A{}^B$ (\ref{osp8osc}) span
$gl(4,{\mathbb R})\subset sp(8,{\mathbb R})$.
$L_{\ga}{}^\gb$ and ${L}_{\pa}{}^\pb$ include the
Lorentz generators, ${\cal H}$ (\ref{hel}) and $\D$ (\ref{dil}).
These span $gl(2; {\mathbb C})= gl(4,{\mathbb R})\cap u(2,2)$.
Note that the generators ${\cal H}$ and $\D$ exchange their roles
compared to the conformal case of $su(2,2)$. Now $\D$
(\ref{dil}) is the central element of $gl(4,{\mathbb R})$,
that characterizes different irreducible subsystems in
the $gl(4,{\mathbb R})$ invariant equations.

In addition, $gl(4,{\mathbb R})$ contains the generators
\be\nn
L_\ga{}^\pa = a_\ga \bar{b}^\pa\q
{L}_\pa{}^\ga = \bar{a}_\pa {b}^\ga\,.
\ee
These are analogues of the translation and special conformal
transformation generators $P_{\ga\pa}$ and $K^{\ga\pa}$
of the conformal algebra. The important difference is, however, that
$P_{\ga\pa}$ and $K^{\ga\pa}$ are self-conjugated while
$L_\ga{}^\pb$ and ${L}_\pa{}^\gb$ are conjugated to each other.
This implies in particular that although
\be\nn
[L_\ga{}^\pa \,,L_\gb{}^\pb ] =0\q
[\bar{L}_\pa{}^\ga\,,\bar{L}_\pb{}^\gb]=0\,,
\ee
neither $L_\ga{}^\pa$ nor ${L}_\pa{}^\ga$ are
translation generators of a Poincare' subalgebra of
$gl(4,{\mathbb R})$.

One reason why $gl(4,{\mathbb R})$ symmetry might have been missed
in field-theoretical models is that it does not allow
 lowest weight unitary modules because neither
$L_\ga{}^\pa$ nor ${L}_\pa{}^\ga$  can serve as
step operators. A related property is that $GL(4,{\mathbb R})$
admits no induced modules to define induced
$GL(4,{\mathbb R})$ action on tensor fields.
On the other hand, $ gl(4,\mR)$ does act on the
lowest weight unitary $sp(8,\mR)$--modules and therefore can
act on relativistic fields and their  single-particle
quantum states.

Now we observe that
\be\nn
{\cal P}_\ga{}_\pa = L_\ga{}_\pa +{L}_\pa{}_\ga\,,\qquad
(L_\ga{}_\pa =L_\ga{}^\pb \epsilon_{\pb\pa}\q
{L}_\pa{}_\ga =\bar{L}_\pa{}^\gb\epsilon_{\gb\ga})
\ee
along with the Lorentz generators
span the $AdS_4$ subalgebra $o(3,2)\sim sp(4,{\mathbb R})\subset
gl(4,{\mathbb R})$ (for simplicity we set
$\lambda =1$ in the rest of this section). This observation leads to
the alternative interpretation of the proposed
$sp(8,{\mathbb R})$ invariant equations,  as  $gl(4,{\mathbb R})$ invariant
equations in $AdS_4$ associated with the embedding
$o(3,2)\sim sp(4,\mR)\subset gl(4,\mR)\subset sp(8,\mR)$.

In the oscillator realization we have
$
{\cal P}_\ga{}_\pa = a_\ga \bar{b}_\pa +\bar{a}_\pa{}b_\ga.
$
So defined ${\cal P}_\ga{}_\pa$ acts in finite dimensional
spaces of homogeneous polynomials of the modules $\go_{ii}$ and
$C_{ii}$. On the other hand, $\go_{i\,1-i}$ and
$C_{i\,1-i}$ now decompose into the infinite sum of
infinite dimensional
$sp(4,\mR)$-modules. This means that
the subsystem of field equations  for $\go_{ii}$ and $C_{i\,1-i}$
now describes massless fields in $AdS_4$ while that
for $\go_{i\,1-i}$ and $C_{ii}$ describes an infinite set
of topological fields. We see that the massless
and topological fields exchange their roles depending on
which $sp(4,\mR)$ subalgebra of $sp(8,\mR)$ is identified with the
$AdS_4$ symmetry.

Note that the original form of the
nonlinear massless field equations of \cite{more} (see also \cite{Gol})
in which the massless fields are self-conjugated as presented in
Section \ref{hsads} is most naturally related to the
 $gl(4,{\mathbb R})$ invariant version of the equations.
 The free $gl(4,{\mathbb R})$ invariant system admits a
 reduction to the subsystem that describes
 a single   massless field of any spin. Algebraically, this
 is because, in the $gl(4,{\mathbb R})$ invariant case, spin is
 characterized by the eigenvalues of the non-compact generator
 ${\cal D}$ (cf (\ref{d10})-(\ref{d11})) that admits one dimensional
 real modules rather than by the compact generator ${\Hh}$
 with minimal two dimensional modules as in the conformal case.
Flat limit
can be taken using the rescaling procedure of
Section \ref{hsmin}. However, analogously to $su(2,2)$,
 the $gl(4,{\mathbb R})$ symmetry does
not survive in the flat limit, thus being invisible in
Minkowski space.

Equivalently, the $gl(4,{\mathbb R})$ covariant description can be obtained
by keeping the same oscillator generators as in the conformal
case but changing the conjugation conditions to
$
\overline {a_\ga} = \bb_{\pa}$, $\overline {b_\ga} = -\ba_{\pa}\,,
$
which implies that
$
\overline{|i,j|}= |1-j,1-i|
$
and, therefore,
$
\overline{\phi_{ij}}= \phi_{1-j,1-i}
$
for $\phi_{ij} =\go_{ij}$ or $C_{ij}$.
With these reality conditions we obtain that
$\overline{\phi_{00}}= \phi_{11}$,
$\overline{\phi_{01}}= \phi_{01}$ and
$\overline{\phi_{10}}= \phi_{10}$.
In this setup, the sets of higher spin and topological
fields remain the same as in the $su(2,2)$ case
but the reality conditions change.

As shown in Subsection \ref{gsfl}, the free equations for a
 single massless field of a fixed spin  admit the action of
 $su(2,2)$ that involves dual gauge potentials in the transformation
 law. This means that, allowing nonlocal field transformations of
 this kind, free field equations of a massless field of a given
 spin are invariant under both $su(2,2)$ and $gl(4,\mR)$
 and, therefore, under their closure $gl(4,\mC)$. Note that
 $gl(4,\mC)$ acts locally on the doubled sets of massless fields
 with the  natural realization of a pair of real fields as a single
 complex field. Let us stress that $gl(4,\mC)$ algebra does not
 belong to $sp(8,\mR)$ that acts individually on every Fock module.
Rather, $gl(4,\mC)\subset sp(8,\mC)$ where $sp(8,\mC)$ mixes two
Fock modules that describe massless fields in a chosen
vacuum realization.

An interesting project for the future is to look for nonlinear
$gl(4,{\mathbb R})$ invariant models in $AdS_4$.
Since $gl(4,{\mathbb R})$ acts
individually on fields of different spins, the problem
can be analyzed, {\it e.g.} for spin two, \ie $AdS_4$
gravity or supergravity.  We hope to come back
to this intriguing question elsewhere.

\section{$\gs_-$ analysis in Minkowski space}
\label{ceq}
\subsection{Grading}
\label{grading}
An appropriate grading $G$ of the Fock
modules in the unfolded HS equations is
$
G = \Big \vert \D \Big \vert=\half \Big \vert a_A b^A \Big \vert\,,
$
where $\Big \vert \D\Big \vert $ results from the dilatation generator
 $\D$ via replacing  its eigenvalues
 by their absolute values. In other words, the grading $G$
of a field equals to the absolute value of its scaling dimension.
So defined $G$ is diagonalizable and bounded from below.
Abusing notation, {}from (\ref{d10})-(\ref{d11}) we obtain
for $\phi_{ij} = C_{ij}$ or $\omega_{ij}$
\be
\label{G00}
G \phi_{00} (b,\bar{b}) = \half \left ( b^\ga \f{\p}{\p b^\ga} +
\bar{b}^\pa \f{\p}{\p \bar{b}^\pa} +2\right ) \phi_{00} (b,\bar{b})\,,
\ee
\be
\label{G11}
G \phi_{11} (a,\bar{a}) = \half \left ( a_\ga \f{\p}{\p a_\ga} +
\bar{a}_\pa \f{\p}{\p \bar{a}_\pa}+2 \right ) \phi_{11} (a,\bar{a})\,,
\ee
\be
\label{G10}
G \phi_{10} (a,\bar{b}) = \half \Big \vert a_\ga \f{\p}{\p a_\ga} -
\bar{b}^\pa \f{\p}{\p \bar{b}^\pa}\Big \vert \phi_{10} (a,\bar{b})\,,
\ee
\be
\label{G01}
G \phi_{01} (b,\bar{a}) = \half \Big \vert b^\ga \f{\p}{\p b^\ga} -
\bar{a}_\pa \f{\p}{\p \bar{a}_\pa} \Big \vert \phi_{01} (b,\bar{a})\,.
\ee

The grading $G$ treats symmetrically the parts $A_+$ and $A_-$
in the decomposition (\ref{dec}) underlying the flat limit
and leads to the standard description
\cite{Frhs,Frfhs} of $4d$ massless fields.

\subsection{0-forms}
Let us analyze the
equation (\ref{4deq}) on the 0-form $C(b|x)$.
The grading (\ref{G00}) implies that
\be\nn
\sigma_- = dx^{\ga\pa} \f{\p^2}{\p b^\ga \p \bar{b}^\pa}\,.
\ee
That $\gs_-^2=0$ is the consequence of
anticommutativity of the differentials $ dx^{\ga\pa}$.
As expected, the dynamical fields in $H^0(\gs_- )$ are
holomorphic and antiholomorphic
\be\nn
H^0(\gs_-) \,: \qquad C(b) + \overline{C}(\bar{b})\,.
\ee
$H^1(\gs_-) $ is of the form
\be
\label{H14}
H^1(\gs_-) \,: \qquad  dx^{\ga\pa} \Big (
 b_\ga E_\pa (b) +\bar{b}_\pa \overline{E}_\ga (\bar{b})
+  b_\ga \bar{b}_\pa E \Big )\,,
\ee
where $ E_\pa (b)$, $ \overline{E}_\ga (\bar{b})$  are arbitrary
polynomials of their arguments and $E$ is a constant.
It is easy to see that the elements in (\ref{H14}) are
$\gs_-$ closed but not exact.
We leave it as an exercise to the reader to check that
(\ref{H14}) describes full $H^1(\gs_-)$.
$ E_\pa (b)$, $ \overline{E}_\ga (\bar{b})$
and $E$ parameterize the \lhss of the equations (\ref{hol}),
(\ref{kg4}). Indeed, the unfolded equations (\ref{4deq})
demand all derivatives of
the 0-form dynamical fields,
that turn out to be $\gs_-$ closed because the dynamical
fields themselves are $\gs_-$ closed, to be zero except
for those that are $\gs_-$ exact to be absorbed by
$\gs_- C^{aux}$ with some auxiliary fields $C^{aux}$ (for
more detail see e.g. \cite{solv}).
This has the consequence that the equations (\ref{hol}), (\ref{kg4})
follow from (\ref{4deq}). Since (\ref{H14}) describes
the full cohomology, they
provide the full list of differential equations imposed
 by the unfolded equations (\ref{4deq}) on the
dynamical fields.

To analyze the content of the $4d$ massless
field equations (\ref{CON12}) and (\ref{CON22}) for gauge
potentials first of all we observe that all fields $C(b,0|x)$ and
$C(0,\bar{b}|x) $, except for the scalar field $C(0,0|x)$ and
spinor field linear in $b^A$, are expressed via the gauge fields
$\go$ by the equation (\ref{CON12}) thus becoming auxiliary fields.
To take this into account it is convenient to redefine $\gs_- \to\sigma^\prime_-$
where $\gs^\prime_-$ acts on the direct sum of spaces of 0-forms
$C$ and 1-forms $\go$ so that the terms on the \rhs of
(\ref{CON12}) become $\sigma^\prime_- (C)$. As a result,
all auxiliary fields $C(x)$ that correspond to spins
$s \geq 1$ disappear from
$H^0(\sigma^\prime_-) $. Correspondingly, their field equations become
consequences of the Bianchi identities for the equations
(\ref{CON12}) for $s>1$, thus disappearing from
$H^1(\sigma^\prime_-) $.

The case of spin one is special. Here, the equation
(\ref{CON12}) is just the definition of the Maxwell field strength
$C_{\ga\gb}$ and $\overline{C}_{\pa\pb}$ in terms of potentials while
the spin one equation is still a part of the cohomology
 (\ref{H14}),
$
 dx^{\ga\pa}( b_\ga E_{\gb\pa} b^\gb - \bar{b}_\pa
 E_{\ga\pb} \bar{b}^\pb)\,,
$
where $E_{\gb\pa}$ is an arbitrary Hermitian bispinor (\ie Lorentz vector)
 that parameterizes
the \lhss of the second pair of the Maxwell equations.
The first pair associated with
$ dx^{\ga\pa}( b_\ga F_{\gb\pa} b^\gb +
\bar{b}_\pa F_{\ga\pb} \bar{b}^\pb)$
becomes the Bianchi identity and disappears from $H^1(\sigma^\prime_-) $.

The sector of spin 0 and 1/2 is unaffected by the transition
from $\gs_-$ to $\gs_-^\prime$. As a result, we conclude that
the relevant part of the $\gs_-^\prime$ cohomology
in the sector of 0-forms $C$ is
\be
\label{h0sp}
H^0(\gs_-^\prime,C ) \,: \qquad C + b^\ga C_\ga +
\bar{b}^\pa \overline{C}_\pa\,,
\ee
\be
\label{h1sp}
H^1(\gs_-^\prime, C ) \,:\qquad
 dx^{\ga\pa}  (
 b_\ga E_\pa  +\bar{b}_\pa \bar{E}_\ga
+  b_\ga \bar{b}_\pa E
 + b_\ga E_{\gb\pa} b^\gb - \bar{b}_\pa E_{\ga\pb} \bar{b}^\pb  )
\,,
\ee
where $C$, $C_\ga$, $\overline{C}_\pa$ and $E$, $E_\ga$, $E_\pa$,
$E_{\ga\pa}$ parameterize, respectively, dynamical fields of spin 0 and 1/2 and
the \lhss of the dynamical equations of spin 0, 1/2 and 1.
 Now we are in a position to analyze the sector of gauge fields.

\subsection{1-forms}
\label{1f}
Dynamics of spins $s\geq 1$ is described by the
gauge 1-forms $\go$.
As shown in  \cite{Fort1,Ann} (without using the
cohomology language, however) the dynamical fields are
$\go_{\ga_1\ldots \ga_n\,,\pa_1\ldots \pa_m }$ with
$n=m$ for bosons and $|n-m|=1$ for fermions, which are the
frame-like counterparts of the Fronsdal's double traceless boson
and triple $\gamma$--transverse fermion metric-like
fields, respectively. Field equations for massless fields of
all spins $s>1$ are contained in the sector of gauge 1-forms.
Let us show how these facts are reproduced in terms of
$\gs_-$ cohomology.

In the case of 1-forms, the grading
operator is of the type (\ref{G10})
\be
\label{Gnbn}
G=\half  \Big | n-\bar{n} \Big |\q
n=y^\gb \f{\p}{\p y^\gb}\q \bar{n}=\by^\pb \f{\p}{\p \by^\pb}\,.
\ee
The operator  $\sigma^\prime_- $ is
\be
\label{gspr}
\sigma_-^\prime (A) = \gs_- A + \gs_-^{weyl} A\q \gs_- A=
 e^{\ga\pb}\Big (y_\ga \frac{\partial}{\partial \bar{y}^\pb}
{A}_-(y,\bar{y} \mid x)
+ \frac{\partial}{\partial {y}^\ga}\bar{y}_\pb {A}_+(y,\bar{y} \mid x)
\Big ) \,,
\ee
where $\gs_-^{weyl}$ is the part of $\gs_-^\prime$ responsible for
gluing the Weyl 0-forms $C$ to the field strengths of the gauge
1-forms via the terms on the \rhs of (\ref{CON12}).
Note that $\gs_-$ defined as the $e$-dependent part of (\ref{adfl})
respects the decomposition (\ref{dec}).
Equivalently,
\be
\label{gsp}
\gs_-^\prime= \rho_- \theta(n-\bar{n}-2) +\overline{\rho}_- \theta(\bar{n}-n-2)
 +\gs_-^{weyl}\,,
\ee
where
\be
\label{taupm}
\rho_- = e^{\ga\pb}\f{\p}{\p y^\ga} \by_\pb\q
\overline{\rho}_- =e^{\ga\pb}\f{\p}{\p \bar{y}^\pb } {y}_\ga \,
\ee
and
\be
\label{theta}
\theta (m)=1\,(0)\q m\geq 0\,\, (m<0)\,.
\ee

Although $\rho_-$ and $\overline{\rho}_-$
do not anticommute, $\gs_-^\prime$
squares to zero because $(\rho_-)^2=(\overline{\rho}_- )^2=0$
and the step functions guarantee that the parts of $\gs^\prime_-$
associated with $\rho_-$ and $\overline{\rho}_-$ act in different spaces.
The nontrivial cohomology of $\gs_-^\prime$ is concentrated in the subspaces
of $G$-grades $0,$ $ 1/2$ and $1 $. This follows from the fact
that the operators $\rho_-$ and $\overline{\rho}_-$ (\ref{taupm})
act as the exterior differentials $\theta^\ga \f{\p}{\p y^\ga}$
and $\overline{\theta}^\pa \f{\p}{\p \by^\pa}$ with
$\theta^\ga = e^{\ga\pa} \bar{y}_\pa$ and
$\bar{\theta}^\pa=e^{\ga\pa} y_\ga$
in the spaces of functions of $y^\ga$ and $\by^\pa$, respectively.
As a result, by Poincare's lemma, the cohomology
is concentrated in the sectors where $\gs^\prime_-$ differs
from $\rho_-$ or $\overline{\rho}_-$, that is where
the step functions differ from a constant.
Also let us note that, in the gauge field sector, the difference
between $\gs_-$ and
$\gs_-^\prime$ due to $\gs_-^{weyl}$ matters only in the computation of
$H^2(\gs_-^\prime)$ because the Weyl 0-forms in (\ref{CON12})
contribute to the sector of 2-forms.

$H^0(\gs^\prime_-)$ is easy to compute. The nontrivial cohomology
appears in the subspaces of grades $G=0$ or $1/2$ where $\gs_-$
acts trivially because of the step functions in (\ref{gsp}).
So,
\be
\label{h0t-}
H^0(\gs^\prime_- ):\qquad
\epsilon (y,\bar{y})
=\sum_{|n-m|\leq 1}
\frac{1}{2\,n!m!}
{y}_{\alpha_1}\ldots {y}_{\alpha_n}{\bar{y}}_{{\pb}_1}\ldots
{\bar{y}}_{{\pb}_m
} \epsilon{}^{\alpha_1\ldots\alpha_n}{}_,{}^{{\pb}_1
\ldots{\pb}_m}\,.
\ee
$
\epsilon{}^{\alpha_1\ldots\alpha_n}{}_,{}^{{\pb}_1
\ldots{\pb}_m}(x)\,,
$
$|n-m|\leq 1$
are  parameters of differential gauge symmetry transformations
of spin $s=1+\half (n+m)$ massless fields.
For integer spins with $n=m=s-1$,
the corresponding spin $s$ gauge symmetry parameter is
equivalent to a rank $s-1$ symmetric traceless Lorentz
tensor. This agrees with the standard Fronsdal formulation
\cite{Frhs}. For half-integer spins, $n=s-3/2, m=s-1/2$ or
$m=s-3/2, n=s-1/2$. The corresponding spin $s$ gauge symmetry
parameter is equivalent to a rank $s-3/2$ symmetric
$\gga$-transversal tensor-spinor in tensor-spinor notation.
This agrees with the Fang-Fronsdal theory \cite{Frfhs}.

Eq. (\ref{h0t-}) implies that all gauge parameters
$
\epsilon{}^{\alpha_1\ldots\alpha_n}{}_,{}^{{\pb}_1
\ldots{\pb}_m}(x)
$
with $|n-m|>1$ are of Stueckelberg type, {\it i.e.,} the
associated
gauge field transformations contain algebraic shifts that
gauge away some  components of the HS connections.
In particular, the linearized local Lorentz symmetry parameters
$\epsilon{}^{\alpha\beta}(x) $ and
$\epsilon{}^{\pa\pb}(x) $, which allow to gauge away the
antisymmetric part of the spin two vierbein, are of this class.

$H^1 (\gs_-^\prime)$ describes  those components of the 1-form
connections that are neither auxiliary ({\it i.e.,} being expressed
in terms of other connections by algebraic constraints
that can be imposed in terms of
HS curvatures, {\it a la} zero-torsion constraint in gravity)
nor Stueckelberg (\ie cannot be gauge fixed to zero
by algebraic shift gauge symmetries).
It is not hard to see that $H^1(\gs_-^\prime)$ is concentrated in the
subspace with  $G=0$ for bosons and  $G= 1/2$ for fermions, where
$\gs_-^\prime$ acts trivially so that all elements are
$\gs_-^\prime$ closed in this sector.
To compute $H^1(\gs_-^\prime)$ it is thus enough to factor out
the $\gs_-^\prime$ exact part, which is equivalent to gauging away the
Stueckelberg components of the gauge fields. This
gives the following results.

In the bosonic case
\be
\label{bh14}
H^{1\,bos}(\gs_-^\prime): \qquad \go(y,\bar{y}) =
e^{\ga\pb}\f{\p^2}{\p y^\ga \p \bar{y}^\pb}\phi (y,\bar{y}) +
e^{\ga\pb} y_\ga  \bar{y}_\pb\phi^\prime (y,\bar{y})\,,\quad
G(\omega)=0\,,
\ee
\ie spin $s\geq 1$ dynamical fields identify with the
0-forms $\phi(y,\by| x)$ and $\phi^\prime (y,\by| x)$ that satisfy
\be\nn
n \phi (y,\bar{y}| x)=
\bar{n} \phi (y,\bar{y}| x)=
s \phi (y,\bar{y})(x)\q
n \phi^\prime (y,\bar{y}| x)=
\bar{n} \phi^\prime (y,\bar{y}| x)=
(s-2) \phi^\prime (y,\bar{y}| x)\,.
\ee
These describe two irreducible components
of the spin $s$ Fronsdal double traceless  symmetric tensor  field.
In particular, in the spin two sector, $H^{1\,bos}(\gs_-^\prime)$
describes a rank two symmetric tensor in terms of two-component spinors.
This is the fluctuational
part of metric equivalent to the fluctuational
part of the vierbein modulo linearized local Lorentz gauge symmetry.

In the fermionic case
\be
\label{hf1}
H^{1\,fer}(\gs_-^\prime): \qquad
\go^+(y,\bar{y}) + \go^-(y,\bar{y})\qquad
(n-\bar{n}) \go^\pm(y,\bar{y})=\pm \go^\pm(y,\bar{y})\,,
\ee
\be\nn
\go^+ (y,\bar{y})=
e^{\ga\pb}\f{\p^2}{\p y^\ga \p \bar{y}^\pb}\psi^+_1 (y,\bar{y}) +
e^{\ga\pb}y_\ga\f{\p}{ \p \bar{y}^\pb}\psi^-_2 (y,\bar{y}) +
e^{\ga\pb} y_\ga  \bar{y}_\pb\psi^+_3 (y,\bar{y})\,,
\ee
\be\nn
\go^- (y,\bar{y})=
e^{\ga\pb}\f{\p^2}{\p y^\ga \p \bar{y}^\pb}\psi^-_1 (y,\bar{y}) +
e^{\ga\pb}\bar{y}_\pb\f{\p}{ \p {y}^\ga}\psi^+_2 (y,\bar{y}) +
e^{\ga\pb} y_\ga  \bar{y}_\pb\psi^-_3 (y,\bar{y})\,.
\ee
Here $\psi^+_1 (y,\bar{y}| x)$, $\psi^-_2 (y,\bar{y}| x)$
 and $\psi^+_3 (y,\bar{y}| x)$ and their
conjugates $\psi^-_1 (y,\bar{y}| x)$,
$\psi^+_2(y,\bar{y}| x)$ and $\psi^-_3(y,\bar{y}| x)$
describe
three irreducible components of the Fang-Fronsdal
triple $\gga$-transverse symmetric tensor-spinor
fermionic field. For a half-integer spin $s$ we have
\be\nn
n \psi^\pm_1 = (s\pm1/2)\psi^\pm_1\,,\qquad
\bar{n} \psi^\pm_1 = (s\mp1/2)\psi^\pm_1\,,
\ee
\be\nn
n \psi^\pm_2 = (s-1 \pm1/2)\psi^\pm_2\,,\qquad
\bar{n} \psi^\pm_2 = (s-1\mp1/2)\psi^\pm_2\,,
\ee
\be\nn
n \psi^\pm_3 = (s-2\pm1/2)\psi^\pm_3\,,\qquad
\bar{n} \psi^\pm_3 = (s-2\mp1/2)\psi^\pm_3\,.
\ee

Finally, $H^2 (\gs_-^\prime)$ classifies differential equations
on the dynamical fields contained in the $4d$ unfolded HS system.
$H^2 (\gs_-)$ consists of the generalized Weyl part parameterized
by the \rhss of (\ref{CON12}) and the Einstein cohomology
that represents the \lhss of the massless field equations.
Since the generalized Weyl tensor part has already been taken into
account by $\gs_-^{weyl}$ in (\ref{gspr}),
$H^2 (\gs^\prime_-)$ consists of the Einstein cohomology.

Let us start with the simpler fermionic case.
$H^{2\,fer} (\gs^\prime_-)$ consists of the grade $1/2$
2-forms $R$, which are automatically closed,
modulo exact 2-forms
\be\nn
R^{exact }=\rho_- W\,,\quad (n-\bar{n} )W =3W\q
\overline{R}^{exact }=\overline{\rho}_-
\overline{W}\,,\quad (\bar{n}-{n} )\overline{W} =3\overline{W}\,.
\ee
Elementary computation shows that $H^{2\,fer}(\gs_-^\prime)$ is
\be
\label{h2pr}
H^{2\,fer} (\gs^\prime)  = E^+(y,\bar{y}) + E^- (y,\bar{y})
\q (n-\bar{n}) E^\pm(y,\bar{y})=\pm E^\pm(y,\bar{y})\,,
\ee
\be\nn
E^+(y,\bar{y})= \overline{H}^{\pa\pb} \left (
\f{\p^2}{\p \bar{y}^\pa \p \bar{y}^\pb} E^-_1 (y,\bar{y})
+ \bar{y}_\pa
\f{\p}{  \p \bar{y}^\pb} E^+_{2} (y,\bar{y})\right ) +
H^{\ga\gb}
{y}_\ga   {y}_\gb E^-_{3} (y,\bar{y})
\ee
\be\nn
{E}^-(y,\bar{y})= H^{\ga\gb} \left (
\f{\p^2}{\p {y}^\ga \p {y}^\gb} {E}^+_1 (y,\bar{y})
+ {y}_\ga
\f{\p}{  \p {y}^\gb} E^-_{2} (y,\bar{y})  \right )
+\overline{H}^{\pa\pb}
\bar{y}_\pa   \bar{y}_\pb E_{3}^+ (y,\bar{y})\,,
\ee
where $H^{\ga\gb}$ and
$\overline{H}^{\pa\pb}$ are the basis 2-forms (\ref{H}) and
\be\nn
n E^\pm_1 = (s\pm1/2)E^\pm_1\,,\qquad
\bar{n} E^\pm_1 = (s\mp1/2)E^\pm_1\,,
\ee
\be\nn
n E^\pm_2 = (s-1 \pm1/2)E^\pm_2\,,\qquad
\bar{n} E_2^\pm = (s-1\mp1/2)E_2^\pm\,,
\ee
\be\nn
n E^\pm_3 = (s-2\pm1/2)E^\pm_3\,,\qquad
\bar{n} E^\pm_3 = (s-2\mp1/2)E^\pm_3\,.
\ee

As a linear space, Einstein cohomology $H^{2\,fer} (\gs^\prime_-)$
is isomorphic to $H^{1\,fer} (\gs^\prime_-)$
(\ref{hf1}). This is expected because  the
massless field equations are Lagrangian,
\ie there are as many equations as field variables.

Let us now consider the bosonic case. Here the sectors of
 $G=0$ and $1$ should be analyzed.

 It is easy to see that
\be
\label{h20pr}
H^2(\gs^\prime_- )\Big |_{G=0}=0\,,
\ee
which means  that any 2-form with $G=0$,
which is automatically $\gs_-^\prime$ closed,
is $\gs_-^\prime$ exact. For spin one
this means that the respective part of the equation
(\ref{CON12}) with $G=0$ is a constraint that
expresses the Maxwell stress tensor $C^{\ga\gb}$,
$\overline{C}^{\pa\pb}$ via derivatives of the spin one
gauge potential. For spins $s \geq 2$, (\ref{h20pr})
implies that the $4d$ 1-form gauge fields
$\go(y,\bar{y})$ with $G = 1$ are
auxiliary being expressed (modulo pure gauge components)
via derivatives of the dynamical fields by the ``zero torsion
condition" $R(y,\bar{y}) |_{G=0}=0$ that
imposes no differential equations on the dynamical Fronsdal
fields. In particular, in the spin two sector, (\ref{h20pr})
allowed to impose the standard zero-torsion condition,
that expresses Lorentz connection via derivatives
of the vierbein imposing no restrictions on the latter.

Now  consider the part of the cohomology $H^2 (\gs_-^\prime)$
in the $G=1$ sector. We have to
find such 2-forms $\Phi$ with $n-\bar{n}=2$ and
$\overline{\Phi}$ with $n-\bar{n}=-2$
that
\be
\label{cans}
\rho_- \Phi +\overline{\rho}_- \overline{\Phi} =0\,.
\ee
The nontrivial cohomology consists of solutions of this
condition with $\rho_- \Phi \neq 0$ and
$\overline{\rho}_- \overline{\Phi} \neq 0$
(otherwise, $\Phi$ and $\overline{\Phi}$ are $\gs_-^\prime$ exact).
The key point is that, to cancel out in (\ref{cans}),
the terms coming from $\Phi$ and $\overline{\Phi}$ should carry
equivalent representations of the Lorentz algebra eventhough they
support polynomials of $y$ and $\bar{y}$ with $n-\bar{n}=2$
and $n-\bar{n}=-2$,
respectively.

It is not hard to see that this is possible if
the cohomology space is spanned by the polynomials
$E_i(y,\bar{y})$ that contain as many $y^\alpha$
as $\bar{y}^\pa$, \ie $
G(E_i(y\,,\by))=0\,.
$
The appropriate Ansatz is
\be
\label{p+4}
\Phi = H_{\ga\gb}y^\ga y^\gb  E_{1}(y\,,\by) +\overline{H}_{\pa\pb}
\frac{\p^2}{\p \by_\pa \p \by_\pb} E_{2}(y\,,\by)\,,
\ee
\be
\label{p-4}
\overline{\Phi} =
{H}^{\ga\gb}
\frac{\p^2}{\p y^\ga \p y^\gb} E_{3}(y\,,\by)+
\overline{H}^{\pa\pb} \by_\pa \by_\pb
E_{4}(y\,,\by)\,.
\ee
Using the identities
\be\nn
H^{\ga\gb}\wedge e^{\gga \pa} = \epsilon^{\ga\gga} \Hh^{\gb\pa}+
\epsilon^{\gb\gga} \Hh^{\ga\pa}\q
\overline{H}^{\pa\pb}\wedge e^{\ga \pga} =-
\epsilon^{\pa\pga} \Hh^{\ga\pb}-
\epsilon^{\pb\pga} \Hh^{\ga\pa}\,,
\ee
where
$
\Hh^{\ga\pb}= -\f{1}{3}e^\ga{}_\pa\wedge e^{\gb\pa}\wedge e_\gb{}^\pb
$
are $4d$ basis 3-forms, we obtain
\be\label{result}
\rho_- \Phi +\overline{\rho}_- \overline{\Phi} =-
\Hh^{\ga\pa}\Big (y_\ga\by_\pa(y^\gb \f{\p}{\p y^\gb}+3)(E_1-E_4)+
\f{\p^2}{\p y^\ga \p \by^\pa}(y^\gb\f{\p}{\p y^\gb}-1)(E_2-E_3)\Big ).
\ee
This implies that (\ref{p+4}) and (\ref{p-4}) describe a nontrivial
cohomology provided that
$
E_2=E_3 = E$ and  $E_1 = E_4 = E^\prime\,,
$
where  $E^\prime (y\,,\by )$ is an arbitrary
grade zero polynomial while  $E(y\,,\by)$ should be at least
of fourth order to contribute.

For any integer spin $s\geq 2$, the Einstein cohomology
\be
\label{ein}
H^{2\,bos}(\gs_-^\prime) =
\left (\overline{H}_{\pa\pb}
\frac{\p^2}{\p \by_\pa \p \by_\pb} +
{H}^{\ga\gb}
\frac{\p^2}{\p y^\ga \p y^\gb}\right ) E(y\,,\by)+
\left (H_{\ga\gb}y^\ga y^\gb +
\overline{H}^{\pa\pb} \by_\pa \by_\pb \right)
E^\prime(y\,,\by)\,
\ee
is described by the polynomials $E(y\,,\by )$
and $E^\prime(y\,,\by )$ that satisfy
\be\nn
 n E(y\,,\by )= \bar{n} E(y\,,\by ) = s E(y\,,\by )\q
 n E^\prime(y\,,\by ) = \bar{n} E^\prime(y\,,\by )
 =(s-2) E^\prime(y\,,\by )\,.
\ee

Einstein cohomology is
responsible for the dynamical field equations for any integer spin
$s\geq 2$: the condition that
the part of the HS curvature 2-form parameterized by
$E(y\,,\by)$ and $E^\prime (y\,,\by )$ is zero, which is
true by the unfolded equations (\ref{CON12}), imposes the
Fronsdal field equations on the dynamical fields of spins $s\geq 2$.
This agrees with the condition
 that there are as many dynamical equations as field variables.
In terms of Lorentz tensors, $E(y\,,\by )$
and $E^\prime(y\,,\by )$ are equivalent to rank $s$ and
rank $s-2$ traceless tensors, respectively.

The spin one part of $E$ is $\gs_-$ exact because
from (\ref{result}) it follows that $\rho_-(\Phi)=
\overline{\rho}_-(\overline{\Phi})=0$ in this case.
This conforms the fact that
the spin one dynamical equations are described by the 0-form
cohomology (\ref{h1sp}).

The pattern of $H^p(\gs_-^\prime)$ proves the so
called Central-On-Shell theorem \cite{Ann} that states
that the equations (\ref{CON12}) and (\ref{CON22})
are equivalent to the standard $4d$ massless field equations
plus an infinite set of constraints that express infinitely
many auxiliary fields via derivatives of the dynamical massless fields.
Let us stress that the Central-On-Shell theorem is true both in
$4d$ Minkowski space and in $AdS_4$. Actually,
although the frame 1-form $e^{\ga\pa}(x)$
that enters $\gs_-^\prime$
does depend on a chosen geometry and coordinate system,
the analysis of the $\gs_-^\prime$ cohomology only uses
that the vacuum connection is flat and the vacuum
frame field is nondegenerate, forming a frame
in the space of $4d$ 1-forms.

\section{$\gs_-$ analysis in $\M_4$}
\label{s-}
\subsection{ $\gs_-$ and $\tau_-$}

The grading operator $G$, which is independent
of the choice of a base manifold, is given by
(\ref{G00})-(\ref{G01}).
 $\sigma_-$ is the grade $-1$
part of the covariant derivative. Let $\sigma_{-ij}$ denote
 $\sigma_-$ in the module $\phi_{ij}$. We obtain
\be
\label{s00}
\sigma_{-00} =\half h^{AB}\f{\p^2}{\p b^A \p b^B}\,,
\ee
\be
\label{s11}
\sigma_{-11} =\half f_{AB}\f{\p^2}{\p a_A \p a_B}\,,
\ee
\bee
\label{s01}
\sigma_{-01} &=& \half \Big (h^{\ga\gb}\f{\p^2}{\p b^\ga \p b^\gb} +
h^{\ga\pb}\f{\p}{\p b^\ga } \bar{a}_\pb +
h^{\pa\pb} \bar{a}_\pa \bar{a}_\pb\Big )\theta (n-\bar{n}-2)\nn\\
&{+}&
\half \Big (f_{\pa\pb}\f{\p^2}{\p \bar{a}_\pa \p \bar{a}_\pb} -
f_{\ga\pb}\f{\p}{\p \bar{a}_\pb } {b}^\ga +
f_{\ga\gb} {b}^\ga {b}^\gb\Big )\theta (\bar{n}-n-2)\,,
\eee
\bee
\label{s10}
\sigma_{-10} &=& \half\Big (f_{\ga\gb}\f{\p^2}{\p a_\ga \p a_\gb} -
f_{\ga\pb}\f{\p}{\p a_\ga } \bar{b}^\pb +
f_{\pa\pb} \bar{b}^\pa \bar{b}^\pb\Big )\theta (n-\bar{n}-2)\nn\\
&{+}&
\half \Big (h^{\pa\pb}\f{\p^2}{\p \bar{b}^\pa \p \bar{b}^\pb} +
h^{\ga\pb}\f{\p}{\p \bar{b}^\pb } {a}_\ga +
h^{\ga\gb} {a}_\ga {a}_\gb\Big )\theta (\bar{n}-n-2)\,,
\eee
where $n$, $\bar{n}$ and the step function $\theta (n)$
are defined in (\ref{Gnbn}) and (\ref{theta}).
Because
$\sigma_{-\,i\,1-i}$ are asymmetric with respect to primed
and unprimed indices, the manifest symmetry is $GL(2,{\mathbb C})$
which consists of the Lorentz symmetry $SL(2,{\mathbb C})$,
dilatations and duality transformations.

Recall that in the case (\ref{sp48}) of $Sp(4,\mR)$,
 both $h^{AB}$ and $f_{AB}$
are nondegenerate. The naive flat
limit with $f_{AB}\to 0$ is degenerate. However, because the operators
$\sigma_{-\,i\,1-i}$
are defined differently in the sectors with
$n-\bar{n}>0$ and $n-\bar{n}<0$, the fields
can be rescaled  so  that
$\sigma_{-\,i1-i}$  remain nondegenerate
in the flat limit. The result extends the $e$-dependent part of
the Minkowski covariant derivative (\ref{adfl}) to $\M_4$.

Once both $h^{AB}$ and $f_{AB}$ are expressed in terms
the vielbein $e^{AB}$ by (\ref{sp48}), from (\ref{s00})-(\ref{s10})
it is clear that, up to
renaming the oscillator variables, there are two essentially different
$\sigma_-$ operators. One in the sector of $\phi_{00}$ and $\phi_{11}$
and another one in the sector of $\phi_{01}$ and $\phi_{10}$.
To simplify notations we call the former $\sigma_-$ and
the latter $\tau_-$. Denoting the respective oscillators as $y^A$
we have
\be
\label{sy}
\sigma_{-} = e^{AB}\f{\p^2}{\p y^A \p y^B}\,,
\ee
\be\label{tm}
\tau_- = t_- \theta (n-\bar{n}-2)+\bar{t}_- \theta(\bar{n}-n-2)\,,
\ee
where
\be
\label{<>}
t_- =\nu^+ +\nu^0 +\nu^{-}\,,\qquad
\bar{t}_- =\bnu^+ +\bnu^0 +\bnu^-\,
\ee
\be
\label{>}
\nu^-=
e^{\ga\gb}\f{\p^2}{\p y^\ga \p y^\gb}\,,\qquad
\nu^0 =e^{\ga\pb}\f{\p}{\p y^\ga } \bar{y}_\pb\,,\qquad
\nu^+=e^{\pa\pb} \bar{y}_\pa \bar{y}_\pb\,,
\ee
\be
\label{<}
\bnu^-= e_{\pa\pb}\f{\p^2}{\p \bar{y}_\pa \p \bar{y}_\pb}\,,\qquad
\bnu^0=e_{\ga\pb}\f{\p}{\p \bar{y}_\pb } {y}^\ga \,,\qquad
\bnu^+ = e_{\ga\gb} {y}^\ga {y}^\gb\,.
\ee
Note that
\be\nn
\{\nu^i\,,\nu^j \} = 0\,,\qquad
\{\bnu^i\,,\bnu^j \} =0\,,\qquad i,j= -,0,+
\ee
and, therefore $(t_-)^2=(\bar{t}_-)^2=0$.
Although $\{t_-\,,\bar{t}_-\}\neq 0$,
$(\tau_-)^2 =0$  because the step functions
in (\ref{tm}) imply that the parts of $\tau_-$
associated with $t_-$ and $\bar{t}_-$ act in different
subspaces.

The dependence on $\lambda$ has been removed by a
field redefinition along the lines of Section
\ref{hsads}, which also was used to adjust
convenient coefficients in (\ref{<>}).
In the rescaled variables $y^A$, $\lambda$
appears in front of the $\sigma_{+ij}$ part
of the covariant derivative.
Correspondingly, the flat space field equations in $\M_4$
result from setting $\lambda=0$ and dropping the terms with
$\sigma_{+ij}$.
In the flat case of $\M_4$ one can use ``Cartesian coordinates"
with
\be
\label{cart}
e^{AB}= dX^{AB}\q D^{tw;fl} = d + \sigma_-\,,\qquad
D^{ad;fl} = d + \tau_-\,.
\ee

The dynamical content of the unfolded field equations in
$Sp(4,\mR)$ and $\M_4$ is determined by the cohomology
of $\sigma_-$ and $\tau_-$. We
have to calculate $H^0(\sigma_-)$
and $H^1(\sigma_-)$ to identify the independent fields and field
equations in the twisted adjoint 0-form sector and
$H^0(\tau_-)$, $H^1(\tau_-)$ and $H^2(\tau_-)$
to identify the gauge parameters, dynamical fields and gauge invariant
combinations of derivatives of the dynamical fields that either
represent the \lhss of field equations or
identify with the  generalized Weyl tensors via the
 Chevalley-Eilenberg cohomology terms.

We start with $H^p(\sigma_-)$
extending the results of \cite{BHS} to the $AdS$-like case
of $Sp(M,\mR)$.

\subsection{Weyl 0-form sector}
\label{w0f}
In the case of $\M_M$, the unfolded equations in question are (\ref{10eq}). They
are equivalent to
$
D^{tw;fl} C =0\,.
$ $H^0 (\gs_-) = C +y^A C_A$  describes the dynamical fields
$C(X)$ and $C_A(X) b^A$ in $\M_M$.
$H^1 (\gs_-)$ is
\be\nn
H^1 (\gs_-) = e^{AB} y^C y^D  E_{AB,CD} + e^{AB} y^C   E_{AB,C}\,,
\ee
where $ E_{AB,CD}$ and $ E_{AB,C}$ represent the \lhss
of the  field equations (\ref{kg}) and (\ref{dir}) and
satisfy
\be\nn
 E_{AB,CD}= E_{BA,CD}= E_{AB,DC}\q  E_{(AB,C)D}=0\q
 E_{AB,C}= E_{BA,C}\q  E_{(AB,C)}=0\,.
\ee

The  equations (\ref{kg}) and (\ref{dir}) in $\M_4$
were originally derived this way in \cite{BHS}.
The analysis in the curved $Sp(M,\mR)$ background is analogous.
$\sigma_-$ is still given by (\ref{sy}),
where $e^{AB}$ is the generalized vielbein of $Sp(M,\mR)$ introduced
in Section \ref{Generalized conformal geometries}.
The symmetry type of the equations and leading
derivative terms remain the same as in $\M_4$.  The exact
form of the equation (\ref{kg}) is deformed by
the $\lambda^2$-dependent lower-derivative terms.
These appear due to the $\sigma_+$ part of the
covariant derivative associated with the nonzero
``special conformal" connection (\ref{sp48})
\be\nn
\sigma_+=\f{1}{4} \lambda^2 e_{AB} b^A  {b}^B\q
e_{AB}=C_{F A}C_{G B} e^{FG} \,,
\ee
where $C_{AB}$ is the $Sp(M,\mR)$ invariant antisymmetric form.

Also, usual derivatives have to be replaced by the
Lorentz-like $Sp(M,\mR)$
covariant derivatives associated with the connection $\omega_{A}{}^B$
\be\nn
\ls D f_A (X)= d f_A (X)+\omega_A{}^B (X) f_B (X)\,,\quad
D=dX^{\underline{A}\underline{B}}D_{\underline{A}\underline{B}}\,,
\quad
e^{AB}=dX^{\underline{A}\underline{B}}
e_{\underline{A}\underline{B}}{}^{AB}\,,\quad
D_{AB}= e_{AB}{}^{\underline{A}\underline{B}}
D_{\underline{A}\underline{B}}\,.
\ee

As a result, the $Sp(2M,\mR)$ invariant
deformation of the equations (\ref{kg}) and (\ref{dir})
to $Sp(M,\mR)$ reads as
\be
\label{kgsp4}
\Big (D_{AB}D_{CD}- D_{CB}D_{AD} \Big )C(X) -\half\lambda^2
\Big (C_{AC}C_{BD}+C_{BC}C_{AD} \Big ) C(X)=0\,,
\ee
\be
\label{dirsp4}
D_{AB}C_C(X) - D_{CB}C_A(X) =0\,.
\ee

\subsection{Gauge 1-form sector}
In this subsection we analyze $H^p(\tau_-)$ that determines  dynamical
content of the field equations in the sector of 1-form
connections.

\subsubsection{Auxiliary Lemmas}
\label{UL}
The following  useful lemmas will be used in what follows.\\
{\it Lemma 1:}\\
$H^p (\bnu^+ ,P)=0$ at $p=0,1,2$ if $P$ is the space of polynomials
of $y^\ga $ with $\ga=1,2\ldots m >1$.\\
{\it Proof:}
$H^0 (\bnu^+ ,P)=0$ because $Ker\, \bnu^+ =0$ on the
space of polynomials (product of any two
nonzero polynomials never gives  zero polynomial).

The case of $H^1 (\bnu^+, P)$ can be analyzed as follows.
Let $\omega (y)$ be a 1-form
$
\omega (y) = e^{\ga\gb} \go_{\ga\gb}{}^{\gb_1\ldots\gb_n}y_{\gb_1}
\ldots y_{\gb_n}.
$
The $\bnu^+$ closedness condition
$
y_{\ga} y_{\ga} \go_{\gb\gb} (y)-
y_{\gb} y_{\gb} \go_{\ga\ga} (y)=0
$
is equivalent to
\be\nn
\delta_\ga^\gga \delta_\ga^\gga \go_{\gb\gb}{}^{\gga(n)} -
\delta_\gb^\gga \delta_\gb^\gga \go_{\ga\ga}{}^{\gga(n)} =0\,,
\ee
where we use the convention that lower (upper) indices
denoted by the same letter are symmetrized and a number of
symmetrized indices is indicated in parentheses.
Contracting twice the indices $\ga$ with $\gga$ we obtain
\be
\label{tr2}
((n+m+1)(n+m) -2) \go_{\gb\gb}{}^{\gga(n)}= 4n \delta^\gga_\gb \go_{\delta\gb}
{}^{\delta \gga(n-1)} +n(n-1)\delta^\gga_\gb\delta^\gga_\gb
\go_{\delta\delta}
{}^{\delta \delta\gga(n-2)}\,.
\ee
One more contraction gives
\be
\label{tr3}
(n+m-2)\go_{\beta\delta}
{}^{ \delta\gga(n-1)} =(n-1)\delta^\gga_\gb
\go_{\delta\delta}
{}^{\delta \delta\gga(n-2)}\,.
\ee
Now one observes that since a $\bnu^+$-exact 1-form has the form
$
 e^{\ga\ga} \f{\p^2}{\p y^\ga \p y^\ga} \xi(y),
$
a cohomology class can be fixed by setting
$
\go_{\delta\delta}
{}^{\delta \delta\gga(n-2)}=0.
$
Then from (\ref{tr3}) and (\ref{tr2}) it follows that
$\go (y)=0$, \ie $H^1 (\bnu^+ ,P)=0$.
The proof that $H^2 (\bnu^+,P)=0$ is analogous. $\square$

{}From {\it Lemma 1}  follows\\
{\it Lemma 2:}
$H^p(t_- ,P)=H^p( \bar{t}_- ,P )=0$ at $p=0,1,2$.\\
{\it Proof:} Consider the sector of $p$-forms $\go_0$ expandable
into the wedge product of $p$ 1-forms $e^{\ga\gb}$. From
the condition that $\omega$ is
$t_- $ closed it follows that $\go_0$ is $\bnu^+$ closed.
By {\it Lemma 1} it follows that $\go_0$ is $\bnu^+$ exact.
Therefore, one can
choose a representative of $H^p( t_- ,P )=0$ with $\go_0=0$
in the purely $e^{\ga\gb}$ sector.
Then one considers the sector of $p$-forms $\go_1$ that contains  $p-1$
 $e^{\ga\gb}$, repeating the analysis.
The process continues till one proves that $H^p( t_- ,P )=0$.
Analogously one proves that $H^p( \bar{t}_- ,P)=0$.
$\square$

{\it Corollary:}
{}As a consequence of {\it Lemma 2} it follows that $H^p(\tau_- ,P)$ is
concentrated in the subspace with $G\leq 1 $
where $\nu^+$ and $\bnu^+$ do not act independently.

Note that the case with $n-\bar{n}=\pm 2$ is still nontrivial
because here both $Im\,t_-$ and $Im\,\bar{t}_-$
belong to the space with $G=0$ and can cancel each other thus
extending $Ker\,\tau_-$ compared to $Ker\,t_-\oplus
Ker\, \bar{t}_-$. (Analogous phenomenon occurred in
the $4d$ analysis of the bosonic sector of $H^2(\gs^\prime_-)$
in the end of Subsection \ref{1f}.)

\subsubsection{$H^0(\tau_-)$}

$H^0(\tau_-)$ is easy to compute. Here the key observation
is that the sectors with $n-\bar{n}=\pm 2$ do not talk to each other.
As a result, $H^0(\tau_- )$ is described by the same formula
(\ref{h0t-}) as in the $4d$ case, \ie  the
true gauge symmetry parameters in the matrix space
are described by the same set of multispinors
as in Minkowski space.
In particular, all gauge parameters
$
\epsilon{}^{\alpha_1\ldots\alpha_n}{}_,{}^{{\pb}_1
\ldots{\pb}_m}(X)
$
with $|n-m|>1$ are of Stueckelberg type with the
field transformations that contain algebraic shifts
gauging away some  components of the HS 1-form connections.

\subsubsection{$H^1(\tau_-)$}

The computation of $H^1(\tau_-)$ is also
 based on the fact that it is concentrated in the
subspace where $\tau_-$ is identically zero,
{ \it i.e.,} $|n-\bar{n}|\leq 1$.

In the bosonic case, it is necessary to
check that the possible extension of $H^1(\tau_-)$ due to
extension of $Ker \, \tau_-$ compared to
$Ker \, t_- \oplus Ker \, \bar{t}_-$ does not
take place. The proof of this fact, which
 is elementary but somewhat lengthy, we leave to the reader.
 The idea of the proof is illustrated below by the analysis
 of $H^2 (\tau_-$) which results in nontrivial cohomology.

Just as in the $4d$ case, the ambiguity in adding $\tau_-$-exact
1-forms is used to get rid
of Stueckelberg components
of the 1-forms of the form $e^{\ga\pa}\go_{\ga\pa} (y,\bar{y})$.
As a result,
\be\nn
H^1(\tau_-):\qquad
\go(y,\bar{y}) = e^{\ga\gb}\go_{\ga\gb} (y,\bar{y})+
e^{\pa\pb}\overline{\go}_{\pa\pb}(y,\bar{y})+
e^{\ga\pb}\go_{\ga\pb}(y,\bar{y})\,,
\ee
where $\go_{\ga\gb} (y,\bar{y})$ and $\overline{\go}_{\pa\pb}(y,\bar{y})$
are arbitrary fields with $|n-\bar{n}| \leq 1$
while $e^{\ga\pb}\go_{\ga\pb}(y,\bar{y})$ has the same
content (\ref{bh14}) and (\ref{hf1}) as in the $4d$ theory.
We conclude that, apart from the $4d$ Fronsdal gauge fields, the
formulation in $\M_4$ requires the additional fields
$\go_{\ga\gb} (y,\bar{y})$ and $\overline{\go}_{\pa\pb}(y,\bar{y})$,
that describe components of the 1-form connection along the spinning
directions in $\M_4$. Since the system in $\M_4$ is by construction
equivalent to that in Minkowski space, the additional
dynamical fields in $\M_4$ are related by their
field equations  to the $4d$ HS fields.

To figure out the form of nontrivial field equations we have to
compute $H^2 (\tau_-)$. We start with the simpler  fermionic
case.

\subsubsection{$H^2(\tau_-)$: fermions}

 $H^{2\,fer}(\tau_-)$ consists of the
2-forms $R^{\pm}= e^{AB}\wedge e^{CD} R_{AB\,,CD}(y,\bar{y})$
with $n-\bar{n}=\pm 1$ (which are all $\tau_-$-closed) modulo
$\tau_-$-exact 2-forms
\be
\label{fex}
R^{exact\,+ }=t_- W^+\,,\qquad
R^{exact \,-}=\bar{t}_- W^-\,,
\ee
where $W^\pm$ are arbitrary 1-forms such that $(n-\bar{n})W^\pm=\pm3 W^\pm$.

In the $e^{\ga\pa}\wedge e^{\gb\pb}$ sector the analysis
repeats that of the $4d$ case.
The nontrivial class is represented by the Einstein
cohomology and  Weyl cohomology. The Weyl
cohomology is represented by the $\bar{y}$ independent 0-form
$C^{+3/2}_{\ga\gb\gga}y^\ga y^\gb y^\gga $ and its conjugate
 that describe spin $3/2$ (cf.  (\ref{CON1}))
Thus, this part of $H^{2\,fer\,+}$ is
\bee
\label{hf0}
H^{2\,fer\,+}_0  &=&
\overline{H}^{ \pa\pb} \left (
\f{\p^2}{\p \bar{y}^\pa \p \bar{y}^\pb} \phi_{-1} (y,\bar{y})
+ \bar{y}_\pa
\f{\p}{  \p \bar{y}^\pb} \phi_{+2} (y,\bar{y})\right )\nn\\
&+&H^{\ga\gb} \left (
 {y}_\ga   {y}_\gb \phi_{-3} (y,\bar{y})+
\f{\p^2}{\p y^\ga \p y^\gb }
C^{+3/2}(y)\right )\,,
\eee
where $(n-\bar{n})(\phi_{\pm i} (y,\bar{y})) =
\pm \phi_{\pm i} (y,\bar{y})$ and the 2-forms $H^{\ga\gb}$ and
$\overline{H}^{ \pa\pb}$ are defined in (\ref{H}).
$H_0^{2\,fer\,-}$ is given by the complex
conjugated expression.

Important difference  compared to the $4d$ case is that,
by {\it Lemma 1}, the Weyl 0-forms associated
with spins $s> 3/2$ do not correspond to a nontrivial
cohomology in the case of $\M_4$.
 This means that although the generalized Weyl tensors
appear on the \rhss of (\ref{CON31}), in $\M_4$ this
is a consequence of Bianchi identities applied to the lower spin
field equations. This conclusion is consistent with the fact
\cite{BHS} that, in $\M_4$, different spins correspond to
modes of the hyperfields $C(X)$ and $C_A(X)$
with respect to extra spinning directions. In other words,
it is not possible to restrict to zero $C(y|X)$ of some power in $y$
without restricting its dependence on the spinning
coordinates.

Having fixed the representatives $H^{2\,fer\,+}_0$ and
$H^{2\,fer\,-}_0 $ in the form (\ref{hf0}) and complex
conjugated, we have fixed the part of $\tau_-$--exact 2-forms
(\ref{fex}) with $W_0^\pm = e^{\ga\pa}W_0{}^\pm_{\ga\pa}$.
The other components $W_1^\pm =
e^{\ga\gb}W_1{}^\pm_{\ga\gb} + e^{\pa\pb}W_1{}^\pm_{\pa\pb}$
remain to be factored out in
the analysis of the remaining sectors of $H^{2\,fer}(\tau_-)$.
The analysis of the sectors  $e^{\ga\gga}\wedge e^{\gb\pb}$ and
$e^{\pb\pa}\wedge e^{\ga\gga^\prime}$ amounts effectively to the
$4d$ analysis of $H^1(\gs_-^\prime)$ where the leftover components
of the 1-forms $e^{\ga\gb}W_1{}^\pm_{\ga\gb}$ and
$e^{\pa\pb}W_1{}^\pm_{\pa\pb}$
 are treated as the 0-form Stueckelberg parameters. As a
result, this part of $H^{2\, fer}(\tau_-)$ is
\be
\label{hf2}
H^{2\,fer\,+}_1=
e^{\ga\pb}\f{\p^2}{\p y^\ga \p \bar{y}^\pb}\Psi^+_1 (y,\bar{y}) +
e^{\ga\pb}y_\ga\f{\p}{ \p \bar{y}^\pb}\Psi^-_2 (y,\bar{y}) +
e^{\ga\pb} y_\ga  \bar{y}_\pb\Psi^{+}_3 (y,\bar{y})\,,
\ee
where
$\Psi_i^\pm (y,\bar{y})=e^{\ga\gb} \Psi_i{}^\pm_{\ga\gb}(y,\bar{y}) +
\Psi_i{}^\pm_{\ga\gb}  e^{\pa\pb}(y,\bar{y}),
$  and analogously for the complex conjugated.

Finally, the sectors of $e^{\ga\gb}\wedge e^{\gga\delta}$ and
$e^{\pa\pb}\wedge e^{\gga^\prime \delta^\prime}$ fully belong
to $H^2(\tau_-)$
\be\nn
H^{2\,fer\,\pm}_2=e^{\ga\gga} \wedge e_{\gga}{}^\gb
\chi^\pm_{1\,\ga\gb}(y,\bar{y}) +
e^{\pa\gga^\prime} \wedge e_{\gga^\prime}{}^\pb
\chi^\pm_{2\,\pa\pb}(y,\bar{y}).
\ee

 To summarize, in the fermionic case
$
H^{2\,fer}(\tau_-) = H^{2\,fer\,+}(\tau_- )\oplus H^{2\,fer\,-}(\tau_- )\,,
$
where
$H^{2\,fer\,+}(\tau_- )$ and $H^{2\,fer\,-}(\tau_- )$ are
complex conjugated and
\be\nn
H^{2\,fer\,\pm}(\tau_- )=H^{2\,fer\,\pm}_2
\oplus H^{2\,fer\,\pm}_1 \oplus
H^{2\,fer\,\pm}_0\,.
\ee
All components of $H^{2\,fer}(\tau_-)$ except for the
$C$--dependent Weyl cohomology term
in (\ref{hf0}) correspond to the \lhss of differential
field equations imposed by the unfolded equations (\ref{CON32})
and (\ref{CON31}) on the dynamical HS gauge fields in $\M_4$.
Because $H^{2\, fer}(\tau_-)$ is concentrated at
the lowest grade, all these equations are of first-order. Their explicit
form results from projecting to zero the components of
the HS field strengths 2-forms in $\M_4$ that belong to
$H^{2\,fer}(\tau_-)$.

\subsubsection{$H^2(\tau_-)$: bosons}

In the bosonic case nontrivial
cohomology can be concentrated
in the sectors with $G=0$ or $1$.
Let us start with the case of $G=0$, denoting this part of
the cohomology $H^{2\, bos}_0 (\tau_-)$.

In the $4d$ analysis, the $H^2_0(\gs^\prime_- )$ was zero.
 The meaning of this result for spin $s> 1$ was that the 1-form gauge connections
$\go(y,\bar{y})$ with $(n-\bar{n}) \go(y,\bar{y})=\pm 2\go(y,\bar{y})$ were
auxiliary fields expressed by the ``zero torsion condition"
$R(y,\bar{y})|_{G=0}=0$ via derivatives of the dynamical fields
module pure gauge Stueckelberg components.
In particular, in the spin two sector,
the fact that $H^2_0(\gs^\prime_- )=0$
makes it possible to impose the zero-torsion condition in
$4d$ gravity.

The situation in  $\M_4$ is different. The part
of $H_0^{2\, bos}(\tau_-)$ in the sector of $e^{\ga\pa}\wedge
e^{\gb\pb}$ is still zero as in four dimensions but the other sectors
are non-zero. Analogously to the fermionic case,
 it is easy to see, that fixing to zero
the cohomology class in the sector of $e^{\ga\pa}\wedge e^{\gb\pb}$
implies that the nonzero class of $H^{2\, bos}_0(\tau_-)$ in the
sectors of $e^{\ga\gga}\wedge e^{\gb\pb}$ and
$e^{\pa\pb}\wedge e^{\gga\gga^\prime}$ has the form analogous to
that of $H^1(\gs^\prime_-)$ (\ref{bh14})
with the dynamical fields $\phi$ and $\phi^\prime$ replaced by
1-forms with coefficients in $e^{\ga\gb}$ and $e^{\pa\pb}$. Finally,
the part of $H^{2\, bos}_0(\tau_-)$ in the sector free of $e^{\ga\pa}$ remains
unrestricted. To summarize,
\be\nn
H^{2\, bos}_0(\tau_-) = \rho (y,\bar{y}) +
e^{\ga\pb}\f{\p^2}{\p y^\ga \p \bar{y}^\pb} W (y,\bar{y}) +
e^{\ga\pb} y_\ga  \bar{y}_\pb W^{\prime} (y,\bar{y})\,,
\ee
where
\be\nn
\rho (y,\bar{y})= e^{\ga\gga}\wedge e^\gb{}_\gga \rho_{\ga\gb}(y,\bar{y})
+e^{\pa\gga^\prime}\wedge e^\pb{}_{\gga^\prime} \rho_{\pa\pb}(y,\bar{y})
+e^{\ga\gb}\wedge e^{\pa\pb} \rho_{\ga\gb\,\pa\pb}(y,\bar{y})\,,
\ee
\be\nn
W^{(\prime)} (y,\bar{y})= e^{\ga\gb}W^{(\prime)}_{\ga\gb}(y,\bar{y})
+e^{\pa\pb} W^{(\prime)}_{\pa\pb}(y,\bar{y})\,.
\ee
This part of $H^{2\, bos}(\tau_-)$ is responsible for the field equations
that relate additional field components in $H^1(\tau_-)$
to the usual $4d$ massless fields associated with $H^1(\gs^\prime_- )$.

Now let us consider the part of $H^{2\, bos} (\tau_-)$
with $G=1$ which we denote $H^{2\,bos}_1(\tau_-)$.
It is non-zero due to the phenomenon discussed in {\it Corollary}.
The analysis of the  $4d$ sector $e^{\ga\pa}\wedge e^{\gb\pb}$
 is identical to that of Subsection \ref{1f}. The
question is how this cohomology extends to $\M_4$. In
principle, it might happen that the Einstein
cohomology disappears or shrinks to a smaller vector space
in $\M_4$ which would imply that most of
the $4d$ field equations become consequences of the Bianchi
identities in the larger space-time, applied to the extended
zero-torsion condition and/or to a subsystem of the field equations.
Indeed, such a phenomenon occurred in the
analysis of the equations on the Weyl 0-forms in \cite{BHS}
where the infinite dimensional cohomology in $4d$
shrinks to a finite dimensional one in $\M_4$
 in agreement with the fact that the infinite set of
$4d$ field equations (\ref{hol}) and (\ref{kg4})
for massless fields of all spins amounts to
the finite system of equations (\ref{kg}) and (\ref{dir})
in $\M_4$. This does not
happen to the Einstein cohomology, however. Namely it is
still parameterized by $E(y\,,\by)$ and $E^\prime (y\,,\by )$ as
in $4d$ Minkowski space. The detailed analysis is straightforward
although somewhat annoying.
The final result is that the nonzero components $\Phi$
of $H_{1}^{2\, bos}(\tau_-)$  are
\be\nn
\Phi= \Phi^+ +\Phi^-\,,
\ee
where
\be\nn
\Phi^+ = \Phi^+_{2}+\Phi^+_1+\Phi^+_0 +\Phi^+_{4d}+
\Phi^+_{-1}+\Phi^+_{-2}\,,
\ee
\be\nn
\Phi^- = \Phi^-_{2}+\Phi^-_1+\Phi^-_0 +\Phi^-_{4d}+
\Phi^-_{-1}+\Phi^-_{-2}\,
\ee
and
\be
\label{Pein}
\Phi^+_{4d} +\Phi^-_{4d}=
\left (\overline{H}_{\pa\pb}
\frac{\p^2}{\p \by_\pa \p \by_\pb} -
{H}^{\ga\gb}
\frac{\p^2}{\p y^\ga \p y^\gb}\right ) E(y\,,\by)+
\left (H_{\ga\gb}y^\ga y^\gb-
\overline{H}^{\pa\pb} \by_\pa \by_\pb \right)
E^\prime(y\,,\by)\,,
\ee
\bee
\label{p1+}
\Phi_1^+ &=& e^{\gga\pb}\wedge e^{\ga\gb}\by_\pb y_\gga y_\ga y_\gb
\frac{2}{n+4}E^\prime +
e^\gb{}_\pb \wedge e^\ga{}_\gb \frac{\p}{\p \bar{y}_\pb } y_\ga
\frac{4}{n} E^\prime\nn\\
&{}&+
e^\gga{}_\pb \wedge e^{\ga\gb} \frac{\p^2}{\p y^\gga \p \by_\pb}
y_\ga y_\gb \frac{2}{n(n+1)(n+2)} (2 (n+2) E^\prime -
n(n-1) E) \nn\\
&{}&+
e_{\ga\pb} \wedge e^\gga{}_\gb y^\ga y^\gb \frac{\p^2}{\p \by_\pb
\p y^\gga} \frac{2(n-1)}{n(n+1)(n+2)}((n+2) E^\prime + n E)\,,\nn
\eee
\bee
\Phi^-_1 &=&-e^\gga{}_\pb \wedge e^{\ga\gb}
\f{\p^4}{\p y^\gga \p y^\ga \p y^\gb \p \by_\pb} \f{2}{n-2} E
+e^\ga{}_\gga \wedge e^{\gga\pb} \by_\pb \f{\p}{\p y^\ga}\f{4}{n+2} E
\nn\\
&{}&
+e^{\gga\pb} \wedge e^{\ga\gb}y_\gga \by_\pb \f{\p^2}{\p y^\ga \p y^\gb}
\frac{2}{n(n+1)(n+2)} ( (n+2)(n+3) E^\prime + 2n E)\nn\\
&{}&
+e^{\ga\pb}\wedge e^\gga{}_\gb \f{\p^2}{\p y^\ga \p y^\gga}
\by_\pb y^\gb \f{2(n+3)}{n(n+1)(n+2)}(n E +(n+2) E^\prime)\,,\nn
\eee
\bee
\Phi^+_{-1} &=&-2 e^\gga{}_\pga \wedge \bar{e}_{\pa\pb}
\f{\p^4}{\p y^\gga \p \by_\pga \p\by_\pa \p\by_\pb}\f{1}{n-2} E
+\bar{e}_{\pb\pga}\wedge e^{\gga\pga} y_\gga \f{\p }{\p \bar{y}_\pb}
\f{4}{n+2} E\nn\\
&{}&
+e^{\gga \pa}\wedge \bar{e}_{\pb\pga} y_\gga \bar{y}_\pa
\f{\p^2}{\p \by_\pb \p \by_\pga }\f{2}{n(n+1)(n+2)} (2 n E +
(n+2)(n+3) E^\prime)
\nn\\
&{}&+e_{\ga\pb} \wedge e^{\pga}{}_\pa \f{\p^2}{\p\by_\pb \p \by_\pa}
y^\ga \by_\pga \f{2(n+3)}{n(n+1)(n+2)}(n E +(n+2)E^\prime )\,,\nn
\eee
\bee
\Phi^-_{-1}&=&e^{\gga \pga}\wedge e^{\pa\pb} y_\gga\by_\pga \by_\pa\by_\pb
\f{2}{n+4} E^\prime -e^\gb{}_\pb \bar{e}^{\pa\pb} \by_\pa
\f{\p}{\p y^\gb}\f{4}{n} E^\prime\nn\\
&{}&
+e^\ga{}_\pga \wedge e^{\pa\pb}\f{\p^2}{\p y^\ga \p \by_\pga} \by_\pa \by_\pb
\f{2}{n(n+1)(n+2)}(2(n+2)E^\prime -n(n-1) E)\nn\\
&{}&
+2e^{\ga\pb}\wedge e^{\pga}{}_{\pa}\by_\pga \by_\pb
\f{\p^2}{\p y^\ga\p \by_\pa}
\f{n-1}{n(n+1)(n+2)} (nE +(n+2) E^\prime)\,,\nn
\eee
\be\nn
\Phi^+_2 = e^\ga{}_\gga \wedge e^{\gb\gga}
y_\ga y_\gb \f{n-1}{(n+1)(n+2)}((n+2)E^\prime +n E )\,,
\ee
\be\nn
\Phi^-_{2} = -e^\ga{}_\gga \wedge e^{\gb\gga} \f{\p^2}{\p y^\ga \p y^\gb}
\f{n+3}{n(n+1)}((n+2)E^\prime +n E )\,,
\ee
\be\nn
\Phi^+_{-2}=\bar{e}_\pa{}_\pga\wedge \bar{e}_\pb{}^\pga
\f{\p^2}{\p \by_\pa \p \by_\pb}
\f{n+3}{n(n+1)}((n+2)E^\prime +n E)\,,
\ee
\be\nn
\Phi^-_{-2}=-
\bar{e}^\pa{}_\pga\wedge \bar{e}^\pb{}^\pga{ \by_\pa  \by_\pb}
\f{n-1}{(n+2)(n+1)}((n+2)E^\prime +n E)\,,
\ee
\bee
\Phi^+_0 &=& 2\bar{e}_{\pa\pb}\wedge e^\gga{}_\gb
\f{\p^3}{\p\by_\pa \p\by_\pb \p y^\gga} y^\gb
\f{n+3}{n(n+2)(n+1)}((n+2)E^\prime +n E)
\nn\\
&{}&+ 2e^{\ga\gb}\wedge \bar{e}^\pa{}_\pb y_\ga y_\gb
\by_\pa \f{\p}{\p \by_\pb}
\f{n-1}{n(n+2)(n+1)}((n+2)E^\prime +n E)\,,\nn
\eee
\bee
\label{p0-}
\Phi_0^-  &=& 2\bar{e}^{\pa\pb}\wedge e^\gga{}_\gb
\by_\pa \by_\pb y^\gb \f{\p}{\p y^\gga}
\f{n-1}{n(n+2)(n+1)}((n+2)E^\prime +n E)\nn\\
&{}&+2e^{\ga\gb}\wedge \bar{e}^\pga{}_\pa \by_\pga
\f{\p^3}{\p y^\ga \p y^\gb \p \by_\pa}
\f{n+3}{n(n+2)(n+1)}((n+2)E^\prime +n E)\,,\nn
\eee
where $n$ is defined in (\ref{Gnbn}). (Note that the signs
in (\ref{Pein}) differ from those in (\ref{ein}) because of
the sign difference in the definition of $\gs_-^\prime$
(\ref{gsp}) and $\tau_-$ (\ref{tm})).

To impose HS field equations one has to set
$E$ and $E^\prime$  to zero.
The structure of $H_1^{2\, bos}(\tau_-)$
suggests that this can be done in a variety of equivalent
ways by setting to zero any of the components of the
curvature 2-forms that contain $E$ and $E^\prime$.  All differently looking
field equations imposed by setting to zero different
components of the curvatures proportional to $E$ and $E^\prime$
are equivalent by virtue Bianchi identities along with
the grade zero field  equation $R_0=0$, $G(R_0)=0$.

Let stress that the $\gs_-$ cohomology analysis in $Sp(4,\mR)$
is identical to that in $\M_4$ because the form of $\gs_-$
at any given point does not change. This means in particular that
one has as many symmetries, dynamical fields and field equations
in $Sp(4,\mR)$ as in $\M_4$. The precise form of the field
equations may of course be different because of the appearance of
the $\gs_+$ operators as discussed in Section \ref{w0f}.

\section{Conclusion}
\label{con}

The main result of this paper is that free equations for
massless fields of all spins in $AdS_4$
admit $sp(8,\mR)$ covariant formulation not only in terms
of gauge invariant field strengths \cite{BHS},
but also in terms of the gauge potentials.
The key point is that the formulation
is well-defined in the $AdS_4$ background but experiences
certain degeneracy in the flat limit that does not allow
$sp(8,\mR)$ and conformal invariant formulations of spin $s>1$
gauge fields in flat Minkowski space. There are two
alternatives for the flat limit procedure. One leads to
standard flat space massless field equations but blowing up special
conformal symmetry transformations. Another one keeps the
conformal transformation well defined, but the
limiting flat space field equations is hard to interpret.

The formulation in terms of gauge potentials is needed to
reach the $sp(8,\mR)$ covariant formulation at the action level
and/or at the interaction level (even on-shell).
The obtained results provide the starting point for the
$sp(8,\mR)$ covariant study of the interacting HS theory.
That $sp(8,\mR)$ symmetry may play a
role in the nonlinear HS theory has been already
observed in \cite{BHS} in the relation with the doubling of
auxiliary spinor variables in the nonlinear HS field equations
of \cite{more, Gol}.
However, the formalism developed in this paper is different
from that of \cite{more, Gol} because physical degrees
of freedom are described here by Fock modules rather than by
twisted adjoint module as in \cite{more,Gol}. In fact, the
proposed formulation is close to that of the $2d$
HS model of \cite{d2} that was also formulated in terms of Fock
modules. It is tempting to extend this analogy
to the $sp(8,\mR)$ invariant formulation of $4d$ HS theory.

An interesting direction for the future investigation
is to study the models in the matrix spaces $\M_M$ with $M>4$.
Note that the  dynamics of $6d$ and $10d$ conformal
free massless fields in terms of gauge invariant
field strengths has been understood \cite{BLS,BHS,Mar,BBAST}
in the matrix spaces $\M_{8}$ and $\M_{16}$, respectively.
The results of this paper indicate how
this theory can be extended to the level of gauge fields,
although the details remain to be elaborated.

Surprisingly, the obtained results may even shed some light on the
structure of conventional field-theoretical models like gravity.
In particular, we have shown that linearized gravity in $AdS_4$
exhibits the $gl(4,\mR)\sim o(3,3)$ symmetry. It
would be interesting to see whether and how the $gl(4,\mR)$
symmetry extends to the nonlinear gravity in $AdS_4$. A promising
starting point is the MacDowell-Mansouri action \cite{MM}.

An interesting feature of the proposed model is that it has
manifest EM duality symmetry along with its HS generalization
as the $u(1)$ part of $sp(8,\mR)$. In the unreduced model, that
describes two infinite
sets of massless fields of all spins, this symmetry rotates
two species of a spin $s$ gauge field as a complex field. The reduction
to the system of massless fields in which every
spin appears once also respects the duality transformation
because it still has two sets of
the gauge fields that are dual to one another by virtue of the
field equations. Hopefully, the proposed formulation
may be helpful for the further analysis of duality in the
models that contain gravity like those studied
in \cite{DT,Nieto,HT,JLR,DS,LP}.

The main tool for the study of HS fields applied in this paper
is the unfolded formulation which is a covariant first-order
reformulation of a dynamical theory in any dimension \cite{Ann,un}.
The unfolded formulation is perfectly suited for elucidation
of symmetries, dynamical content and  equivalent formulations
of a theory, including those in extended (super)spaces
with extra (super)coordinates. In this paper, the unfolded
formulation insures the $sp(8,\mR)$ invariance,
determines the precise form of $sp(8,\mR)$ field transformations
and allows the straightforward extension of the
$sp(8,\mR)$ invariant HS gauge theory in $AdS_4$
to the ten dimensional  space-time with the coordinates
$X^{AB}$ ($A,B=1,\ldots 4$), which is the group manifold
$Sp(4,\mR)$ in the $AdS$ like case.

The original ten dimensional formulation of HS theory
in terms of gauge invariant Weyl 0-forms \cite{BHS} was manifestly
covariant under $GL(4,\mR)$ transformation of the spinor indices
$A,B\ldots 1,\dots 4$
so that all coordinates $X^{AB}$  appeared on equal footing.
The manifest symmetry of the proposed extension to the
HS model formulated in terms
gauge fields turns out to be reduced to
$GL(2,\mC)\subset GL(4,\mR)$. This happens because different sectors
of fields in the theory respect different subalgebras
$gl(4,\mR)\subset sp(8,\mR)$
which have $gl(2,\mC)$ as the maximal common subalgebra.
As a result, the space-time coordinates
$X^{\ga\pa}$ and spinning coordinates $X^{\ga\gb}$ and
$X^{\pa\pb}$ have different appearance in the theory. Nevertheless,
$sp(8,\mR)$
acts geometrically  on $X^{AB}$ and remains a symmetry of
the system. Let us note that the fact that the manifest symmetry
between space-time and spinning coordinates is lost in the
full theory indicates that a generalized holonomy group
 in the space-times with matrix coordinates should be
$GL(2,\mC)$ rather than $GL(4,\mR)$ which case was tested in
\cite{BPST}.

As a first step towards a new version of nonlinear
HS theory it is interesting to check whether
the proposed formulation exhibits an infinite
dimensional conformal HS symmetry
that contains $sp(8,\mR)$ as a finite dimensional
subalgebra. As explained in Subsection \ref{ff},
one way to check this is to extend the $sp(8,\mR)$
Chevalley-Eilenberg cohomology to the full
HS symmetry. Technically, this is equivalent to the
conformal extension of the $AdS_4$ analysis of \cite{Ann}
to the first order in the Weyl 0-forms,
that accounts all HS 1-form connections.

More generally, a nonlinear extension of the free
HS theory  formulated in
this paper can go far beyond the original $4d$ HS gauge theory
we started with
just because it will allow a formulation in the ten
dimensional space-time. Actually, as observed in \cite{GV},
different types of $sp(8,\mR)$ invariant fields
in $\M_M$ are visualized as
usual fields that live in space-times of different
dimensions $1\leq 10$ so that the resulting theory
may provide a dynamical theory of different types of branes
in the ten dimensional space-time.
Let us note that, for higher rank (brane)
solutions, $sp(8,\mR)$ extends to higher $sp(2^n,\mR)$ including
$sp(32,\mR)$ and $sp(64,\mR)$ that have been argued long ago to
be symmetries of $M$ theory \cite{T,B,GM,G,FP,H}. A
deep relationship between HS theories and $M$ theory
is also indicated by the recent papers \cite{west,HK}.

\section*{Acknowledgement}
This research was supported in part by INTAS Grant No 03-51-6346,
RFBR Grant No 05-02-17654, LSS No 4401.2006.2. The author is
grateful to J.de Azcarraga,
I.Bandos, E.Skvortsov, D.Sorokin, K.Stelle and S.Theisen for useful discussions.
Also I am grateful to M.Eastwood for hospitality at IMA during
the 2006 summer programme on Symmetries and Overdetermined
Systems of Partial Differential Equations where this work was initiated and to
H.Nicolai and M.Staudacher for hospitality at AEI Max Planck Institute,
Golm, where most of this work has been done.

\end{document}